%% file: main.tex
\documentclass[10pt]{article} 
\usepackage[preprint]{tmlr}
\usepackage{bm}
\usepackage[normalem]{ulem}
\useunder{\uline}{\ul}{}
\usepackage{multirow}
\usepackage{amssymb}
\usepackage{pifont}
\usepackage{booktabs} 
\usepackage{amsmath}
\usepackage{enumitem}
\usepackage{hyperref}
\setlist[itemize]{leftmargin=*}

\usepackage{amssymb}
\usepackage{multirow}
\usepackage{graphicx}
\usepackage{amsfonts}
\usepackage{balance}
\usepackage{pifont}
\usepackage[most]{tcolorbox}
\usepackage{xcolor, colortbl}
\usepackage{wrapfig}
\usepackage{float}
\usepackage{subfig}

\newtheorem{remark}{Remark}

\usepackage{listings}
\usepackage{color}
\usepackage[table]{xcolor}
\definecolor{dkgreen}{rgb}{0,0.6,0}
\definecolor{gray}{rgb}{0.5,0.5,0.5}
\definecolor{mauve}{rgb}{0.58,0,0.82}

\newtheorem{definition}{Definition}
\newtheorem{theorem}{Theorem}
\newtheorem{corollary}{Corollary}
\newtheorem{lemma}{Lemma}

\usepackage{etoc}
\etocdepthtag.toc{mtchapter}
\etocsettagdepth{mtchapter}{subsection}
\etocsettagdepth{mtappendix}{none}

\title{Understanding Embedding Scaling in Collaborative Filtering}

\author{
\name Yicheng He \email yh84@uiuc.edu \\
\addr University of Illinois Urbana-Champaign
\AND
\name Kaiyu Zhou \email kaiyu001@e.ntu.edu.sg \\
\addr Nanyang Technological University
\AND
\name Haoyue Bai \email baihaoyue621@gmail.com \\
\addr Arizona State University
\AND
\name Fengbin Zhu \email  zhfengbin@gmail.com \\
\addr National University of Singapore
\AND
\name Yonghui Yang \email yh\_yang@nus.edu.sg \\
\addr National University of Singapore
}

\begin{document}
\maketitle
\begin{abstract}
Scaling recommendation models into large recommendation models has become one of the most widely discussed topics. Recent efforts focus on components beyond the scaling embedding dimension, as it is believed that scaling embedding may lead to performance degradation. Although there have been some initial observations on embedding, the root cause of their non-scalability remains unclear. Moreover, whether performance degradation occurs across different types of models and datasets is still an unexplored area. 
Regarding the effect of embedding dimensions on performance, we conduct large-scale experiments across $10$ datasets with varying sparsity levels and scales, using $4$ representative classical architectures. We surprisingly observe two novel phenomena: \textit{double-peak} and \textit{logarithmic}. For the former, as the embedding dimension increases, performance first improves, then declines, rises again, and eventually drops. For the latter, it exhibits a perfect logarithmic curve. Our contributions are threefold. First, we discover two novel phenomena when scaling collaborative filtering models. Second, we gain an understanding of the underlying causes of the double-peak phenomenon. Lastly, we theoretically analyze the noise robustness of collaborative filtering models, with results matching empirical observations.
\end{abstract}

\input{Alltex/1-intro}

\input{Alltex/3-prelimilaries}

\input{Alltex/4-model}

\input{Alltex/5-experiments}


\input{Alltex/6-Conclusion}



\bibliography{main}
\bibliographystyle{unsrtnat} 
\newpage

\input{Alltex/7-appendix}

\end{document}

%% file: Alltex/1-intro.tex
\section{Introduction}

Scaling up Transformer~\citep{transformer} parameters to boost performance in NLP tasks is a key reason behind the success of large language models. Therefore, a natural question arises: can we replicate similar success in collaborative filtering? It is widely agreed that embedding dimension is one of the key controllable parameters in collaborative filtering. However, there is also a consensus that scaling the embedding dimension may leads to performance degradation~\citep{lms, scalingrec1, wukong, 251, 252, 253}. For example, \cite{lms} argues: ``\textit{Worsely, increasing the embedding dimension does not sufficiently improve the performance or even hurts the model}'', and \cite{wukong} believes: ``\textit{\ldots Merely expanding the sparse component of a model does not enhance its ability to capture the complex interactions~\ldots}''. As a result, recent works~\citep{wukong,lms} on scaling recommendation models bypass the seemingly unviable path of expanding embedding dimension.


A further question arises: why do scaling embedding dimensions fail to improve recommendation performance? \cite{lms} observe that when the embedding dimensions are scaled, embedding will collapse. Specifically, they conduct a spectral analysis based on singular value decomposition of the learned embedding matrices and find that most of the singular values are extremely small. Furthermore, several studies on efficient recommendation~\cite{nas1, nas2, dim1, dim2,dim3} treat the embedding dimensionality as a learnable parameter and optimize it through predefined objectives, thereby enhancing the model's overall effectiveness. Because, they think that there exists an optimal dimensionality for embeddings. In other words, performance exhibits a single-peak phenomenon as the embedding is scaled. Overall, previous works observed that increasing the dimensionality not only worsens performance but also causes changes in the singular values of the embedding. 

In this paper, we are more interested in understanding why the performance deteriorates with scaling and what the underlying causes are. Would it follow the previously observed single-peak phenomenon?

\begin{wrapfigure}{r}{0.4\textwidth} 
  \centering

  \includegraphics[width=0.38\textwidth]{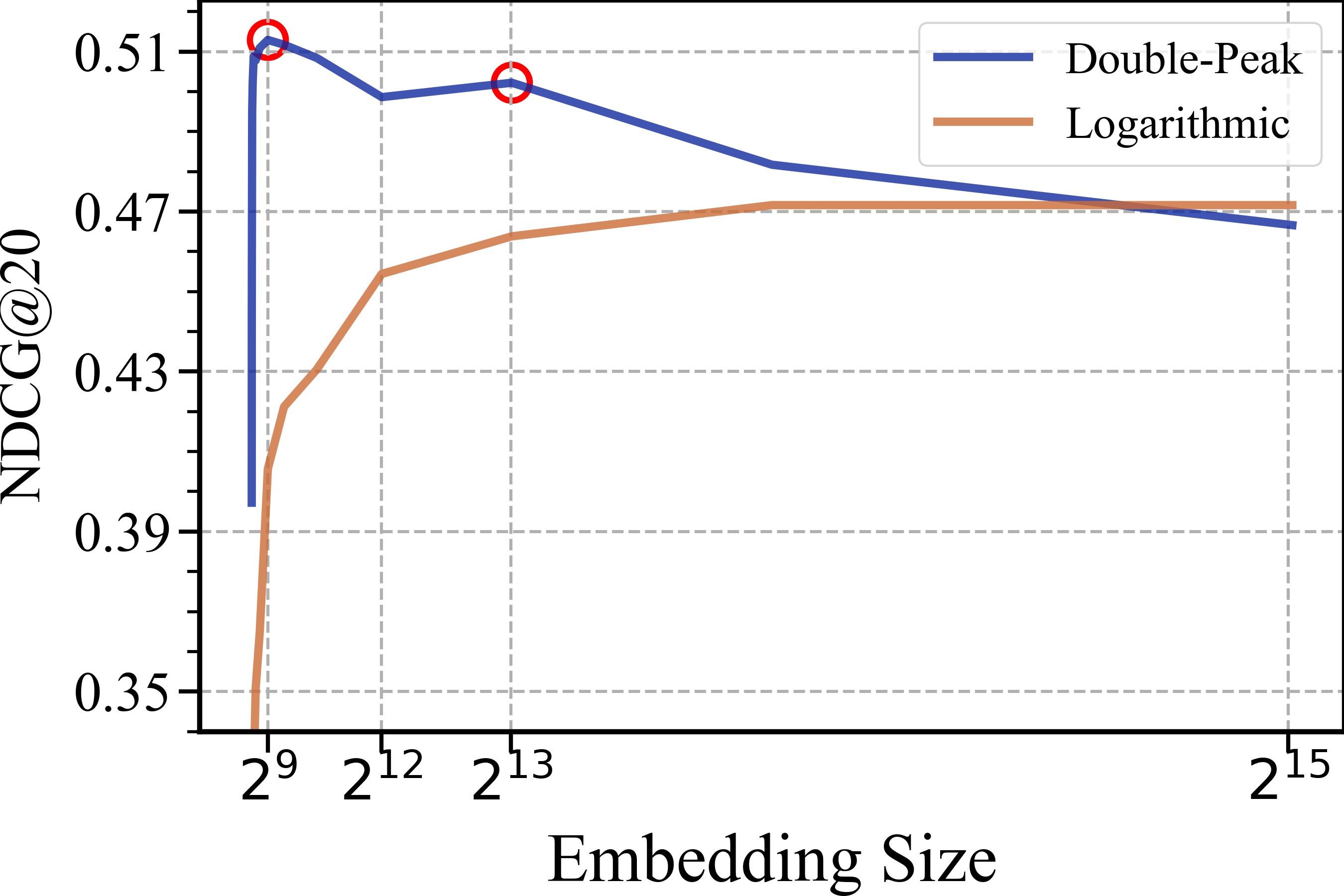}
  \caption{Double-peak: first rising, then falling, followed by another rise and decline; logarithmic, performance follows a logarithmic increase.}
\label{intro}
\end{wrapfigure}

To enable a comprehensive investigation, we conducted large-scale experiments across $10$ datasets with varying sizes and sparsity levels (Table ~\ref{dataset}), using 4 representative collaborative filtering models (reason at~\ref{reason}). Surprisingly, we observe two novel phenomena: \textit{double-peak} and \textit{logarithmic} (Figure~\ref{intro}), apart from the common single-peak. Additionally, we have two observations, for the same model, some datasets exhibit a double-peak trend, while others show a logarithmic increase. More interestingly, we also find that even within the same dataset, the phenomena are not always consistent. For example, BPR~\citep{bpr} exhibits a double-peak pattern, while SGL~\citep{sgl} follows a logarithmic trend.

Clearly, we prefer the logarithmic scaling for its expected stable performance, while the emergence of the double-peak phenomenon indicates a limitation in scalability. Since both phenomena appear across different datasets within the same model, we boldly speculate that this is due to certain negative factors in the interactions. Inspired by~\cite{sparse_dscent}, we concluded that the issue originates from the noise present in the data, and we conducted a thorough analysis of the causes of noise at each stage. We also conducted an in-depth analysis of the noise resistance and robustness of various collaborative filtering models and the results are generally consistent with the empirical evidence.




\textbf{Our contributions are summarized as follows}.
\vspace{-5pt}
\begin{itemize}
  \item To the best of our knowledge, our work is the first report of the double-peak and logarithmic phenomenon. In the realm of collaborative filtering, there has long been a perplexing question: why does expanding embedding dimensions not always result in performance enhancements? Our research make further additions to it.
  \item Through in-depth analysis, we identified interaction noise, in addition to overfitting, as the key factors causing the double-peak phenomenon. To test this idea, we propose a simple yet effective sample drop strategy that helps collaborative filtering models scale more effectively. For instance, the embedding dimensions can be expanded up to 32,768 without significant risk of collapse.
  \item We theoretically analyzed the noise robustness of BPR, NeuMF, LightGCN, and SGL, with results largely aligning with empirical observations, indirectly validating our analysis. For instance, BPR shows poor noise robustness and more frequently exhibits the double-peak phenomenon, while SGL is more robust and tends to follow a logarithmic pattern.
\end{itemize}
\textbf{The implications for future research}.
\vspace{-5pt}
\begin{itemize}
\item Transformer-based LLMs have garnered significant attention. Most of researches~\citep{wukong,meta} aim to bring the success of the Transformer architecture into recommendation models to achieve superior scaling performance. However, in this paper, we explore a different path: finding the "Transformer" within collaborative filtering models. We discovered that SGL possesses structural advantages in this regard. Therefore, future research directions may focus on better filtering clean interactions and further scaling up SGL for enhanced performance.
\end{itemize}

%% file: Alltex/3-prelimilaries.tex
\section{Background: Collaborative Filtering}
\textbf{Notations}. 
We use lowercase letters for scalars. Denote the set of real numbers by $\mathbb{R}$. We use bolded lowercase letters $\mathbf{p}$ to represent the entire embedding, and subscripts to refer to specific embedding $\mathbf{p}_u$. We use calligraphic letters e.g., $\mathcal{N}$ to denote the value spaces (other notions in Appendix~\ref{notions}). 

\textbf{User and Item Embedding}.
In collaborative filtering, the core components involve representing users and items using embedding vectors. The user embedding contains the user's preference information, which can better capture the similarity between different users. Similarly, the item embedding contains the item's feature information. Let us define: $\mathbf{p}_u \in \mathbb{R}^k$ as the embedding vector for user $u$, where $u = 1, 2, \dots, m$. $\mathbf{q}_i \in \mathbb{R}^k$ as the embedding vector for item $i$, where $i = 1, 2, \dots, n$. $k$ is the dimensionality of the embedding.

\textbf{Collaborative Filtering}.
To compute the preference score $\hat{r}_{ij}$ of user $u$ for item $i$, many powerful and representative collaborative filtering models have been proposed, such as BPR~\citep{bpr}, NeuMF~\citep{neumf}, LightGCN~\citep{he2020lightgcn} and SGL~\citep{sgl}. 

\ding{182}~{BPR}~\citep{bpr} calculates the preference score as the inner product of user and item embeddings, or employs a neural network to capture nonlinear interactions:
$$
    \hat{r}_{ij} = \mathbf{p}_u^\top \mathbf{q}_i = \sum_{l=1}^{k} u_{il} \cdot v_{jl}, \quad
    \hat{r}_{ij} = f_{\text{MLP}}(\mathbf{p}_u, \mathbf{q}_i),
$$
where $u_{il}$ and $v_{jl}$ are the $l$-th components of $\mathbf{p}_u$ and $\mathbf{q}_i$, respectively. \\

\ding{183}~{NeuMF}~\citep{neumf} is a representative deep learning-based collaborative filtering model with a classic two-tower architecture. It fuses Generalized Matrix Factorization (GMF) and Multi-Layer Perceptron (MLP) to capture both linear and nonlinear user–item interactions:
$$
    \hat{r}_{ij}^{\text{GMF}} = \mathbf{p}_u^\top \mathbf{q}_i, \quad
    \hat{r}_{ij}^{\text{MLP}} = f_{\text{MLP}}([\mathbf{p}_u \| \mathbf{q}_i]),
$$
where $[\cdot \| \cdot]$ denotes vector concatenation. The final prediction $\hat{r}_{ij}$ is computed by combining the outputs from GMF and MLP, typically through a fully connected layer.

\ding{184}~{LightGCN}~\citep{he2020lightgcn} simplifies GCN by removing feature transformation and nonlinear activation to better capture high-order connectivity in user-item graphs. It adopts symmetric normalization in the graph convolution process, aggregating embeddings from neighboring nodes using a degree-aware weighting scheme:
$$
    \mathbf{p}_u^{(l+1)} = \sum_{i \in \mathcal{N}_u} \frac{1}{\sqrt{|\mathcal{N}_u||\mathcal{N}_i|}} \mathbf{q}_i^{(l)}, \quad
    \mathbf{q}_i^{(l+1)} = \sum_{u \in \mathcal{N}_i} \frac{1}{\sqrt{|\mathcal{N}_i||\mathcal{N}_u|}} \mathbf{p}_u^{(l)}.
$$
This symmetric normalization allows the model to more accurately propagate information and better capture user interests.

\ding{185}~SGL~\citep{sgl} introduces self-supervised learning by augmenting graph structure and node features, incorporating contrastive objectives to improve robustness and alleviate sparsity: 
$$
    \mathcal{L} = \underbrace{\mathcal{L}_{\text{BPR}}}_{\text{collaborative filtering}} + \gamma \underbrace{\mathcal{L}_{\text{cont}}}_{\text{contrastive}}, \label{eq:total_loss}
$$
where $\mathcal{L}_{\text{BPR}}$ is the Bayesian Personalized Ranking loss $-\sum_{(u,i,j) \in \mathcal{D}} \ln \sigma\left( \mathbf{p}_u^\top \mathbf{q}_i - \mathbf{p}_u^\top \mathbf{q}_j \right)$ and $\mathcal{L}_{\text{contrastive}}$ is the contrastive loss encouraging consistency between augmented views. In addition, \( \gamma \) controls the contrastive task's contribution.

%% file: Alltex/4-model.tex
\section{Analysis of Scaling: Scale it up! Scale it up again!}
We focus on three points: First, what strategy we use to increase embedding size. Second, what phenomenon we observe in experiments. Third, why these phenomena happen.

\subsection{Observation Settings}
Our experiments are based on the RecBole~\footnote{https://www.recbole.io/docs/index.html}~\citep{recbole}. We select four popular collaborative filtering models: BPR~\citep{bpr}, NeuMF~\citep{neumf}, LightGCN~\citep{he2020lightgcn} and SGL~\citep{sgl}, along with ten classic datasets, covering different level of scale and sparsity. These datasets are described in the Appendix~\ref{dataset}. To ensure a thorough investigation, our experiments were conducted on a setup equipped with 8*80G GPUs. Other settings regarding all model parameters, optimizer, learning rate, etc. are in the Appendix~\ref{detail_setting}.

\subsection{Double-peak and Logarithmic Phenomenon}
\begin{figure*}[t!]
\centering
\subfloat[\textbf{ModCloth - BPR}]
{\includegraphics[width=0.32\linewidth]{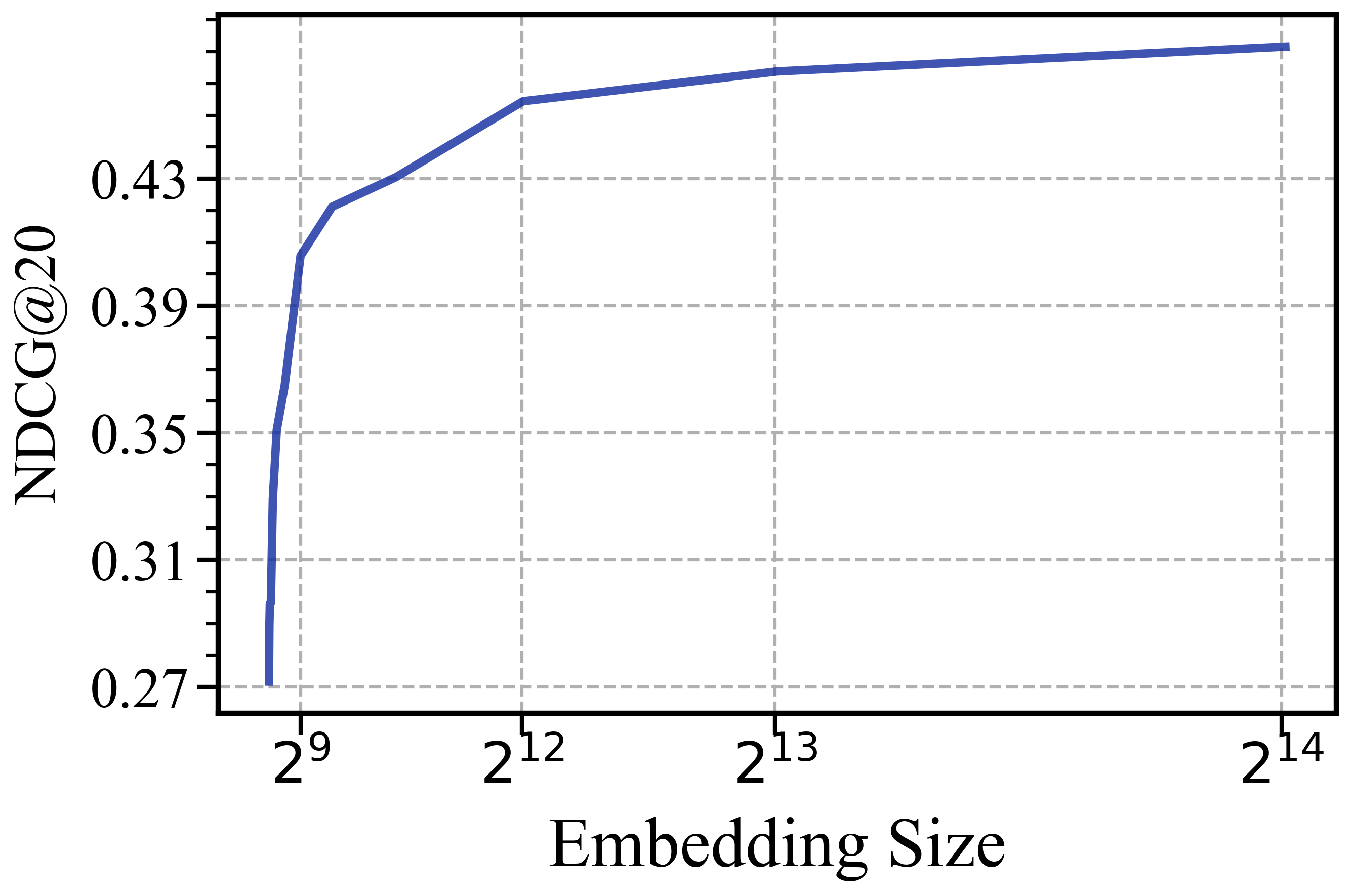}} \hfill
\subfloat[\textbf{Douban - BPR}]
{\includegraphics[width=0.32\linewidth]{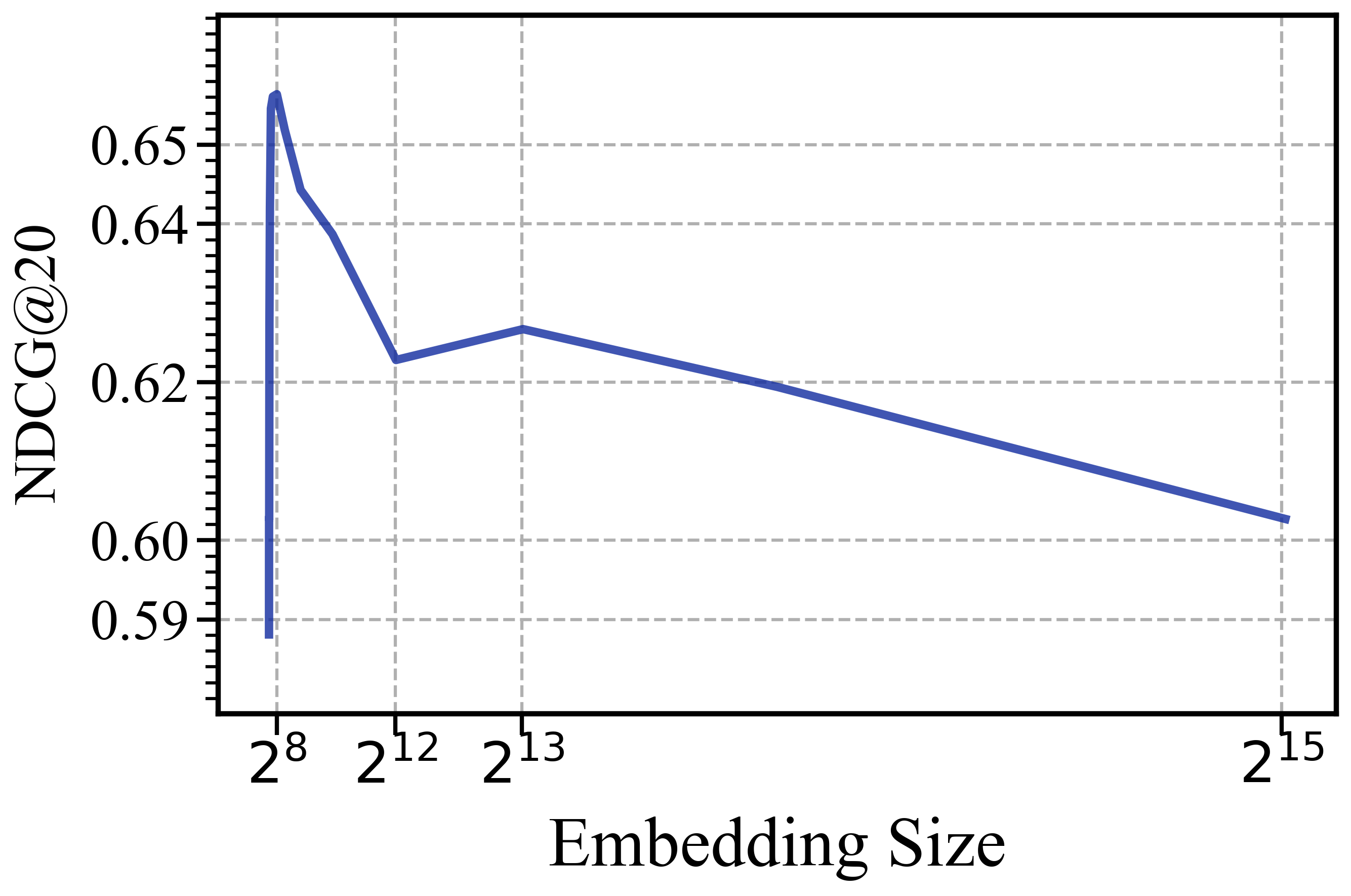}} \hfill
\subfloat[\textbf{ML-100k - BPR}]
{\includegraphics[width=0.32\linewidth]{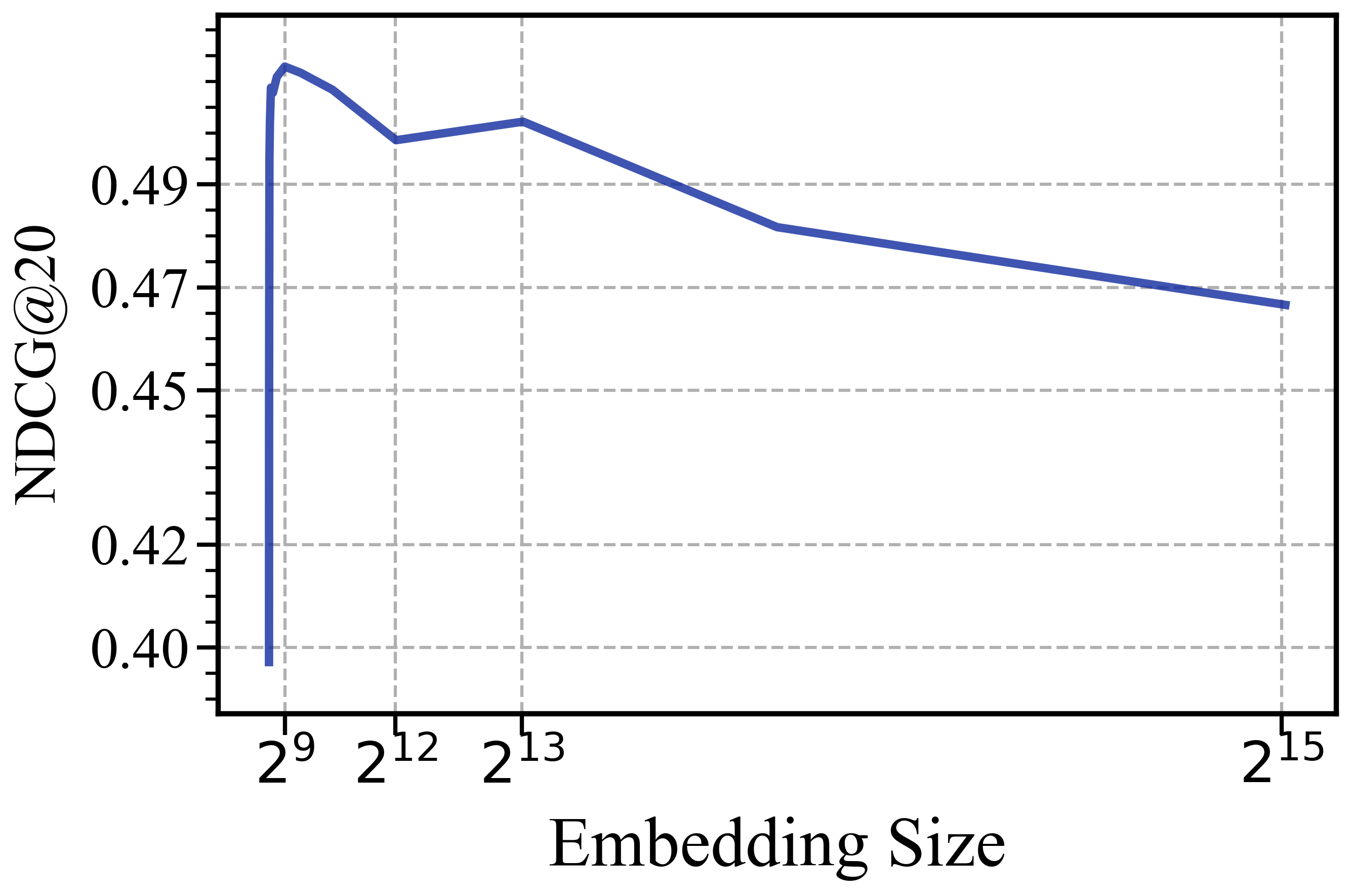}} \
\vspace{-10pt}
\subfloat[\textbf{ModCloth - NeuMF}]
{\includegraphics[width=0.32\linewidth]{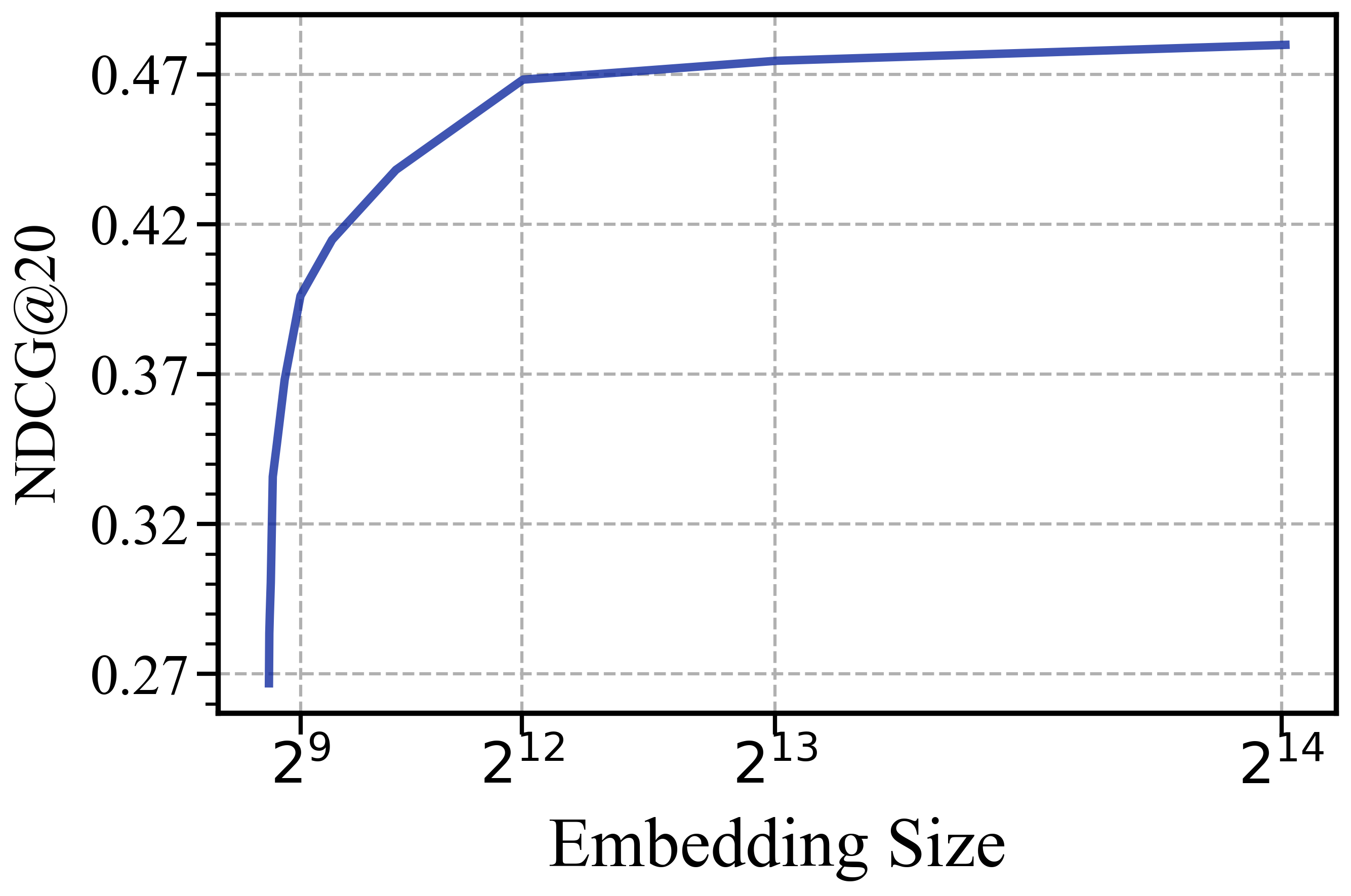}} \hfill
\subfloat[\textbf{Douban - NeuMF}]
{\includegraphics[width=0.32\linewidth]{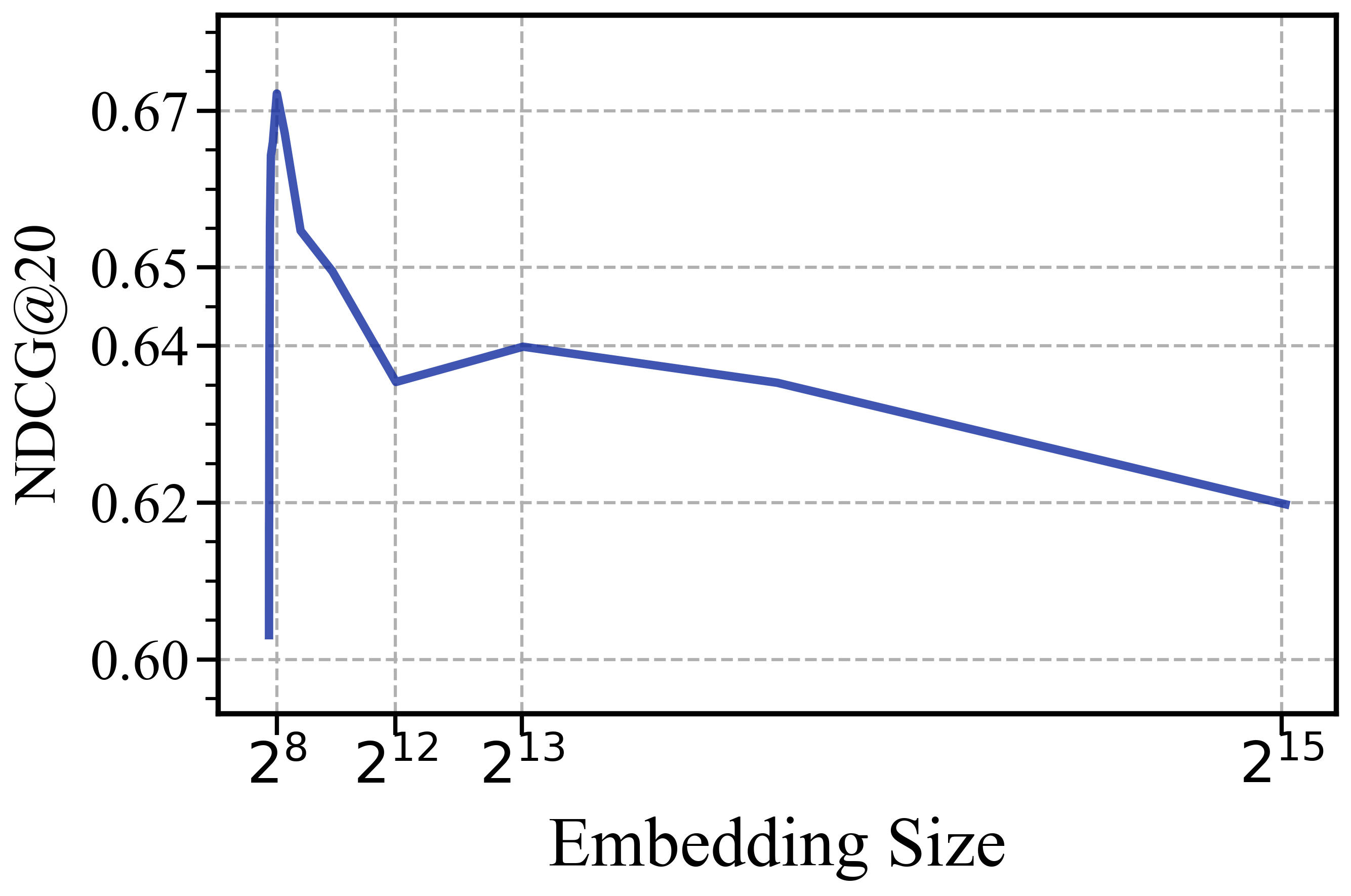}} \hfill
\subfloat[\textbf{ML-100k - NeuMF}]
{\includegraphics[width=0.32\linewidth]{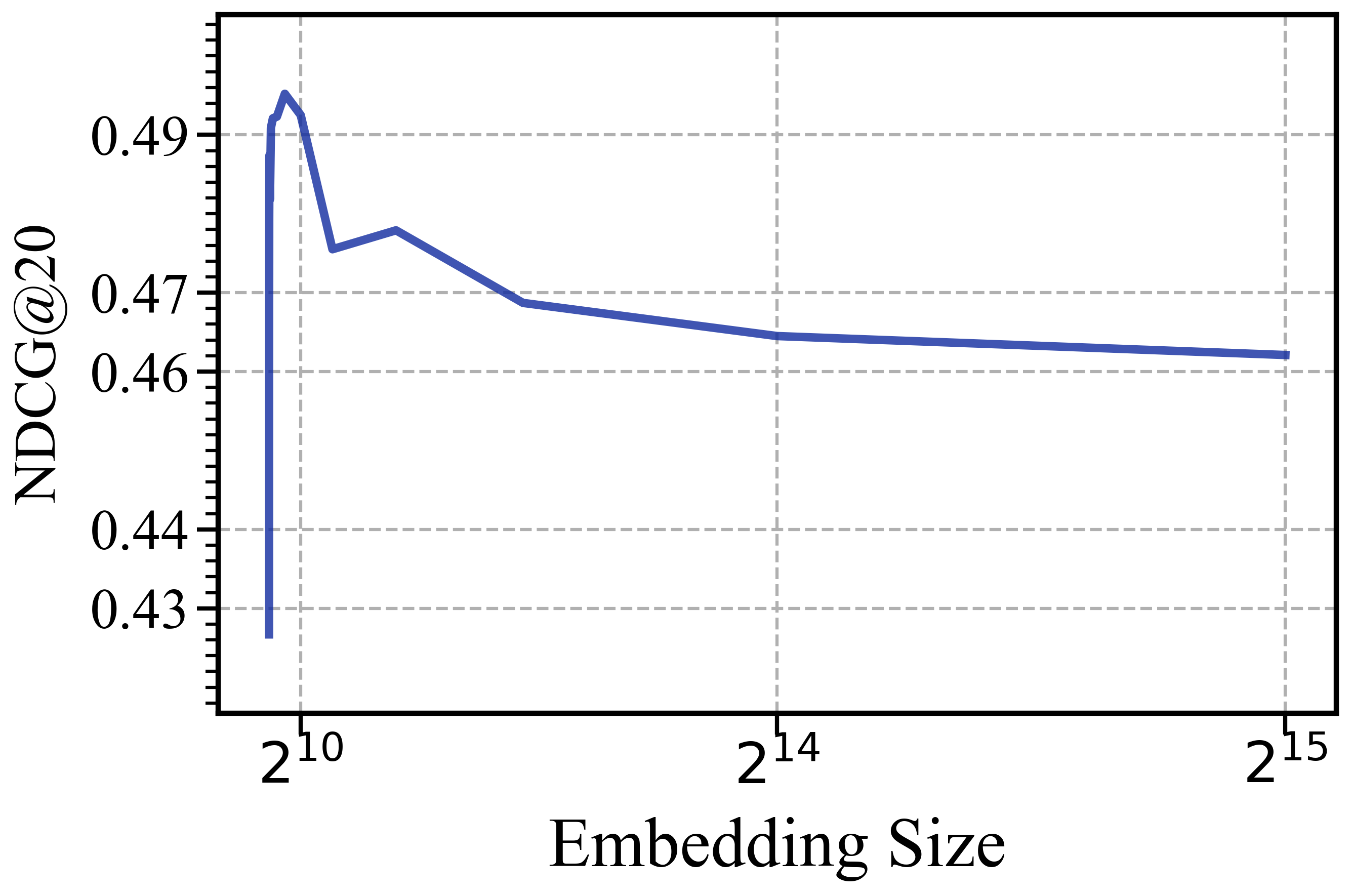}} \
\vspace{-10pt}
\subfloat[\textbf{ModCloth - LightGCN}]
{\includegraphics[width=0.32\linewidth]{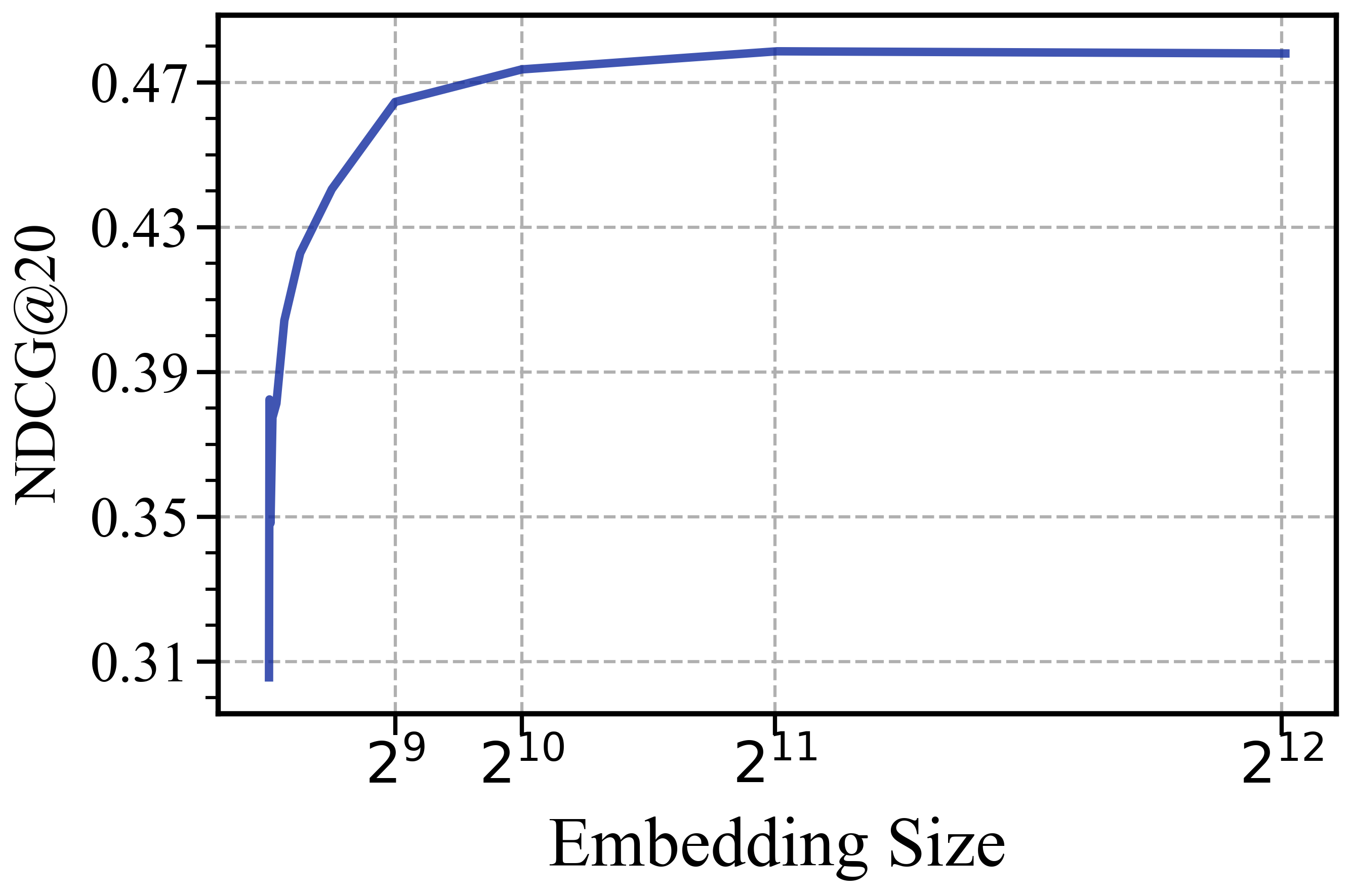}} \hfill
\subfloat[\textbf{Douban - LightGCN}]
{\includegraphics[width=0.32\linewidth]{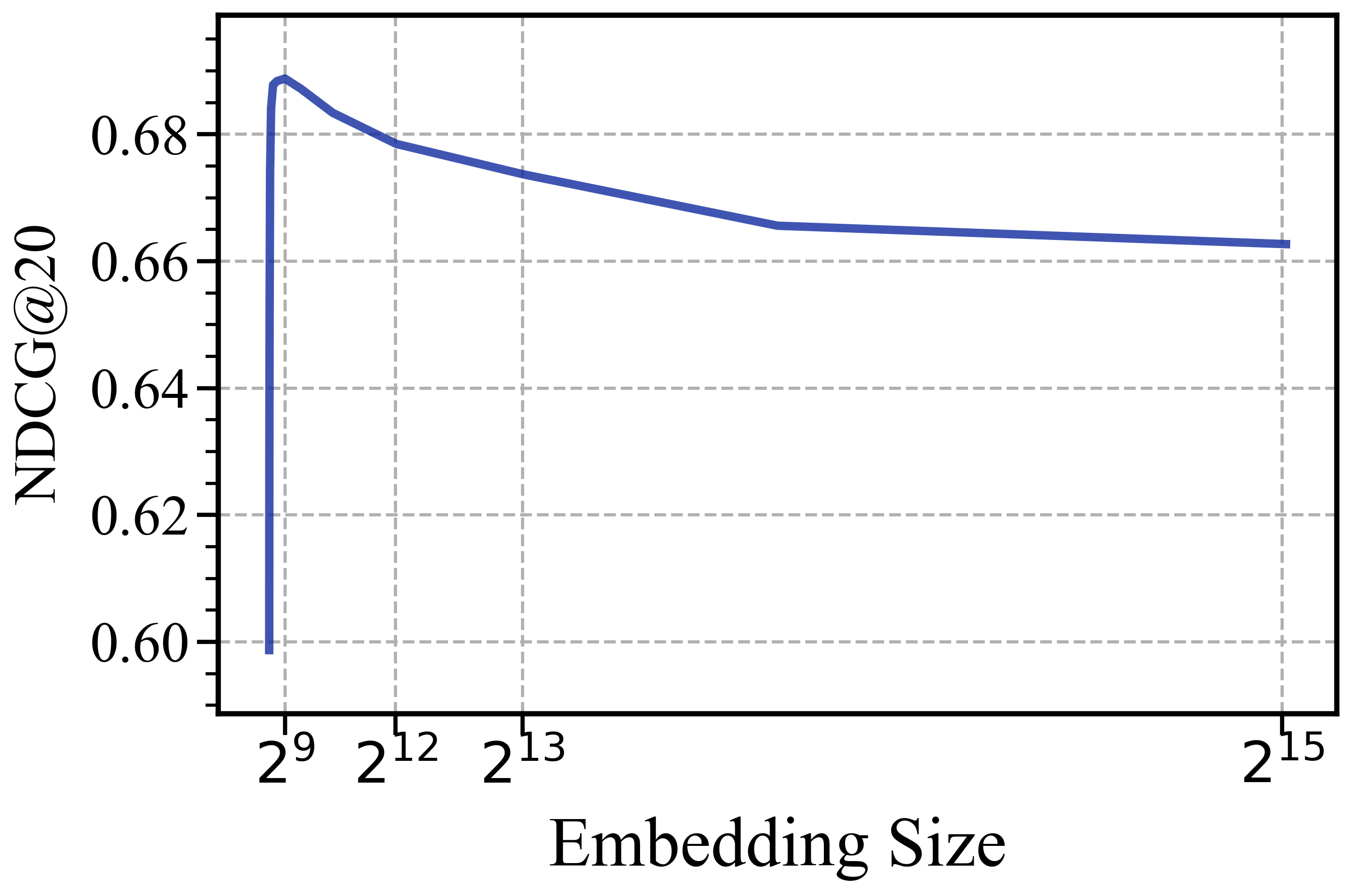}} \hfill
\subfloat[\textbf{ML-100k - LightGCN}]
{\includegraphics[width=0.32\linewidth]{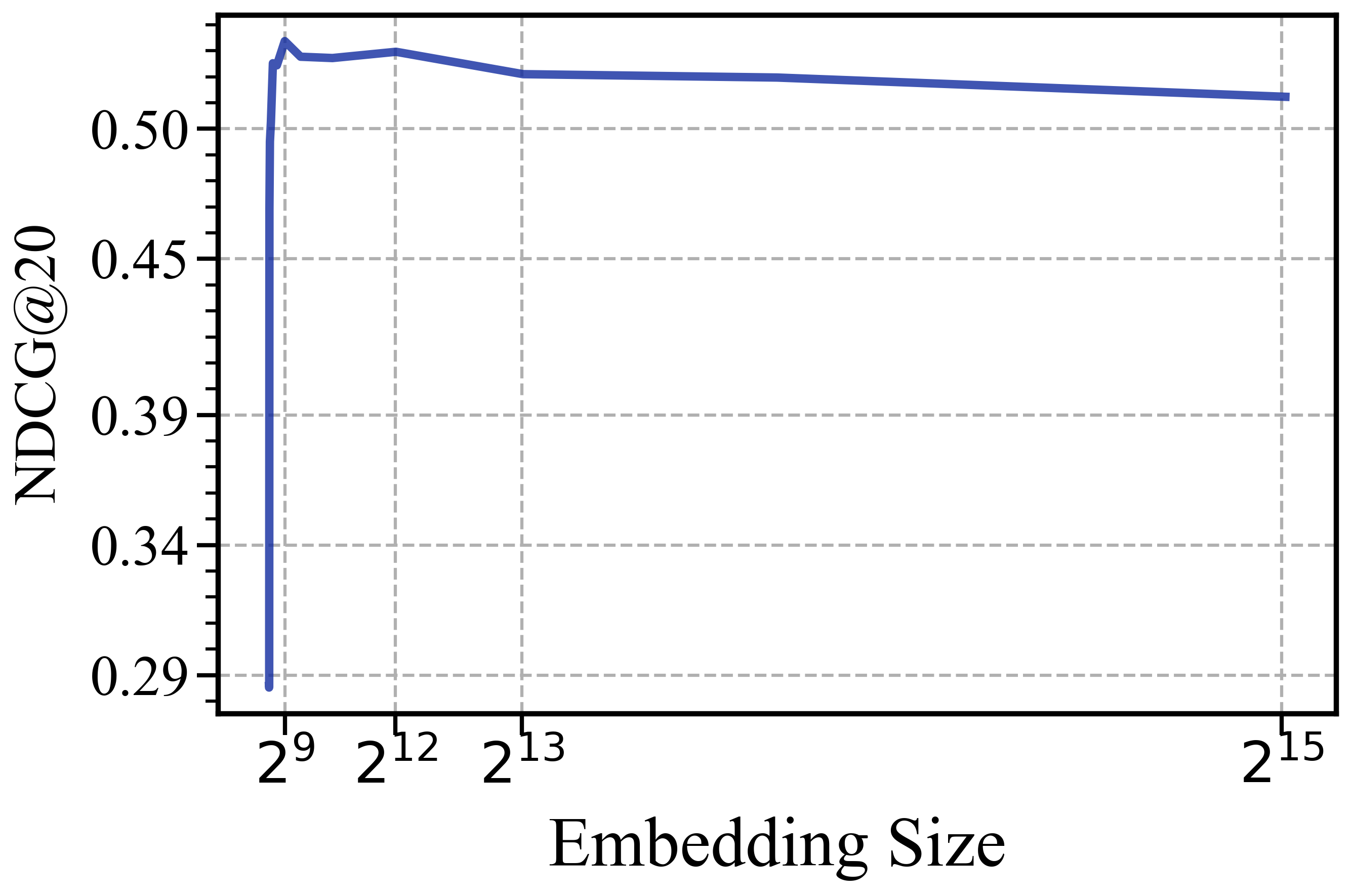}} \
\vspace{-10pt}
\subfloat[\textbf{ModCloth - SGL}]
{\includegraphics[width=0.32\linewidth]{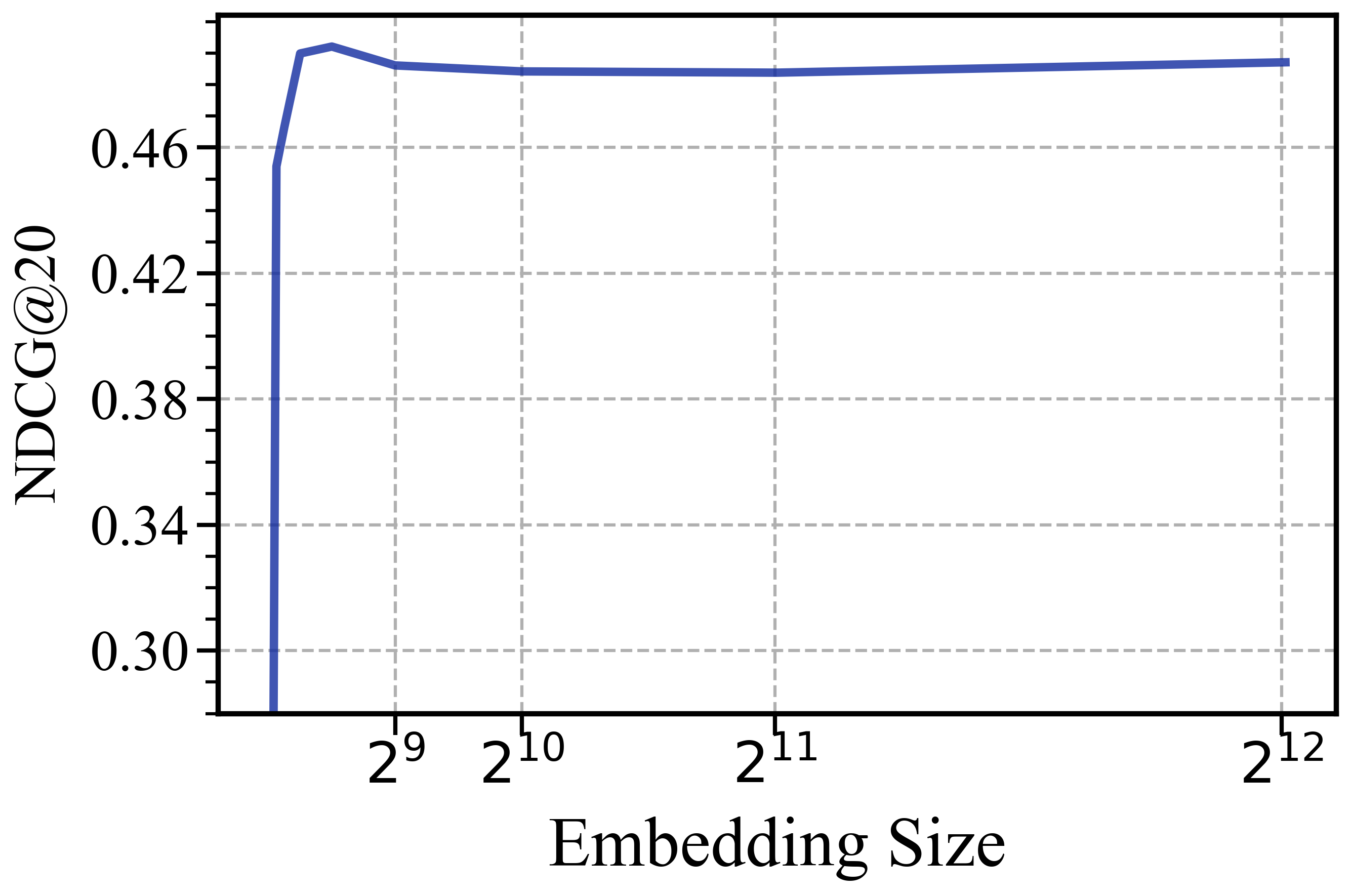}} \hfill
\subfloat[\textbf{Douban - SGL}]
{\includegraphics[width=0.32\linewidth]{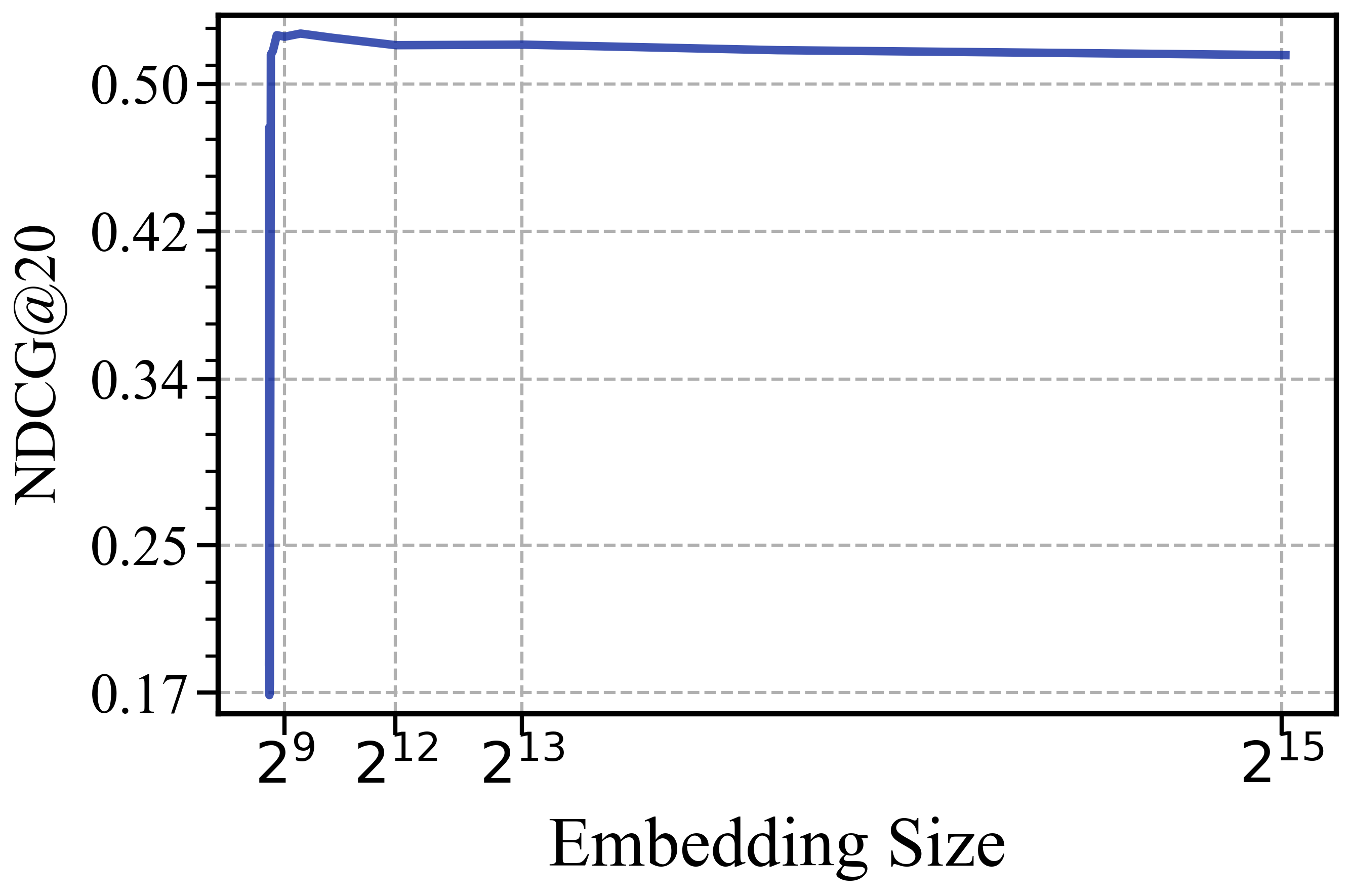}} \hfill
\subfloat[\textbf{ML-100k - SGL}]
{\includegraphics[width=0.32\linewidth]{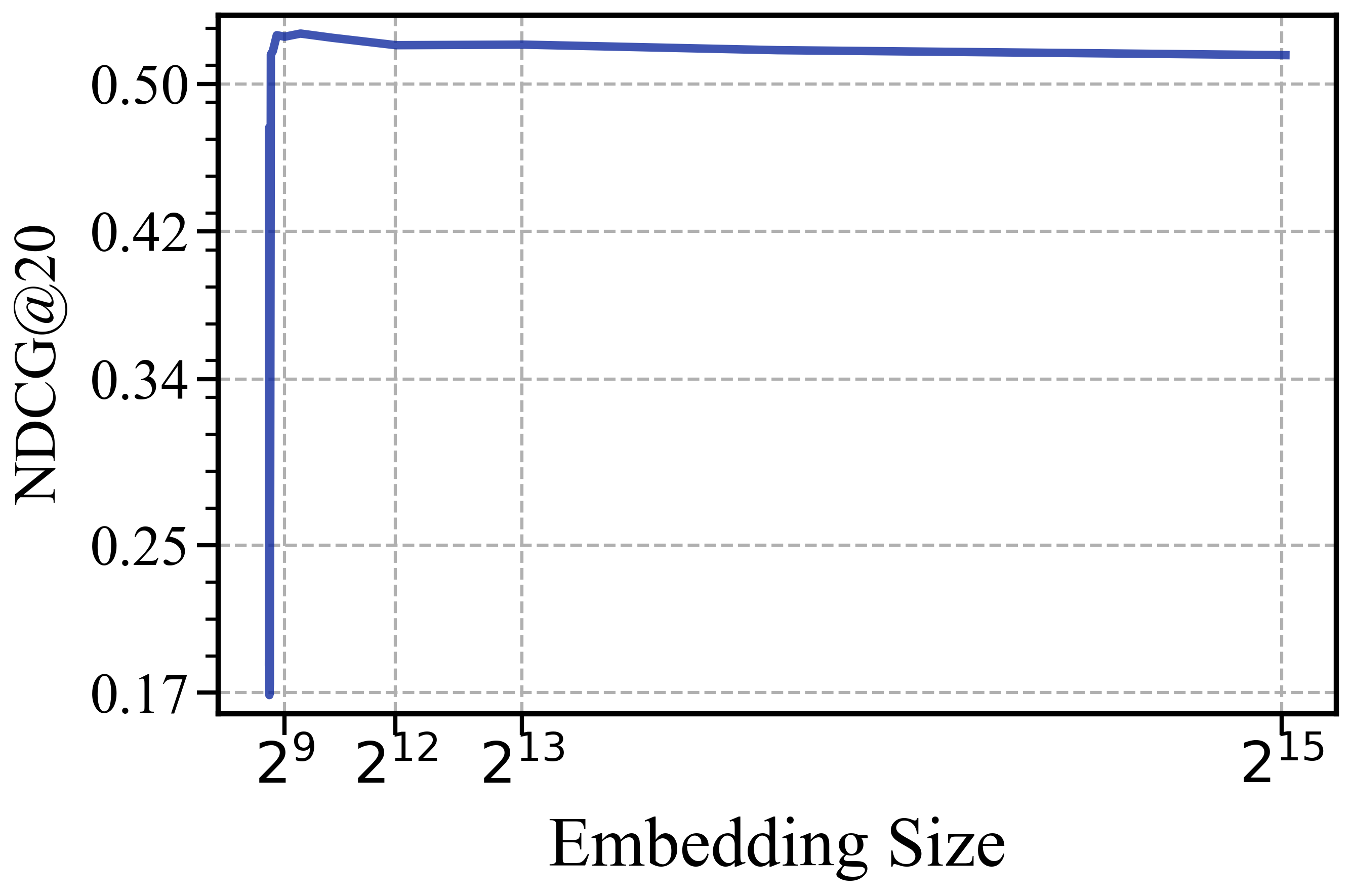}}
\vspace{5pt}
\caption{Scale the embedding dimension exponentially by a factor of 2 across different collaborative filtering models and datasets. Each row corresponds to a model (BPR, LightGCN, SGL, NeuMF), and each column represents a dataset (Modcloth, Douban, ML-100k).}
\label{scale1}
\end{figure*}

In the past understanding, when the embedding dimension increases, the performance shows a single-peak phenomenon of first increasing and then decreasing. However, we believe that this understanding lacks microscopic and extensive observations. Through large-scale experiments (Figure~\ref{scale1}), we find that on the ML-100K dataset, when the embedding dimension increases, the overall performance of the BPR model is indeed consistent with previous studies, showing an overall trend of first increasing and then decreasing. However, we discovere more details for the first time: as the embedding dimension increases, the recommendation performance initially increases significantly, then decreases, then increases slightly, and finally decreases again. We call this new trend \textit{\textbf{double-peak phenomenon}} (as shown in Figure~\ref{scale1} (b, c, e, f)). Moreover, we noticed that the most of peak performance occurs at $2^9$, which are different from the traditional settings of 32, 64, and 128. But unexpectedly, in LightGCN and SGL, this phenomenon is alleviated a lot, and SGL alleviates it better than LightGCN (as shown in Figure~\ref{scale1} (k, h)). In addition, our experiments show that on the Modcloth dataset, BPR, NeuMF, LightGCN, and SGL all show a continuous upward trend. For example, even when the dimension is set to $2^{14}$, the performance of LightGCN is still rising. We call this ideal trend the \textit{\textbf{logarithmic phenomenon}} (as shown in Figure~\ref{scale1} (a)), that is, as the embedding dimension increases, the performance will continue to rise, but the increase amplitude will gradually decrease. The logarithmic phenomenon completely exceeds the past understanding. Compared with the NDCG@20 at 128 dimensions, the performance has increased by 25.57\%. On the contrary, if we only set the dimension to 128, we will lose 34.36\% of the potential performance improvement. Currently, most research in the recommendation field, such as debiasing~\citep{bias2,bias1} and denoising~\citep{tce,dcf}, believes that a 10\% performance improvement is meaningful. However, in our observations, simply increasing the dimension can achieve a performance improvement far exceeding that of elaborately designed methods. 

To ensure the generality of the phenomenon, we design a series of experiments, conducting observations across other 7 datasets (\ref{e1}) and also analysis the results in detail. \uline{Empirical results confirm that the phenomenon is indeed widespread}.

\textbf{Two Observations}.
\ding{182}~\uline{The same model exhibits different phenomenons across different datasets}. As shown in Figure~\ref{scale1} (a, d), BPR exhibits a double-peak phenomenon on the ML-100K dataset, while showing logarithmic growth on the Modcloth dataset. This seems to be due to some negative factor in ML-100K that causes scaling to fail.
\ding{183}~\uline{The same dataset exhibits different phenomena across different models}. As shown in Figure~\ref{scale1} (a, b, c), we observe that BPR exhibits a double-peak phenomenon, while LightGCN and SGL show logarithmic growth. Moreover, we believe this might be because the structures of LightGCN and SGL incorporate mechanisms that mitigate negative factors. Moreover, the logarithmic growth of SGL appears to be more ideal.  

\subsection{Why Two Phenomena Occur} Based on the previous two observations, we believe this is due to the presence of certain negative factors in the dataset that BPR and NeuMF cannot resist, while LightGCN and SGL can. When these negative factors are mitigated, a logarithmic trend emerges, whereas if influenced by these factors, a double-peak pattern is observed. In addition, the difficult-to-understand aspect of the double-peak phenomenon lies in the third stage, where performance rises again. If this stage were absent, with performance simply rising and then falling, it could easily be explained as overfitting. However, the appearance of the third stage prevents us from attributing it to overfitting, and we believe this is the key to understanding this phenomenon. For logarithmic phenomenon, this is an ideal trend. If every model can exhibit this trend on every dataset, it would be wonderful. 

Upon close examination of Figures~\ref{scale1},~\ref{scale2},~\ref{scale3}, we observe that SGL tends to exhibit such a logarithmic trend. We attribute this behavior to the contrastive learning framework in SGL, which provides both regularization and robustness enhancement. In contrast, other methods lack such a mechanism, which may explain their deviation from this desirable pattern.


\textbf{Noisy Interactions Induces the Double-peak Phenomenon}. For the double-peak phenomenon, this is a novel phenomenon. Our approach to analyzing this phenomenon was inspired by two phenomena in machine learning: the memorization effect on noisy labels~\citep{noise_first} and the sparse double descent~\citep{sparse_dscent} which found that the sparsity of model parameters and the level of label noise jointly affect the model's performance (more about noisy interactions at~\ref{noisy_interactions}).

\begin{definition}[Sparse Double Descent]~\citep{sparse_dscent} is a phenomenon that occurs during model pruning. It is characterized by the model's test performance first decreasing due to overfitting as sparsity increases, then improving once overfitting is alleviated after reaching a certain level of sparsity, and finally decreasing again as sparsity increases, due to the loss of useful information.
\end{definition}

We think that scaling the embedding dimension is related to the sparsification of neural networks in recommendation tasks, as increasing the dimension effectively reduces the sparsity of the neural network. \textit{In the first stage}, the cleaning learning stage, the model initially learns from clean interactions. As the dimensionality of the embedding increases, the model has more space to accommodate user interests, leading to performance improvements. \textit{In the second stage}, the noise memorization stage, the presence of interaction noise, combined with a dimensionality large enough to accommodate noisy interactions, causes the model to begin learning from these noisy interactions, resulting in performance degradation. \textit{In the third stage}, the sweet stage, the embedding space enlarges to adapt to interaction noise, thereby enhancing the model's robustness. \textit{In the fourth stage}, the overfitting stage, the continuous increase in embedding dimensions leads to a higher fit to the training dataset, which subsequently reduces generalization performance.


\begin{figure*}[t!]
\centering
\subfloat[\textbf{ML-100K}]
{\includegraphics[width=0.43\linewidth]{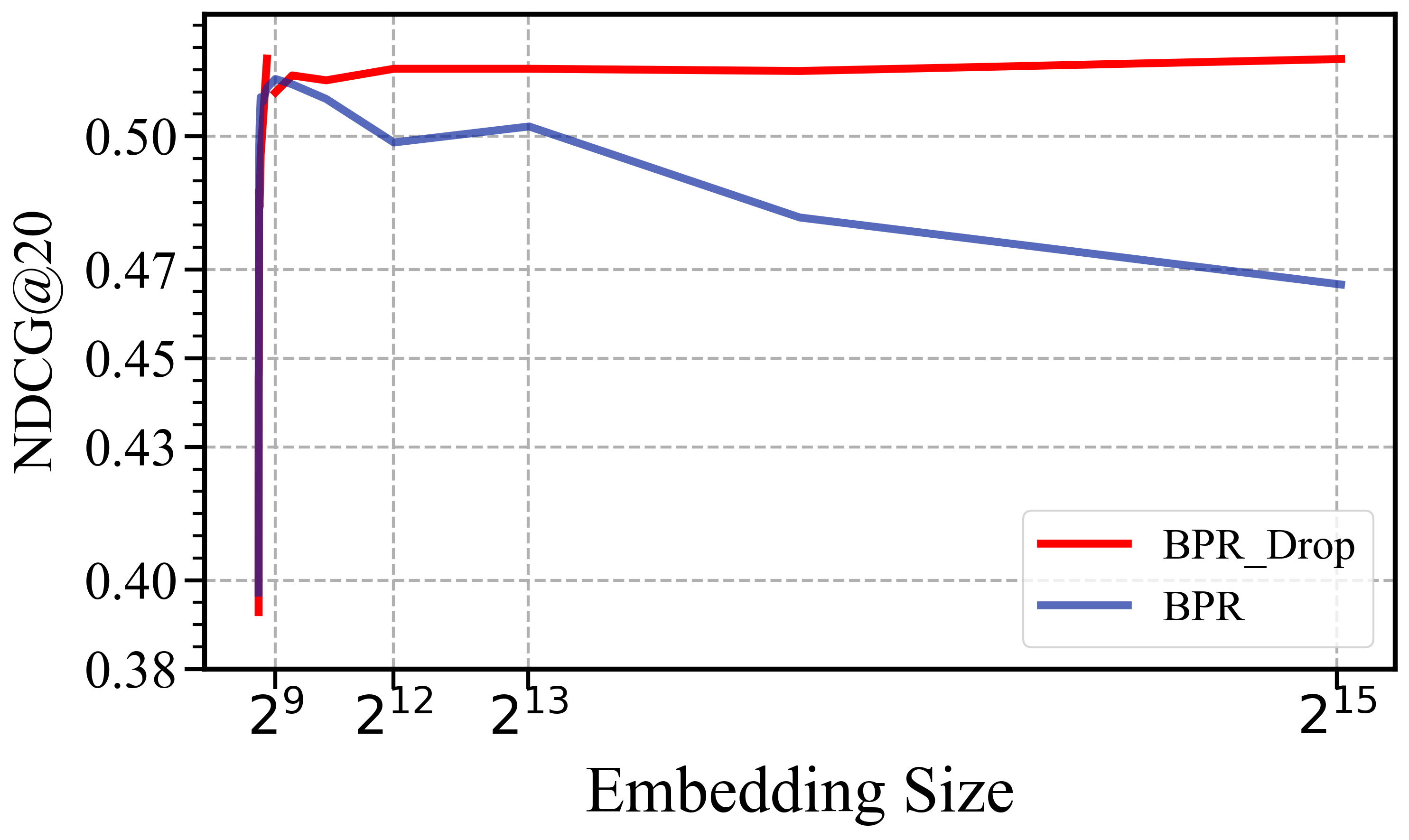}} \hspace{10pt}
\subfloat[\textbf{Douban}]{\includegraphics[width=0.43\linewidth]{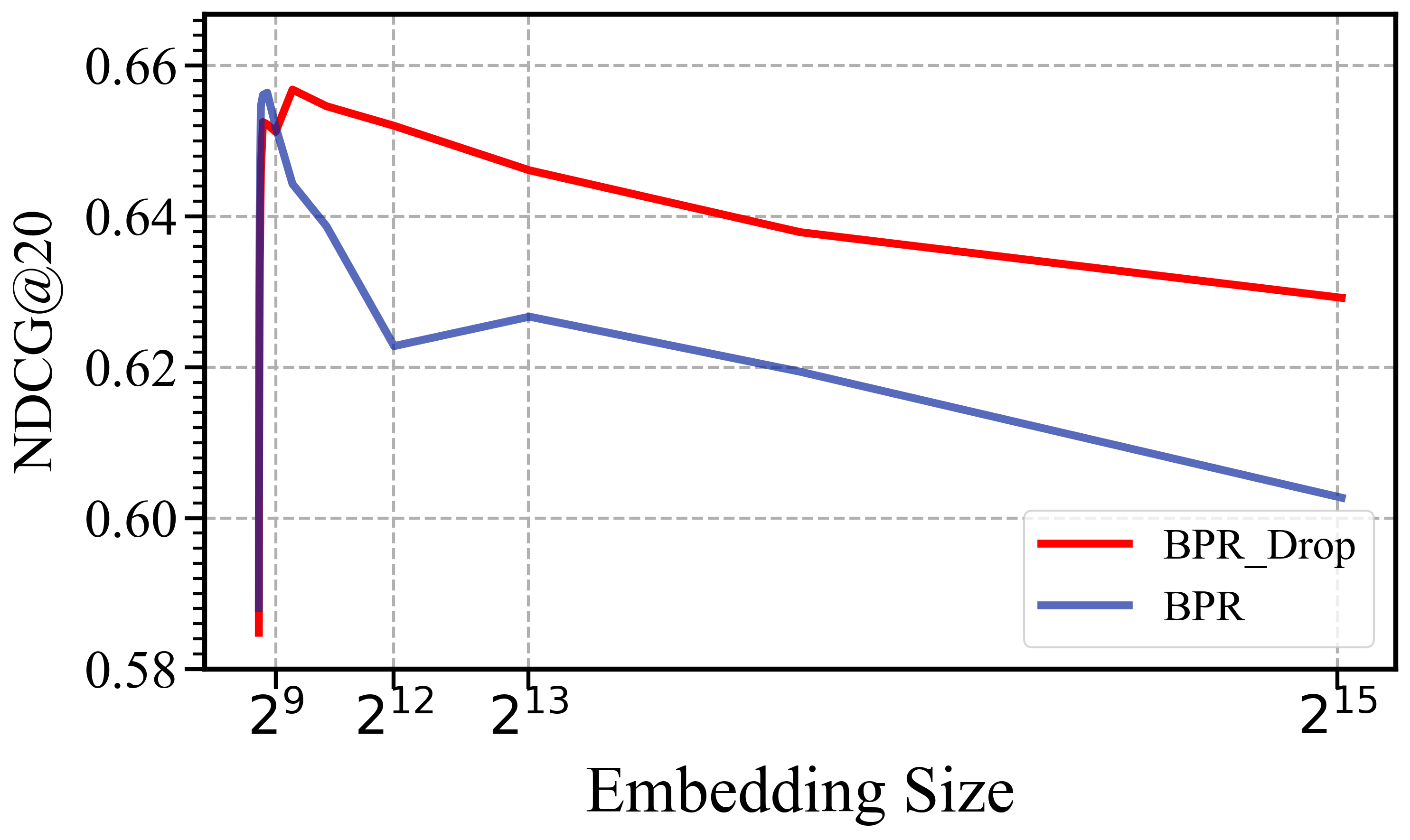}}
\caption{Comparing the standard BPR with the denoising strategy-based BPR\_Drop in ML-100K and Douban, we can clearly observe that the performance degradation has significantly improved.}
\label{drop}
\end{figure*}

\section{One of the Keys to Scaling: Improve Robustness to Noisy Interactions}
In this section, we conduct a detailed analysis based on the models and explore the characteristics of each model, focusing on three main questions: Why are BPR and NeuMF easily affected by noise? Why does graph collaborative filtering exhibit noise resistance? Why does graph self-supervised learning perform better than graph collaborative filtering?

\subsection{Analysis of the Resistance to Noise of BPR}

We first analyze why the BPR model is particularly sensitive to noisy interactions. Our analysis is based on the degradation of representation quality under noisy preference distributions.

\begin{definition}[Representation Quality]
Let \( \Theta^* = \{\mathbf{p}_u^*, \mathbf{q}_i^*\} \) denote the ideal latent representations learned from clean data \( \mathcal{D}_0 \), and \( \Theta = \{\mathbf{p}_u, \mathbf{q}_i\} \) be any learned parameters. The representation quality is defined as:
\begin{equation}
 Q(\Theta) := \|\Theta - \Theta^*\|_F^2 = \sum_u \|\mathbf{p}_u - \mathbf{p}_u^*\|_2^2 + \sum_i \|\mathbf{q}_i - \mathbf{q}_i^*\|_2^2.   
\end{equation}
\end{definition}

\begin{theorem}[Representation Perturbation Under Noisy Interactions]
\label{thm:bpr_noise}
Let \( \Theta_0 \) be the minimizer of the clean empirical loss \( \mathbb{E}_{x \sim \mathcal{D}_0}[\ell(\Theta; x)] \), and consider a noisy interaction distribution \( \mathcal{D}_\delta := (1 - \delta) \mathcal{D}_0 + \delta \mathcal{N} \), where \( \mathcal{N} \) is a noise distribution and \( \delta \in [0,1] \) is the noise ratio. Define the noise gradient shift as \( \Delta_{\mathrm{noise}} := \mathbb{E}_{x \sim \mathcal{N}}[\nabla_\Theta \ell(\Theta_0; x)] - \mathbb{E}_{x \sim \mathcal{D}_0}[\nabla_\Theta \ell(\Theta_0; x)] \). Assume the loss \( \ell(\Theta; x) \) is twice differentiable and the expected clean loss is locally strongly convex around \( \Theta_0 \). Then, for sufficiently small \( \delta \), the representation quality satisfies:
\begin{equation}
Q(\Theta_\delta) := \|\Theta_\delta - \Theta_0\|_F^2 \leq \delta^2 \cdot \|H^{-1} \Delta_{\mathrm{noise}}\|_F^2 + o(\delta^2),
\end{equation}
where \( \Theta_\delta \) is the minimizer under \( \mathcal{D}_\delta \), and \( H := \nabla^2_\Theta \mathbb{E}_{x \sim \mathcal{D}_0}[\ell(\Theta; x)]|_{\Theta = \Theta_0} \) is the Hessian at the clean optimum. The complete proof can be found in~\ref{Proof of Theorem 1}
\end{theorem}

\begin{remark}
Theorem~\ref{thm:bpr_noise} reveals that the degradation in representation quality under noisy interactions is quadratically proportional to the noise ratio \( \delta \), and linearly controlled by \( \|\Delta_{\mathrm{noise}}\| \), which is dictated by the model architecture.
\end{remark}

\begin{lemma}[High Gradient Sensitivity in BPR]
In BPR, the scoring function is defined as \( f(u,i,j) = \mathbf{p}_u^\top (\mathbf{q}_i - \mathbf{q}_j) \), which is linear in its parameters. Therefore, the per-sample gradient is:
\begin{equation}
\nabla_\Theta \ell(\Theta; x) \propto (\mathbf{q}_j - \mathbf{q}_i, \mathbf{p}_u, -\mathbf{p}_u),
\end{equation}
with unbounded norm when embeddings grow. As a result, \( \|\Delta_{\mathrm{noise}}\| \) tends to be large, making BPR more vulnerable to noisy samples.
\end{lemma}

\begin{remark}
This explains the double-peak phenomenon observed in BPR training, especially under large embedding dimensions, due to overfitting to noise.
\end{remark}
\textbf{Experimental Evidence}. To validate the robustness of BPR, we adopted a simple strategy: we applied denoising collaborative filtering methods to observe whether there would be an improvement in performance, particularly when the embedding dimensions are high (Code at~\ref{code}).
\begin{equation}
\Theta^{t+1} \leftarrow \Theta^t-\eta \nabla \ell_{\rm \small BPR} \left(\Theta^t ;\left\{x_i \mid \mathcal{L}\left(x_i\right) \leq \tau\right\}\right).
\end{equation}
This equation illustrates a dynamic sample-dropping strategy~\citep{tce} to enhance BPR's noise robustness. During gradient updates, only samples with losses below threshold $\tau$ are retained, filtering out high-loss interactions likely to be noise. Experiments show this approach effectively mitigates noise in high-dimensional embedding spaces while improving recommendation performance (as shown in Figure~\ref{drop}). 
The threshold generally depends on the dropout rate: higher dropout leads to a lower threshold, and vice versa. We conducted extra experiments on Figure~\ref{scale5} with different dropout rates for further evidence.
\subsection{Analysis of the Resistance to Noise of NeuMF}

We now analyze the noise sensitivity of NeuMF from a Jacobian-based perspective. Despite the inclusion of nonlinear components, the gradient propagation through multiple nonlinear layers can amplify noise, depending on the Jacobian of the model output with respect to the parameters.

\begin{lemma}[Noise-Induced Gradient Amplification]
\label{lemma:jacobian}
Let \( f(u,i) \) be the NeuMF prediction function, and let \( \Theta \in \mathbb{R}^d \) denote all model parameters. Define the Jacobian matrix of the scoring function with respect to parameters as \( J_x := \nabla_\Theta f(u, i) \in \mathbb{R}^{1 \times d} \) for input \( x = (u, i) \). 
Then for a noisy interaction \( x \sim \mathcal{N} \), the norm of the loss gradient is bounded by:
\begin{equation}
   \|\nabla_\Theta \ell(\Theta; x)\| \leq L_\ell \cdot \|J_x\|, 
\end{equation}
where \( L_\ell \) is the Lipschitz constant of the loss function. Hence, the gradient norm is directly governed by the spectral norm \( \|J_x\| \) of the Jacobian.
\end{lemma}

\begin{theorem}[Gradient Instability in NeuMF]
Let \( f(u, i) = \mathrm{MLP}([\mathbf{p}_u; \mathbf{q}_i]) + \mathbf{p}_u^\top \mathbf{q}_i \) be the NeuMF scoring function, where \( \mathrm{MLP} \) has \( L \) layers with weight matrices \( W^{(1)}, \dots, W^{(L)} \), each with spectral norm \( \|W^{(l)}\|_2 \). Then the Jacobian satisfies:
\begin{equation}
 \|J_x\| = \left\| \frac{\partial f(u, i)}{\partial \Theta} \right\| \leq \left( \prod_{l=1}^{L} \|W^{(l)}\|_2 \right) \cdot \|\nabla_z \ell(f(u,i), y)\|,   
\end{equation}
which grows multiplicatively with depth, implying NeuMF can have exponentially amplified gradient sensitivity under noisy interactions. The proof precess are at~\ref{Proof of Theorem 2}.
\end{theorem}

\begin{corollary}[Comparative Noise Sensitivity: NeuMF vs. BPR]
Although BPR has unbounded inner product gradients w.r.t. embeddings, its Jacobian norm is linear in embedding size, i.e.,
$\|J_x^{\mathrm{BPR}}\| \propto \|\mathbf{p}_u\| + \|\mathbf{q}_i\|.$
In contrast, NeuMF's Jacobian grows with the product of layer spectral norms, potentially leading to
\begin{equation}
  \|J_x^{\mathrm{NeuMF}}\| > \|J_x^{\mathrm{BPR}}\|,  
\end{equation}
in deep or poorly regularized MLPs. Hence, NeuMF may not exhibit superior robustness, and may in fact be more susceptible to gradient perturbations from noise.
\end{corollary}

\subsection{Analysis of the Resistance Noise of Graph Collaborative Filtering} LightGCN exhibits a relatively expected logarithmic growth pattern. We believe this is likely due to the inherent noise-resistant structure of LightGCN. We think that the process of aggregating information from a user's neighboring nodes in graph convolution~\citep{he2020lightgcn} is highly consistent with Mixup~\citep{mixup}. 
\begin{equation}
\begin{aligned}
(\tilde{\mathbf{u}}, \tilde{\mathbf{v}})&=\underbrace{\left(\frac{1}{\left|\mathcal{N}_i\right|} \sum_{k \in \mathcal{N}_i} \mathbf{u}_k,  \mathbf{v}_i\right)}_{graph \, convolution}; \\
(\tilde{\mathbf{u}}, \tilde{\mathbf{v}})&=\underbrace{\left(\sum_{i=1}^N \lambda_i \mathbf{u}_i, \sum_{i=1}^N \lambda_i \mathbf{v}_i\right)}_{Mixup}, \text { s.t. } \sum_{i=1}^N \lambda_i=1.
\end{aligned}
\end{equation}
Specifically, first-order neighbors aggregate information about items, while second-order neighbors aggregate information from other users. Since Mixup is regarded as noise-resistant, we hypothesize that graph-based collaborative filtering possesses similar noise-resistant capabilities. Moreover, this idea aligns with findings in~\cite{gcnemixup}.
\begin{equation}
\begin{aligned}
(\tilde{\mathbf{u}}, \tilde{\mathbf{v}})&=\left(\frac{1}{\left|\mathcal{N}_i\right|} \sum_{k \in \mathcal{N}_i} \mathbf{u}_k, \mathbf{v}_i\right) \\
&=\left(\sum_{k \in \mathcal{N}_i} \frac{1}{\left|\mathcal{N}_i\right|} \mathbf{u}_k, \sum_{k \in \mathcal{N}_i} \frac{1}{\left|\mathcal{N}_i\right|} \mathbf{v}_i\right) \\
& \stackrel{\lambda_i=\frac{1}{\left|\mathcal{N}_i\right|}}{=} \left(\sum_{k \in \mathcal{N}_i} \lambda_i \mathbf{u}_k, \sum_{k \in \mathcal{N}_i} \lambda_i \mathbf{v}_i\right).
\end{aligned}
\end{equation}
This makes it easy to understand why LightGCN is noise-resistant, as Mixup, being a classic and popular method for domain generalization, is effective in mitigating noise disturbances. Next, we analyze from the spectral perspective.

\begin{definition}[Spectral Representation of Graph Convolution]
    Let the adjacency matrix of the user-item interaction graph $\mathcal{G}$ be $A \in \mathbb{R}^{(m+n)\times(m+n)}$, and the degree matrix be $D$. The normalized adjacency matrix is $\tilde{A} = D^{-1/2} A D^{-1/2}$. Performing eigendecomposition on $\tilde{A}$:
\begin{equation}
   \tilde{A} = U \Lambda U^T, \quad \Lambda = \text{diag}(\lambda_1, \lambda_2, \dots, \lambda_{m+n}), 
\end{equation}
where $|\lambda_1| \geq |\lambda_2| \geq \dots \geq |\lambda_{m+n}|$, and $U$ is the orthogonal eigenvector matrix.
\end{definition}

\begin{definition}[Low-Rank Property of Graph Convolution]
The embedding propagation in LightGCN (Eq.2) can be represented as a spectral filtering process. After $L$ layers of propagation, the final user/item embeddings are:
\begin{equation}
\mathbf{p}_u = \sum_{l=0}^L \beta_l \tilde{A}^l \mathbf{p}_u^{(0)}, \quad \mathbf{q}_i = \sum_{l=0}^L \beta_l \tilde{A}^l \mathbf{q}_i^{(0)},
\end{equation}
where $\beta_l = \frac{1}{L+1}$. In the spectral domain, this is equivalent to applying a low-pass filter to the initial embeddings:
\begin{equation}
\mathbf{p}_u = U \left( \sum_{l=0}^L \beta_l \Lambda^l \right) U^T \mathbf{p}_u^{(0)}.
\label{eqthm2}
\end{equation}
The filter coefficient matrix $\sum_{l=0}^L \beta_l \Lambda^l$ attenuates high-frequency components (small $|\lambda_k|$) while preserving low-frequency components (large $|\lambda_k|$), leading to low-rank embeddings. The complete proof can be found in~\ref{derivation 4}.
\end{definition}


\subsection{Analysis of the Resistance Noise of Self-supervised Graph Collaborative Filtering}
Self-supervised graph collaborative filtering (SGL) is a subset of graph-based collaborative filtering. Naturally, we can also infer that self-supervised graph collaborative filtering inherently possesses the ability to resist interaction noise due to the graph convolution mechanism. Moreover, in most cases, SGL outperforms LightGCN. So, what additional module plays a crucial role in this difference?


SGL adopts self-supervised contrastive learning.
SGL generates multi-view representations via stochastic graph perturbations and optimizes contrastive loss. For view generation, we create augmented graphs \( \mathcal{G}' \) and \( \mathcal{G}'' \) by:
edge dropout: Remove edges with probability \( \rho \) and feature masking: zero out embedding dimensions with probability \( \rho \). For contrastive loss, we maximize consistency between views for the same node:
\begin{equation}
    \mathcal{L}_{\text{cont}} = -\sum_{u \in \mathcal{U}} \log \frac{\exp\left( \text{sim}(\mathbf{z}_u', \mathbf{z}_u'') / \tau \right)}{\sum_{v \in \mathcal{U}} \exp\left( \text{sim}(\mathbf{z}_u', \mathbf{z}_v'') / \tau \right)}, \label{eq:contrastive_loss}
\end{equation}
where \( \text{sim}(\cdot) \) is cosine similarity, \( \tau \) is a temperature hyperparameter, and \( \mathbf{z}_u', \mathbf{z}_u'' \) are augmented embeddings of \( u \).

\begin{definition}[Mutual Information Objective in Contrastive Learning]
   SGL maximizes the mutual information $I(z_u'; z_u'')$ between augmented views $z_u'$ and $z_u''$ via the contrastive loss $\mathcal{L}_{\text{cont}}$. By the information bottleneck principle:
\begin{equation}
    I(z_u'; z_u'') = H(z_u') - H(z_u' | z_u''),
\end{equation}
where $H(\cdot)$ denotes entropy. Minimizing $\mathcal{L}_{\text{cont}}$ forces $z_u'$ and $z_u''$ to share noise-invariant information. 
\end{definition}

\begin{theorem}[Embedding Space Constraint via Contrastive Learning]
    Let the noise perturbation $\delta_u$ satisfy $\delta_u \perp \mathcal{S}$, where $\mathcal{S}$ is the subspace of clean interaction signals. The contrastive loss in SGL implicitly constrains the embedding vectors $z_u$ to lie within $\mathcal{S}$:
\begin{equation}
\forall u, \quad \| \text{Proj}_{\mathcal{S}^\perp}(z_u) \|_2 \leq \epsilon,
\end{equation}
where $\epsilon$ is a constant related to noise level, and $\mathcal{S}^\perp$ is the orthogonal complement of $\mathcal{S}$. The complete proof can be found in~\ref{Proof of Theorem 3}.
\end{theorem}

\begin{corollary}[Noise Filtering in SGL]
As demonstrated in Theorem 3, SGL achieves noise resistance by employing low-pass filtering in graph convolution to attenuate high-frequency noise while retaining essential low-frequency components, and by imposing a subspace constraint in contrastive learning to further reduce noise variance and improve noise robustness.
Mathematically, SGL embeddings are:
\begin{equation}
z_u = \underbrace{\text{Proj}_{\mathcal{S}}(z_u^*)}_{\text{low-rank signal}} + \underbrace{\text{Proj}_{\mathcal{S}^\perp}(\delta_u)}_{\text{noise}},
\end{equation}
where $\|\text{Proj}_{\mathcal{S}^\perp}(\delta_u)\|_2 \leq \epsilon$, and $\epsilon$ decreases with contrastive learning strength $\gamma$.
\end{corollary}

%% file: Alltex/5-experiments.tex


%% file: Alltex/6-Conclusion.tex
\section{Discussion}
\textbf{The quality of the recommendation dataset is more important than we thought}. The quality of the dataset, especially noisy interactions, yields a substantial influence on the scaling recommendation performance. When the dataset contains less noise, the model can scale more smoothly, with algorithms adapting and expanding more efficiently to handle larger volumes of data. Conversely, a high level of noise data muddles the underlying patterns, undermining the model's ability to scale effectively and leading to suboptimal performance as the scale increases.

\textbf{Collaborative filtering models with noise-resistant structures have more advantages when it comes to scaling}. We can clearly see that, compared to BPR, NeuMF and LightGCN, SGL demonstrates stronger scaling capabilities. Even when the embedding dimensions are significantly increased, SGL maintains stable performance or stays close to peak performance, unlike BPR, which experiences a sharp decline. This advantage stems from the inherent noise resistance of SGL's structure. We believe that further exploration of this structure could potentially lead to the emergence of a "Transformer" in the field of collaborative filtering.

\section{Conclusion and Limitations}
\textbf{Conclusion}. Our contributions lie in the discovery of two new phenomena (double-peak and logarithmic), the interpretation of their causes, and the theoretical analysis of noise robustness in representative collaborative filtering models. Specifically, to investigate whether scaling embedding dimensions can improve collaborative filtering models, we conducted extensive observational experiments across 10 datasets and 4 representative models. Compared to the previously recognized single-peak phenomenon, we identified two novel phenomena: double-peak and logarithmic. We conducted a detailed analysis of the underlying cause of the double-peak effect and found it to result from noise interactions. To empirically validate our hypothesis, we introduced a denoising loss function, which supported our explanation. Furthermore, we provided a theoretical analysis of the models’ robustness to noise with results aligned with our empirical observations.

\textbf{Limitations}. Due to limitations in available computational resources, we were unable to conduct experiments on a broader range of collaborative filtering models. Additionally, our observations were based primarily on the Top@K metric with K set to 20. With sufficient resources, future work could explore additional values of K (e.g., 5, 10) and incorporate more fine-grained evaluation metrics to further validate the generality of the observed phenomena.

%% file: Alltex/7-appendix.tex
\newpage
\appendix
\onecolumn

\begin{center}
	\LARGE \bf {Appendix of ``Understanding  Embedding Scaling''}
\end{center}

\etocdepthtag.toc{mtappendix}
\etocsettagdepth{mtchapter}{none}
\etocsettagdepth{mtappendix}{subsection}
\tableofcontents

\newpage

\section{Notions}
\label{notions}
\begin{table}[h]
    \centering
    \small
    \caption{Symbol Definitions}
    \begin{tabular}{ll}
        \toprule
        \textbf{Symbol} & \textbf{Description} \\
        \midrule
        \midrule
        \( u, v \) & Indices representing users. \\
        \( i, j \) & Indices representing items. \\
        \( \mathbf{p}_u \in \mathbb{R}^k \) & Latent representation vector for user \( u \). \\
        \( \mathbf{q}_i \in \mathbb{R}^k \) & Latent representation vector for item \( i \). \\
        \( \Theta = \{\mathbf{p}_u, \mathbf{q}_i\} \) & Set of all user and item latent representations. \\
        \( \Theta^* = \{\mathbf{p}_u^*, \mathbf{q}_i^*\} \) & Ideal set of user and item latent representations satisfying true preference relations. \\
        \( \Theta_0 \) & Optimal representations by minimizing the loss function without noise. \\
        \( \Theta_\eta \) & Optimal representations by minimizing the loss function with noise proportion \( \eta \). \\
        \( \mathcal{D}_0 \) & Set of authentic (noise-free) interaction triples \( (u, i, j) \). \\
        \( \mathcal{D}_\eta \) & Set of noisy interaction triples \( (u, j, i) \). \\
        \( \mathcal{D} = \mathcal{D}_0 \cup \mathcal{D}_\eta \) & Combined set of authentic and noisy interaction triples. \\
        \( \eta \in (0,1) \) & Proportion of noisy interactions in the dataset. \\
        \( \lambda > 0 \) & Regularization parameter in the loss function. \\
        \( \sigma(z) = \frac{1}{1 + e^{-z}} \) & Sigmoid function. \\
        \( \mathcal{L}_0(\Theta) \) & Loss function without noise. \\
        \( \mathcal{L}_\eta(\Theta) \) & Loss function component due to noisy interactions. \\
        $\mathcal{L}(\Theta)$ & Total loss function with noise \(  = \mathcal{L}_0(\Theta) + \mathcal{L}_\eta(\Theta) \). \\
        \( Q(\Theta) \) & Representation quality metric. \\
        \( \|\cdot\|_F \) & Frobenius norm. \\
        \( \|\cdot\|_2 \) & Euclidean (L2) norm. \\
        \( \mu > 0 \) & Strong convexity parameter. \\
        \( c > 0 \) & Lower bound constant related to noisy interactions. \\
        \bottomrule
    \end{tabular}
    \label{tab:symbols}
\end{table}

\section{Related Work}
\subsection{Scaling Recommendation.}
Large language models (LLM) have shone brightly in today's era, and one of the key reasons is that the performance of the foundational Transformer~\citep{transformer} model can scale with parameter expansion. Recently, a few efforts~\citep{scalingrec1,scaling4seqrec,wukong,lms,meta,251,252,253,jumingxuan} in the recommendation field have explored a similar idea: whether the performance of recommendation models can increase with parameter scaling. \citep{scalingrec1} explores the scaling properties of recommendation models across data, compute, and parameters, revealing that parameter scaling has limited impact on performance compared to data scaling, and highlights the need for future advancements in model architecture and system design. \citep{lms} identifies the embedding collapse phenomenon as the inhibition of scalability, where the embedding matrix leans towards occupying a low-dimensional subspace. It demonstrates a two-sided effect of feature interaction in recommendation models, with collapsed embeddings restricting learning and worsening the collapse.

\subsection{Efficient Recommendation.} 
How to set the dimension of embeddings is a common question for almost everyone studying collaborative filtering. This is because one of the core aspects of collaborative filtering is representing users or items with embeddings, and the quality of these embeddings directly determines the performance of the recommendation system. Reflecting on the efforts of recent years, to determine the optimal embedding dimension, many researchers have learned the embedding dimension automatically from the perspectives of neural architecture search~\citep{nas1, nas2}, weight pruning~\citep{dim1, dim2,dim3,ef25} or exploring mixed dimension~\citep{mixed}. These methods seem to have yielded notable results. The implicit assumption of these methods is that there exists an optimal dimension, meaning the performance exhibits a single-peak phenomenon. 

\subsection{Denoising Recommendation}
\label{noisy_interactions}
In recommendation systems based on implicit feedback, user behaviors (e.g., clicks, views) often contain \textit{noisy interactions}—signals that do not truly reflect user preferences, such as clicks driven by item position or popularity rather than genuine interest~\cite{chen2025sparsemoe,yang2024graph,yang2025invariance}. These biases degrade model generalization and make learning reliable user preferences challenging.
One of the simplest and most practical approaches is the drop-based method, which identifies noisy samples based on their loss values: samples with higher loss are more likely to be noisy, while those with lower loss are considered clean. drop-based method filters out noisy interactions before training. For example, T-CE~\citep{tce} finds that noisy samples tend to have higher loss values, which can be used as a denoising signal. AutoDenoise~\citep{AutoDenoise-reinforce} models denoising as a search-and-decision process and applies reinforcement learning to automate it. DCF~\citep{dcf} further improves performance by distinguishing between noisy interactions and hard samples, and by relabeling the noisy ones.


\section{Theoretical Discussions}
\subsection{Proof of Theorem 1}
\label{Proof of Theorem 1}
\textit{Proof.} 
Since \( \Theta_0 \) minimizes the clean objective, we have the first-order optimality condition:
\begin{equation}
  \nabla_\Theta \mathbb{E}_{x \sim \mathcal{D}_0}[\ell(\Theta_0; x)] = 0.  
\end{equation}
Let us denote the noisy objective as:
\begin{equation}
\mathcal{L}_\delta(\Theta) := \mathbb{E}_{x \sim \mathcal{D}_\delta}[\ell(\Theta; x)] = (1 - \delta) \mathbb{E}_{x \sim \mathcal{D}_0}[\ell(\Theta; x)] + \delta \mathbb{E}_{x \sim \mathcal{N}}[\ell(\Theta; x)].
\end{equation}
Then the gradient of the noisy objective at the clean optimum \( \Theta_0 \) is:
\begin{align*}
\nabla_\Theta \mathcal{L}_\delta(\Theta_0) 
&= (1 - \delta) \nabla_\Theta \mathbb{E}_{x \sim \mathcal{D}_0}[\ell(\Theta_0; x)] + \delta \nabla_\Theta \mathbb{E}_{x \sim \mathcal{N}}[\ell(\Theta_0; x)] \\
&= \delta \cdot \Delta_{\mathrm{noise}}.
\end{align*}

Assume the loss is twice continuously differentiable and the clean loss is locally \( \mu \)-strongly convex. Let \( H := \nabla^2_\Theta \mathbb{E}_{x \sim \mathcal{D}_0}[\ell(\Theta; x)]|_{\Theta = \Theta_0} \succ 0 \) be the positive definite Hessian. Then, using Taylor expansion:
\begin{equation}
\nabla_\Theta \mathcal{L}_\delta(\Theta_\delta) \approx \nabla_\Theta \mathcal{L}_\delta(\Theta_0) + H (\Theta_\delta - \Theta_0) + R,
\end{equation}
where \( R \in \mathcal{O}(\|\Theta_\delta - \Theta_0\|^2) \).

At the noisy optimum, the gradient vanishes:
\begin{equation}
0 = \nabla_\Theta \mathcal{L}_\delta(\Theta_\delta) \Rightarrow \Theta_\delta - \Theta_0 \approx -H^{-1} (\delta \cdot \Delta_{\mathrm{noise}}).
\end{equation}

Now compute:
\begin{align*}
Q(\Theta_\delta) &= \|\Theta_\delta - \Theta_0\|_F^2 \\
&\approx \|\delta H^{-1} \Delta_{\mathrm{noise}}\|_F^2 \\
&= \delta^2 \cdot \|H^{-1} \Delta_{\mathrm{noise}}\|_F^2 + o(\delta^2).
\end{align*}

\subsection{Proof of Theorem 2}
\label{Proof of Theorem 2}
\textit{Proof.} 
Let \( f(u,i) = \mathrm{MLP}([\mathbf{p}_u; \mathbf{q}_i]) + \mathbf{p}_u^\top \mathbf{q}_i \), and define the loss \( \ell(\Theta; x) := \ell(f(u,i), y) \) for a sample \( x = (u, i, y) \). The gradient of the loss with respect to model parameters \( \Theta \) satisfies the chain rule:
\begin{equation}
\nabla_\Theta \ell(\Theta; x) = \nabla_z \ell(f(u,i), y) \cdot \nabla_\Theta f(u, i),
\end{equation}
where \( z := f(u, i) \in \mathbb{R} \) is the scalar output score.

Let \( J_x := \nabla_\Theta f(u, i) \) denote the Jacobian of the scoring function. Taking norms on both sides yields:
\begin{equation}
\|\nabla_\Theta \ell(\Theta; x)\| \leq \|\nabla_z \ell(f(u,i), y)\| \cdot \|J_x\|.
\end{equation}

We now analyze \( \|J_x\| \). The scoring function can be decomposed as:
\begin{equation}
f(u, i) = \mathrm{MLP}([\mathbf{p}_u; \mathbf{q}_i]) + \mathbf{p}_u^\top \mathbf{q}_i.
\end{equation}
The Jacobian of the inner product term is linear in \( \mathbf{p}_u \) and \( \mathbf{q}_i \), and contributes only a bounded term proportional to \( \|\mathbf{p}_u\| + \|\mathbf{q}_i\| \). We focus on the MLP term, which dominates in depth.

Let the MLP be defined recursively as:
\[
\begin{aligned}
h^{(0)} &= [\mathbf{p}_u; \mathbf{q}_i], \\
h^{(l)} &= \sigma^{(l)}(W^{(l)} h^{(l-1)}), \quad \text{for } l = 1, \dots, L,
\end{aligned}
\]
where \( \sigma^{(l)} \) is a pointwise activation function (e.g., ReLU or sigmoid), and \( W^{(l)} \in \mathbb{R}^{d_l \times d_{l-1}} \) is the weight matrix at layer \( l \). Then the gradient of the output with respect to parameters is:
\begin{equation}
\frac{\partial f(u, i)}{\partial \Theta} = \frac{\partial \mathrm{MLP}(h^{(0)})}{\partial \Theta} + \frac{\partial (\mathbf{p}_u^\top \mathbf{q}_i)}{\partial \Theta}.
\end{equation}

Focusing on the MLP term, its Jacobian w.r.t. \( h^{(0)} \) is:
\begin{equation}
\frac{\partial \mathrm{MLP}(h^{(0)})}{\partial h^{(0)}} = \prod_{l=1}^L D^{(l)} W^{(l)},
\end{equation}
where \( D^{(l)} := \mathrm{diag}(\sigma'^{(l)}(W^{(l)} h^{(l-1)})) \) is a diagonal matrix of activation derivatives.

Taking operator norms, we get:
\begin{equation}
\left\| \frac{\partial \mathrm{MLP}(h^{(0)})}{\partial h^{(0)}} \right\| \leq \prod_{l=1}^L \|D^{(l)}\|_2 \cdot \|W^{(l)}\|_2.
\end{equation}

Since \( \|D^{(l)}\|_2 \leq \max_x |\sigma'^{(l)}(x)| \), which is bounded for standard activations like ReLU (bounded by 1) or sigmoid (bounded by 1/4), we conclude:
\begin{equation}
\left\| \frac{\partial f(u,i)}{\partial \Theta} \right\| \leq C_\sigma \cdot \prod_{l=1}^L \|W^{(l)}\|_2,
\end{equation}
for some constant \( C_\sigma > 0 \) depending on activation functions.

Combining this with the earlier chain rule, we obtain:
\begin{equation}
\|\nabla_\Theta \ell(\Theta; x)\| \leq \|\nabla_z \ell(f(u,i), y)\| \cdot C_\sigma \cdot \prod_{l=1}^L \|W^{(l)}\|_2.
\end{equation}

Hence, the gradient norm can grow multiplicatively with the spectral norms of the MLP layers, implying exponential sensitivity in depth.

\subsection{Proof of Theorem 3}
\label{Proof of Theorem 3}
\textit{Proof.}
Decompose the embedding as $z_u = z_u^* + \delta_u$, where $z_u^* \in \mathcal{S}$ is the clean signal and $\delta_u \in \mathcal{S}^\perp$ is noise. The contrastive loss requires:
\begin{equation}
\cos(z_u', z_u'') = \frac{(z_u^* + \delta_u')^T (z_u^* + \delta_u'')}{\|z_u^* + \delta_u'\| \|z_u^* + \delta_u''\|} \approx 1.
\end{equation}
Expanding this approximation:
\begin{equation}
1 - \frac{\|\delta_u' - \delta_u''\|^2}{2\|z_u^*\|^2} \approx 1 \implies \|\delta_u' - \delta_u''\|^2 \approx 0.
\end{equation}
Since $\delta_u'$ and $\delta_u''$ are i.i.d., the expected variance is:
\begin{equation}
\mathbb{E}[\|\delta_u' - \delta_u''\|^2] = 2\mathbb{E}[\|\delta_u\|^2] - 2\mathbb{E}[\delta_u'^T \delta_u'']] = 2\text{Var}(\delta_u).
\end{equation}
Thus, $\text{Var}(\delta_u) \leq \epsilon^2$, constraining noise to low variance and ensuring embeddings primarily reside in $\mathcal{S}$.

\subsection{Derivation of Definition 4}
\label{derivation 4}
\textit{Derivation.}  
We begin by recursively applying the propagation rule described in Eq.~(\ref{eqthm2}). The embeddings of the $l$-th layer are given by:
\begin{equation}
\mathbf{P}^{(l)} = \tilde{A} \mathbf{P}^{(l-1)},
\end{equation}
where $\mathbf{P}^{(0)}$ represents the initial embeddings, and $\tilde{A}$ is the normalized adjacency matrix. This recurrence relation reflects the embedding propagation at each layer of the graph convolutional network.

The final embeddings after $L$ layers of propagation are obtained by taking the weighted sum across all layers:
\begin{equation}
\mathbf{P} = \sum_{l=0}^L \beta_l \tilde{A}^l \mathbf{P}^{(0)},
\end{equation}
where $\beta_l = \frac{1}{L+1}$ is the weight for the $l$-th layer. 

Substituting the spectral decomposition of $\tilde{A}$ into this expression, we get:
\begin{equation}
\tilde{A} = U \Lambda U^T,
\end{equation}
where $U$ is the matrix of eigenvectors and $\Lambda$ is the diagonal matrix of eigenvalues of $\tilde{A}$. Therefore, we can rewrite the embeddings as:
\begin{equation}
\mathbf{P} = \sum_{l=0}^L \beta_l (U \Lambda^l U^T) \mathbf{P}^{(0)}.
\end{equation}
Using the associativity of matrix multiplication, this simplifies to:
\begin{equation}
\mathbf{P} = U \left( \sum_{l=0}^L \beta_l \Lambda^l \right) U^T \mathbf{P}^{(0)}.
\end{equation}

Now, let us analyze the behavior of the filter coefficients $\sum_{l=0}^L \beta_l \Lambda^l$ in the spectral domain. For each eigenvalue $\lambda_k$ of $\tilde{A}$, the corresponding coefficient behaves as follows:

For $\lambda_k \approx 1$ (low-frequency components), we have:
\begin{equation}
\sum_{l=0}^L \beta_l \lambda_k^l \approx \sum_{l=0}^L \frac{1}{L+1} \cdot 1^l = 1.
\end{equation}
This means that low-frequency components are preserved.

For $\lambda_k \approx 0$ (high-frequency components), we have:
\begin{equation}
\sum_{l=0}^L \beta_l \lambda_k^l \approx \sum_{l=0}^L \frac{1}{L+1} \cdot 0^l = 0.
\end{equation}
This means that high-frequency components are attenuated, effectively suppressing noise.

Thus, the filter $\sum_{l=0}^L \beta_l \Lambda^l$ acts as a low-pass filter that suppresses high-frequency components and retains the low-frequency components of the embeddings.

As a result, the final embedding matrix $\mathbf{P}$ is dominated by the low-frequency components, and the rank of $\mathbf{P}$ is bounded by the number of significant eigenvalues (those corresponding to low-frequency components). Specifically, if $r$ is the number of significant eigenvalues, we can conclude that the rank of $\mathbf{P}$ is bounded by $r$:
\begin{equation}
\text{rank}(\mathbf{P}) \leq r.
\end{equation}

This establishes the low-rank property of the embeddings after $L$ layers of propagation in the LightGCN model.

\section{Experimental Settings}
\subsection{Hyperparameters}
\label{detail_setting}
We pick the most commonly used Top-$k$ metrics NDCG@$k$ to measure performance, as well as for a comprehensive view of the phenomenon, we set $k$ to be 20. We set the embedding dimensions as powers of 2, starting from 2 and going up to $2^{16}$.
We train the model for 300 epochs with a batch size of 2,048 using the Adam optimizer. The learning rate is fixed at 0.001 without weight decay (L2 regularization), and we disable gradient norm clipping. For negative sampling during training, we adopt a static strategy with 1 negative item per positive interaction, sampled uniformly from the item space. We evaluate the model after every training epoch and implement early stopping if validation performance does not improve for 10 consecutive epochs. All models operate in essentially the same way; apart from differences in architecture, aspects like training loss are nearly identical. 

\subsection{Pytorch Code of BPR\_Drop}
\label{code}
\begin{lstlisting}
class BPRdropLoss(nn.Module):
    """BPRLoss, based on Bayesian Personalized Ranking

    Args:
        - gamma(float): Small value to avoid division by zero

    Shape:
        - Pos_score: (N)
        - Neg_score: (N), same shape as the Pos_score
        - Output: scalar.

    Examples::

        >>> loss = BPRLoss()
        >>> pos_score = torch.randn(3, requires_grad=True)
        >>> neg_score = torch.randn(3, requires_grad=True)
        >>> output = loss(pos_score, neg_score)
        >>> output.backward()
    """

    def __init__(self, save_ratio, get_low, gamma=1e-10):
        super(BPRdropLoss, self).__init__()
        self.gamma = gamma
        self.save_ratio = save_ratio
        self.get_low = get_low

    def forward(self, pos_score, neg_score):
        # loss = -torch.log(self.gamma + torch.sigmoid(pos_score - neg_score)).mean()
        per_sample_losses = -torch.log(self.gamma + torch.sigmoid(pos_score - neg_score))

        num_samples = per_sample_losses.size(0)
        k = int(num_samples * self.save_ratio)

        k = max(k, 1)

        if self.get_low == 1:
            _, indices = torch.topk(per_sample_losses, k, largest=False)
        else: 
            _, indices = torch.topk(per_sample_losses, k, largest=True)

        selected_losses = per_sample_losses[indices].mean()
        return selected_losses
\end{lstlisting}
\subsection{Detailed description of the datasets}
In our research, we utilized four distinct datasets to ensure the comprehensiveness and robustness of our analysis. Each dataset contributes unique characteristics to our study. Table~\ref{dataset} summarizes the statistics of the 10 datasets used in our experiments, demonstrating the comprehensiveness of our evaluation. The datasets span a wide range of interaction scales, from Small (e.g., ML-100K) to X-Large (e.g., Amazon\_Books), and cover various sparsity levels, from relatively dense (e.g., ML-1M) to extremely sparse (e.g., Amazon\_Beauty). Notably, the majority of datasets fall into the Ultra-sparse category, reflecting the real-world challenge of data sparsity in recommendation. This diversity ensures a thorough and robust evaluation of model performance across different data regimes. The color-coded annotations provide intuitive insights into dataset characteristics.

\subsection{Reason for Choosing Models}
\label{reason}
BPR~\citep{bpr} is a classic collaborative filtering model known for its simplicity and ease of use. Many subsequent influential methods have been developed based on it. NeuMF~\citep{neumf}, in contrast, is a highly representative deep learning-based model that adopts a classic two-tower architecture, combining linear modeling via matrix factorization with nonlinear interaction modeling through neural networks. As research progressed, it became clear that the relationships between users and items can be naturally modeled as graph structures. This insight led to the emergence of numerous graph-based collaborative filtering approaches, such as LightGCN~\citep{he2020lightgcn}. Building upon graph structures, graph self-supervised models—such as SGL~\citep{sgl}—further enhance performance by leveraging both feature and structural information, attracting significant attention. These are the reasons why we chose to focus on these three categories of models. The more specific information is at Table~\ref{tab:rec_models}.

\begin{table}[ht]
\centering
\small
\caption{collaborative filtering models and their publication details}
\begin{tabular}{@{} l p{8cm} l l l @{}}
\toprule
\textbf{Model} & \textbf{Paper Title} & \textbf{Year} & \textbf{Venue} & \textbf{Citation} \\
\midrule
BPR & BPR: Bayesian Personalized Ranking from Implicit Feedback & 2009 & UAI & 8,232 \\
NeuMF & Neural Collaborative Filtering & 2017 & WWW & 8,052 \\
LightGCN & LightGCN: Simplifying and Powering Graph Convolution Network for Recommendation & 2020 & SIGIR & 4,358 \\
SGL & Self-supervised Graph Learning for Recommendation & 2021 & SIGIR & 1,423 \\
\bottomrule
\end{tabular}
\label{tab:rec_models}
\end{table}

\begin{table*}[ht]
\begin{center}
\caption{
Statistics of the 10 benchmark datasets used in our experiments, highlighting both the dataset scale and sparsity level. 
\textbf{Scale} is categorized based on the number of interactions: 
\cellcolor{cyan!10}Small ($<$1M), 
\cellcolor{cyan!20}Medium (1M–5M), 
\cellcolor{cyan!30}Large (5M–10M), and 
\cellcolor{cyan!40}X-Large ($>$10M). 
\textbf{Sparsity Level} is categorized based on the sparsity ratio: 
\cellcolor{red!50}Dense ($<$0.6), 
\cellcolor{red!25}Semi-sparse (0.6–0.95), 
\cellcolor{red!20}Sparse (0.95–0.99), and 
\cellcolor{red!10}Ultra-sparse ($\geq$0.99). 
Colored backgrounds provide intuitive cues for comparison across datasets.
}
\label{dataset}
\resizebox{\textwidth}{!}{
\begin{tabular}{crrrcclc}
  \toprule
  \textbf{Dataset} & \textbf{\#Users} & \textbf{\#Items} & \textbf{\#Interactions} & \textbf{Sparsity} & \textbf{Scale} & \textbf{Sparsity Level} \\
  \midrule
  ML-100K       & 943        & 1,682      & 100,000       & 0.93695       & \cellcolor{cyan!10}Small       & \cellcolor{red!25}Semi-sparse \\
  Modcloth      & 9,000      & 10,000     & 500,000       & 0.99945       & \cellcolor{cyan!10}Small       & \cellcolor{red!10}Ultra-sparse \\
  Amazon\_Baby  & 531,890    & 64,426     & 915,446       & 0.99997       & \cellcolor{cyan!10}Small       & \cellcolor{red!10}Ultra-sparse \\
  ML-1M         & 6,040      & 3,706      & 1,000,209     & 0.97570       & \cellcolor{cyan!20}Medium      & \cellcolor{red!20}Sparse \\
  Douban        & 11,257     & 2,281      & 1,330,000     & 0.99945       & \cellcolor{cyan!20}Medium      & \cellcolor{red!10}Ultra-sparse \\
  Pinterest     & 55,187	   & 9,911	    & 1,445,622   & 0.99736       & \cellcolor{cyan!20}Medium      & \cellcolor{red!10}Ultra-sparse \\
  Amazon\_Beauty& 1,210,271  & 249,274    & 2,023,070     & 0.99999       & \cellcolor{cyan!20}Medium      & \cellcolor{red!10}Ultra-sparse \\
  Yelp          & 1,326,101  & 174,567    & 5,261,669     & 0.99998       & \cellcolor{cyan!30}Large       & \cellcolor{red!10}Ultra-sparse \\
  Gowalla       & 107,092    & 1,280,969  & 6,442,892     & 0.99990       & \cellcolor{cyan!30}Large       & \cellcolor{red!10}Ultra-sparse \\
  Amazon\_Books & 8,026,324  & 2,330,066  & 22,507,155    & 0.99990       & \cellcolor{cyan!40}X-Large     & \cellcolor{red!10}Ultra-sparse \\
  \bottomrule
\end{tabular}
}
\end{center}
\end{table*}

\section{Experiments}
\subsection{All Observations}
\label{e1}
\begin{figure*}[t!]
\centering
\subfloat[\textbf{Amazon Beauty- BPR}]
{\includegraphics[width=0.32\linewidth]{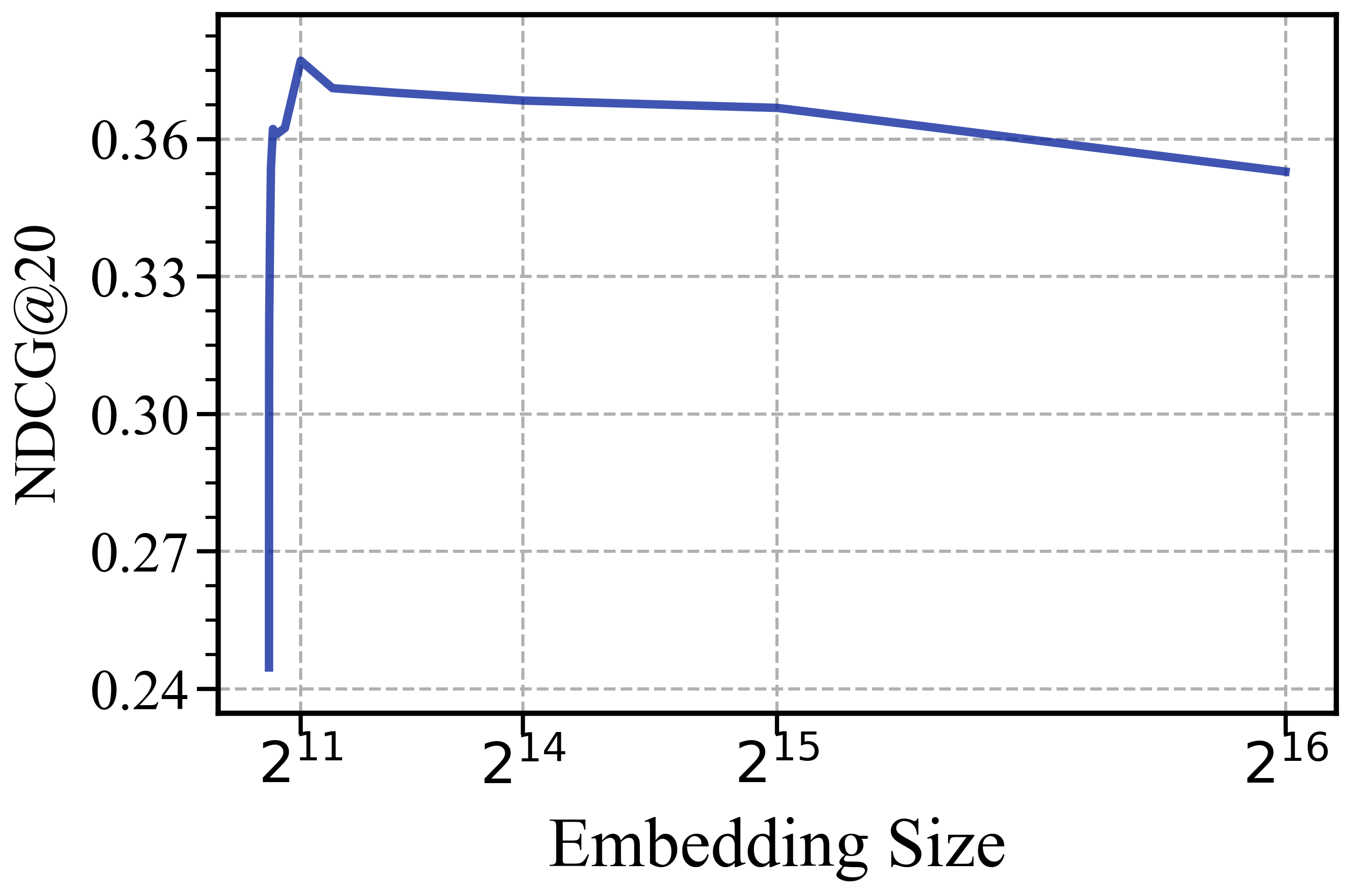}} \hfill
\subfloat[\textbf{Amazon Baby - BPR}]
{\includegraphics[width=0.32\linewidth]{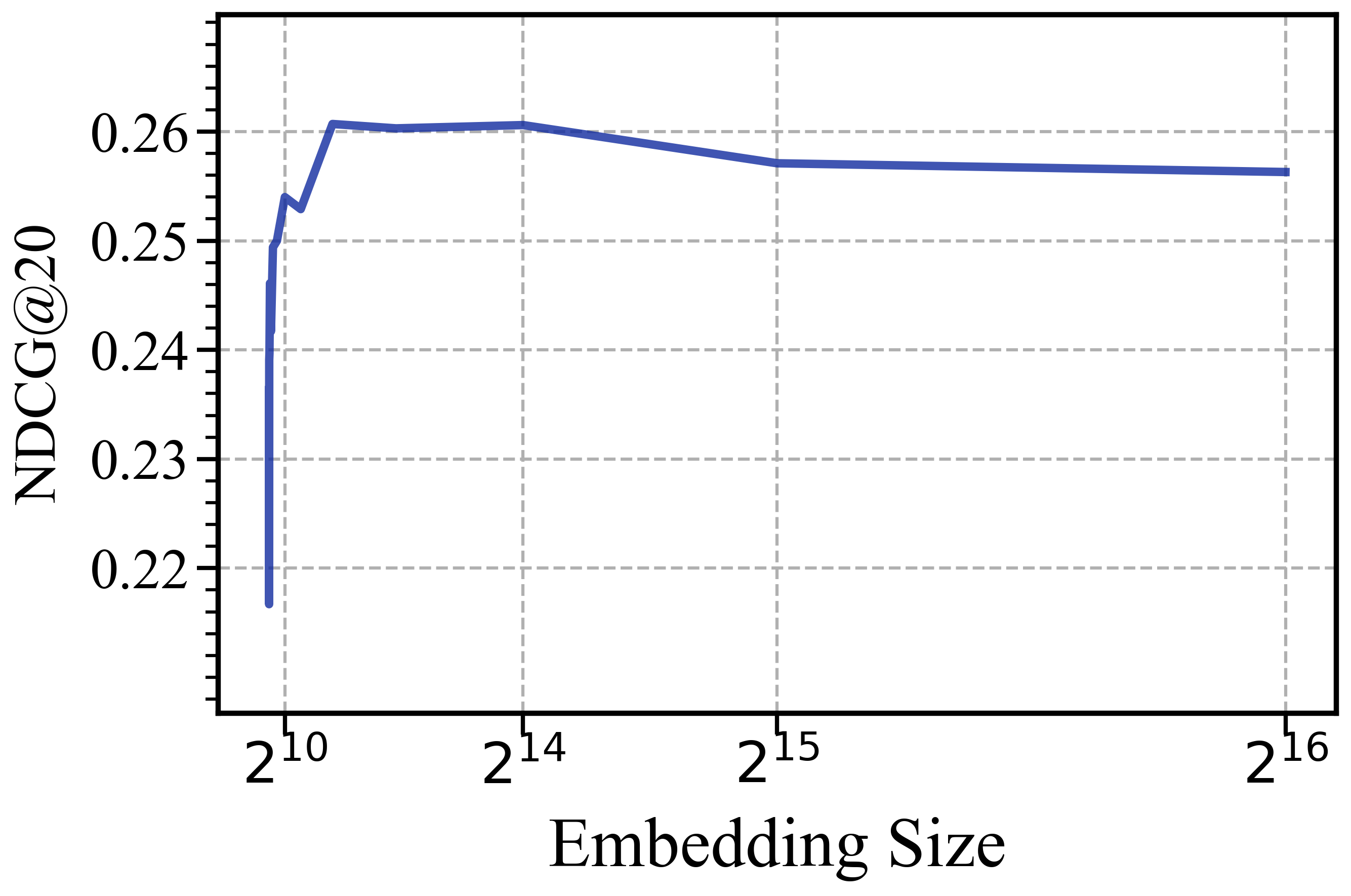}} \hfill
\subfloat[\textbf{Amazon Books - BPR}]
{\includegraphics[width=0.32\linewidth]{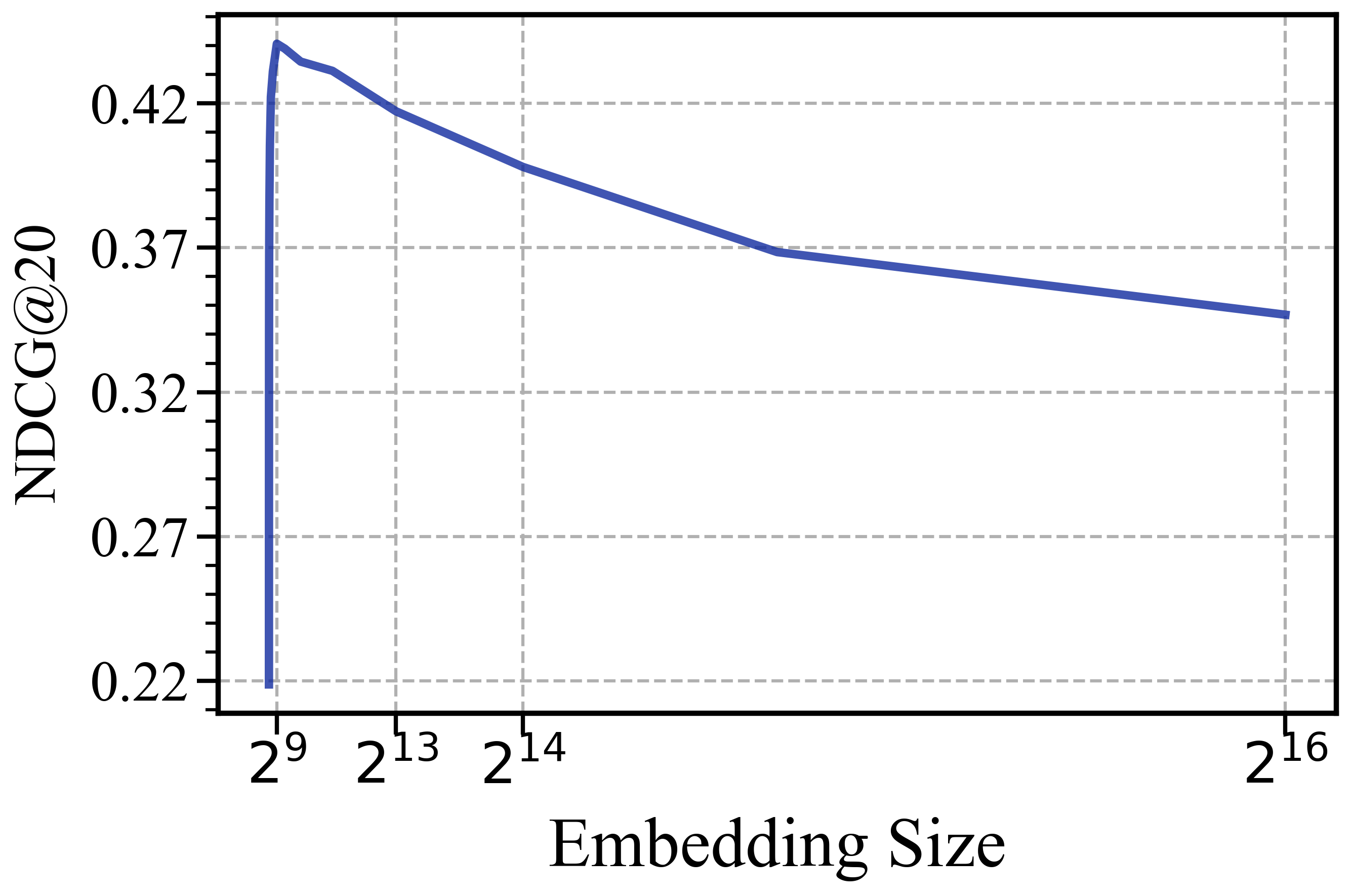}} \
\vspace{-10pt}
\subfloat[\textbf{Amazon Beauty - NeuMF}]
{\includegraphics[width=0.32\linewidth]{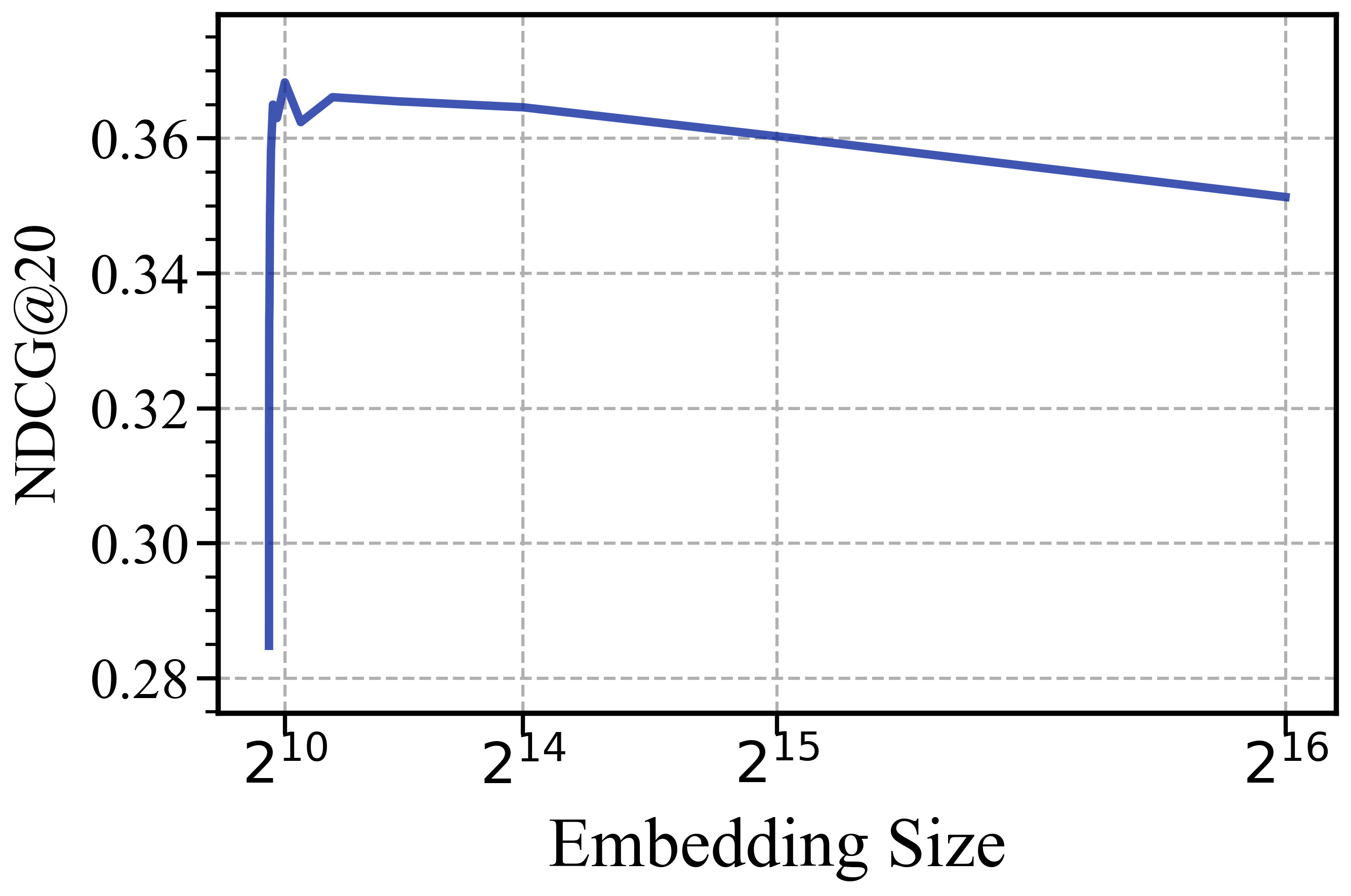}} \hfill
\subfloat[\textbf{Amazon Baby - NeuMF}]
{\includegraphics[width=0.32\linewidth]{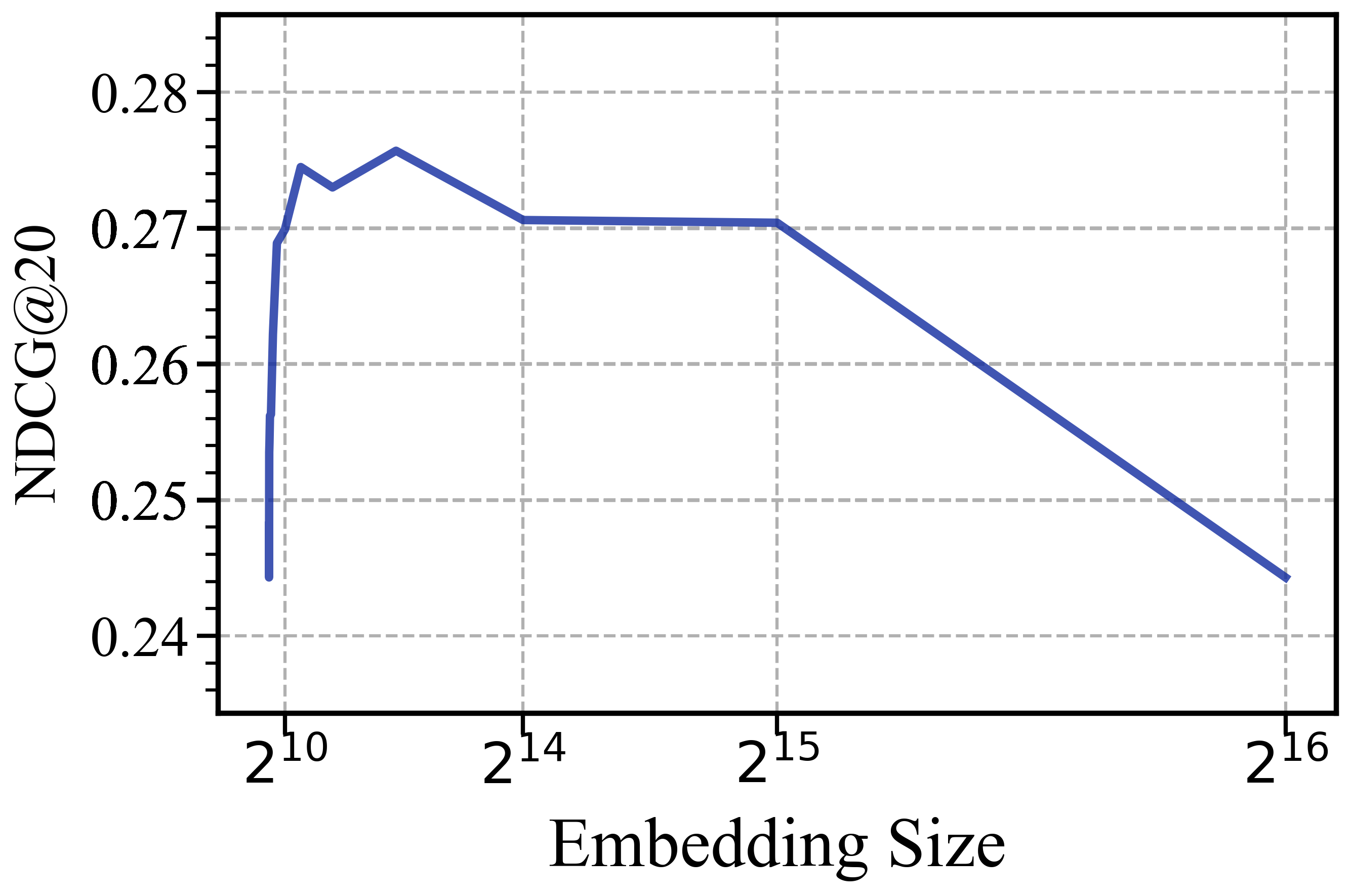}} \hfill
\subfloat[\textbf{Amazon Books - NeuMF}]
{\includegraphics[width=0.32\linewidth]{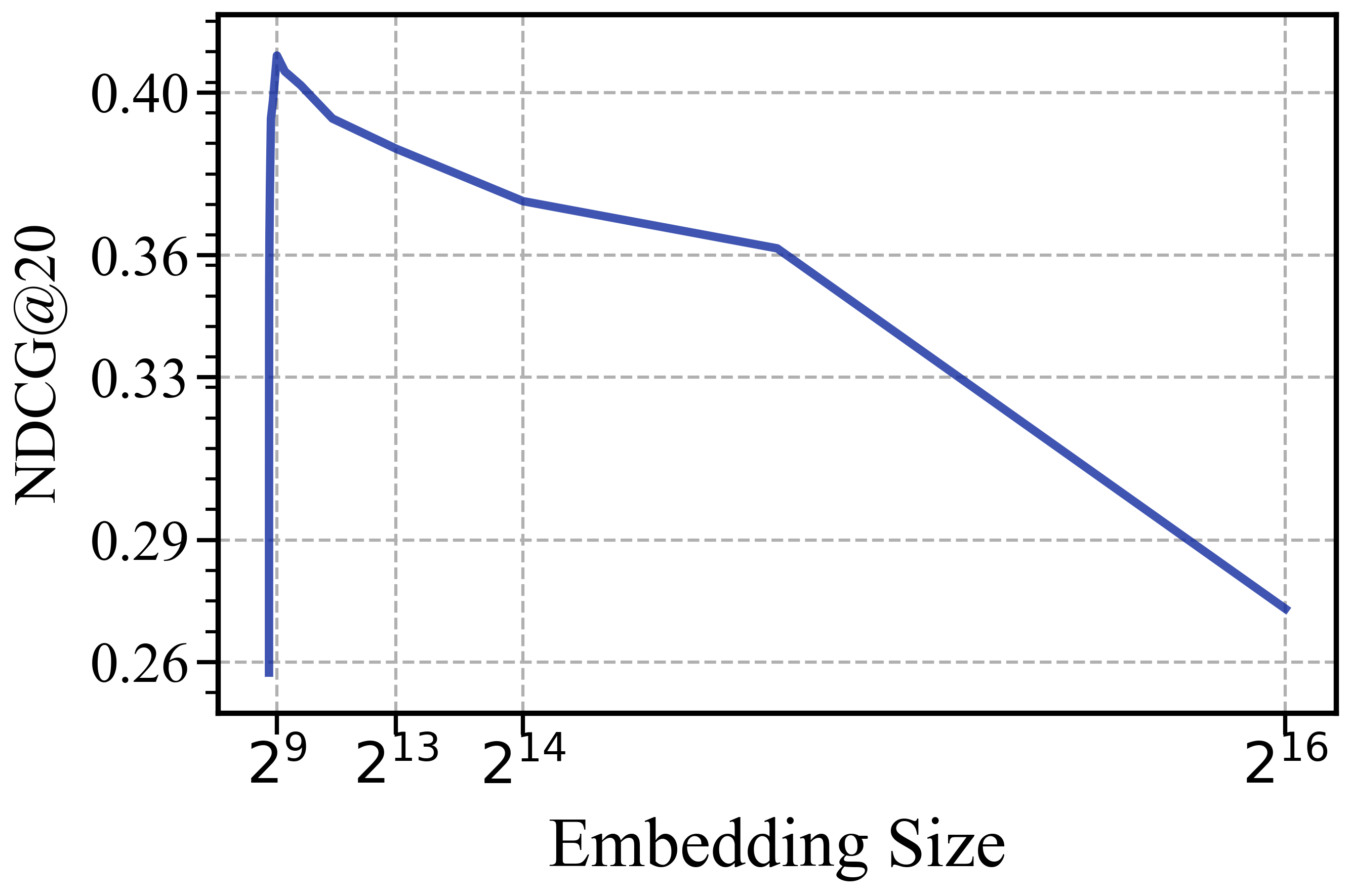}} \
\vspace{-10pt}
\subfloat[\textbf{Amazon Beauty - LightGCN}]
{\includegraphics[width=0.32\linewidth]{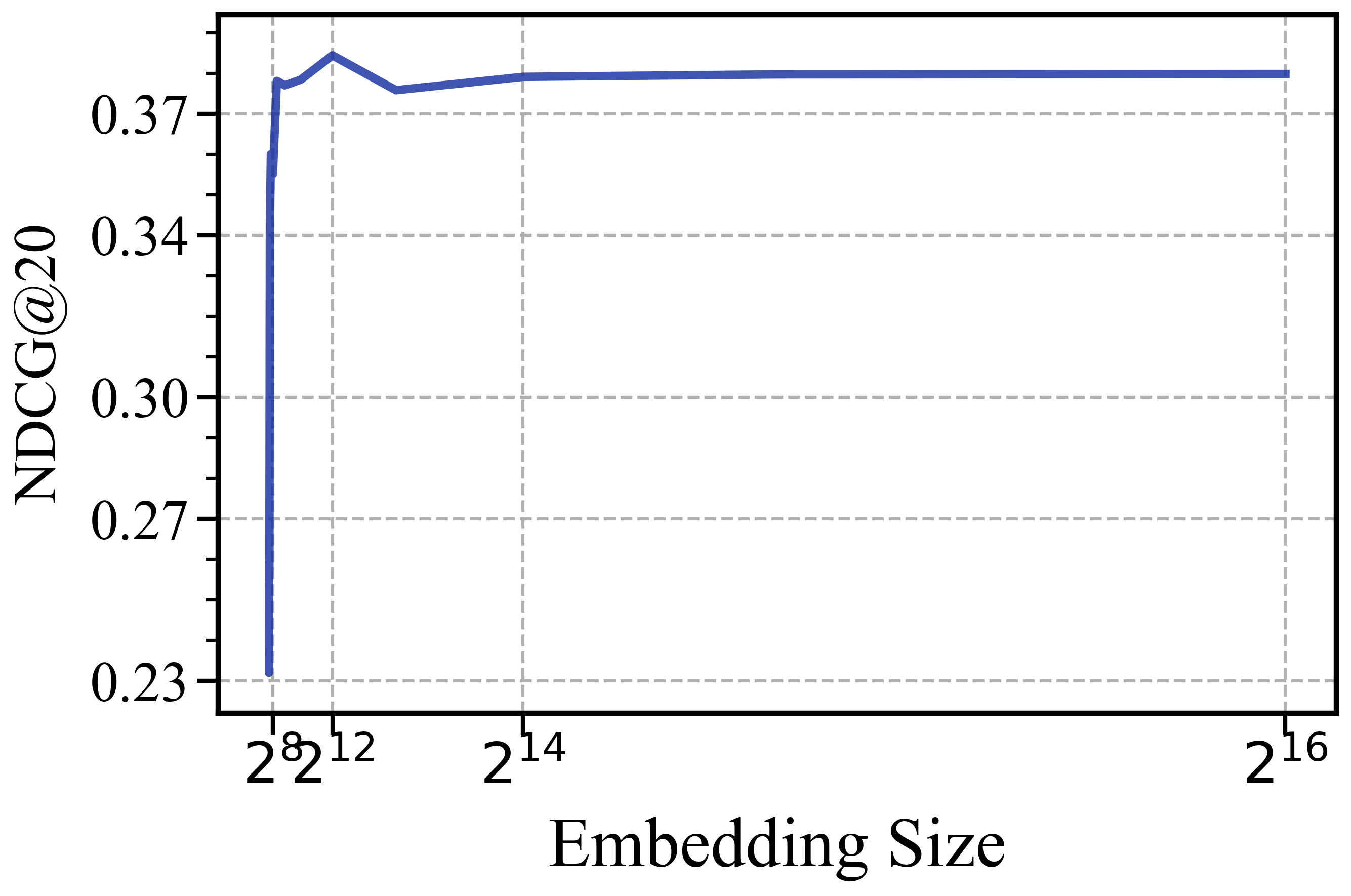}} \hfill
\subfloat[\textbf{Amazon Baby - LightGCN}]
{\includegraphics[width=0.32\linewidth]{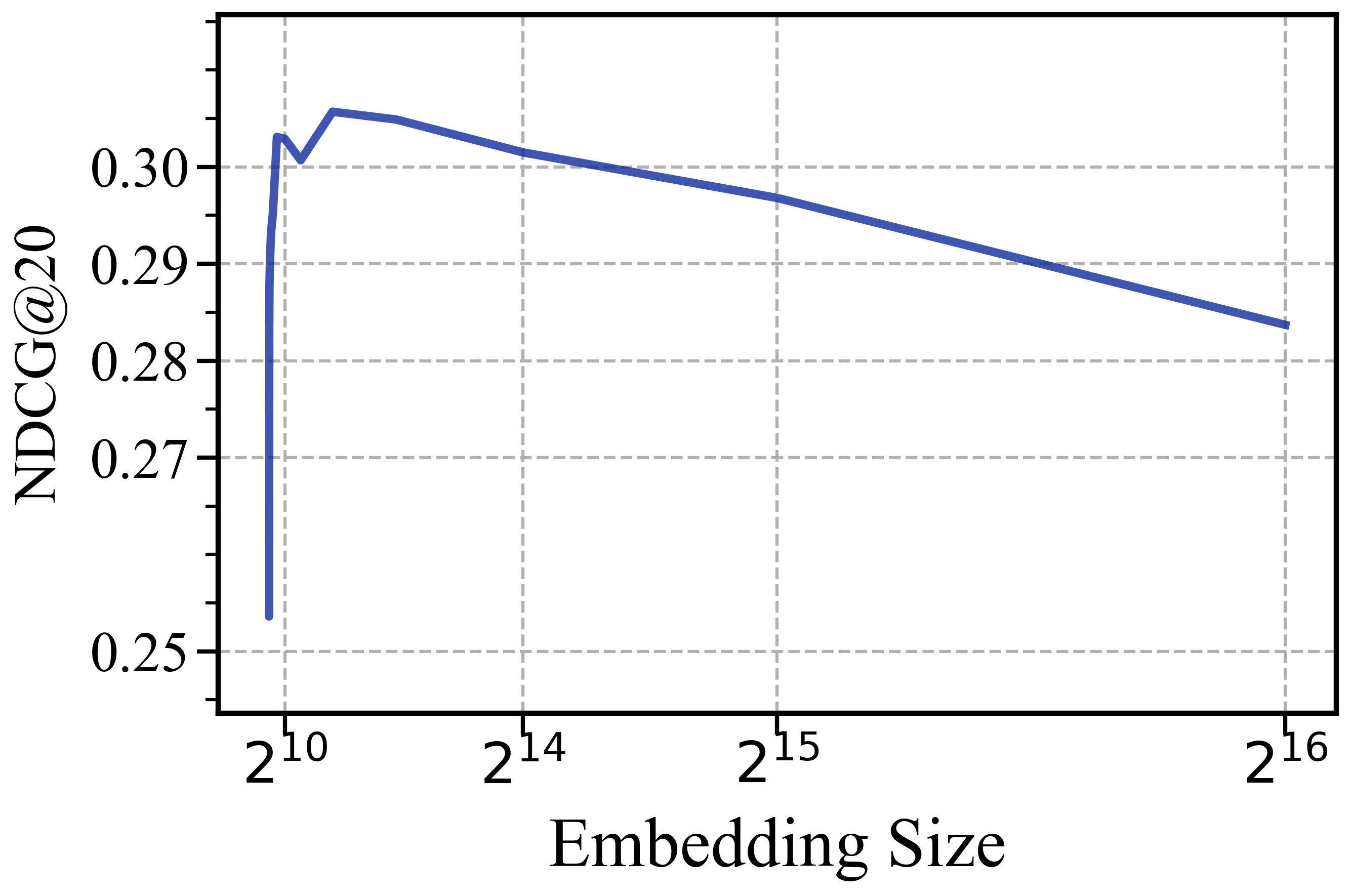}} \hfill
\subfloat[\textbf{Amazon Books - LightGCN}]
{\includegraphics[width=0.32\linewidth]{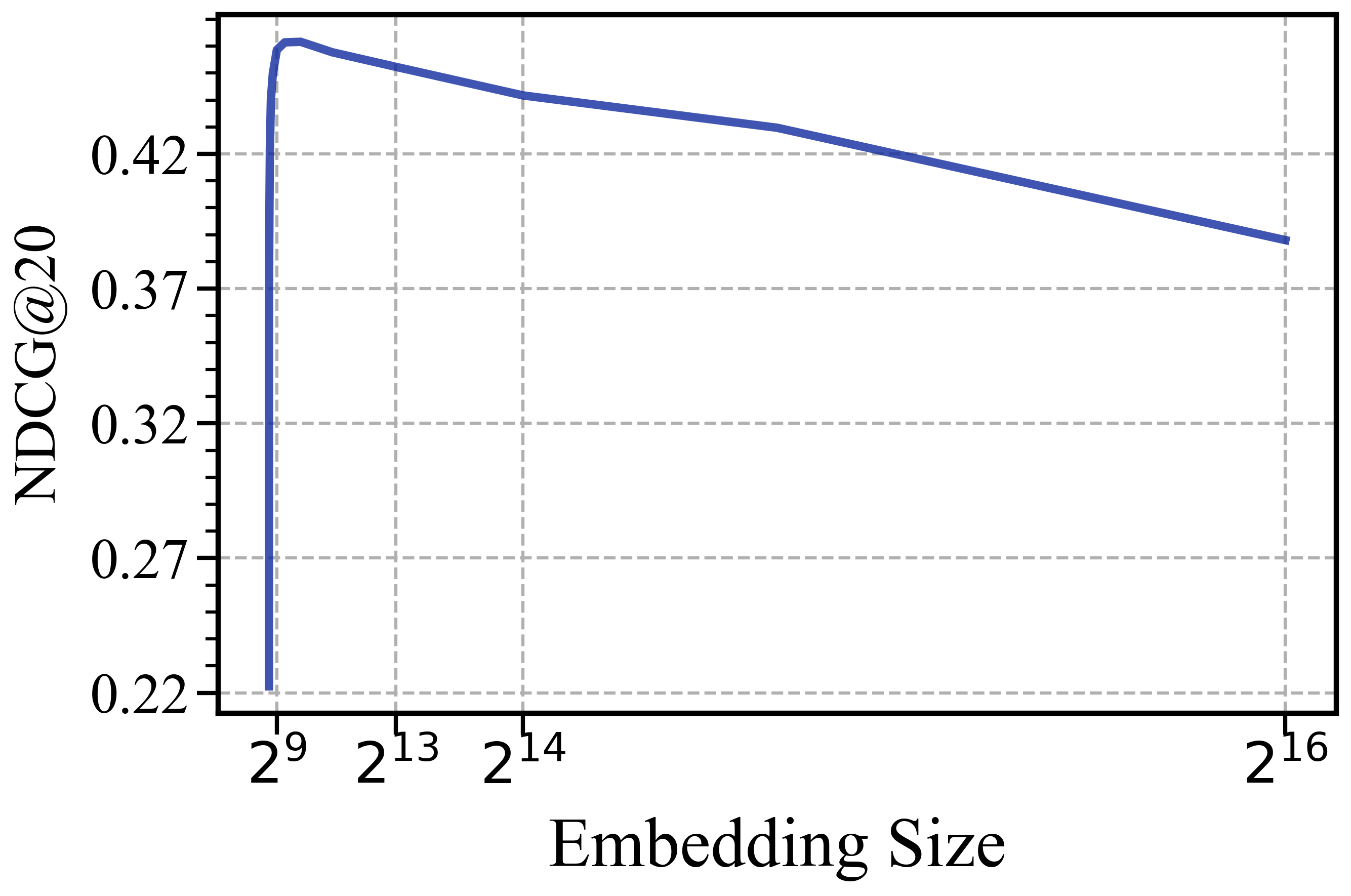}} \
\vspace{-10pt}
\subfloat[\textbf{Amazon Beauty - SGL}]
{\includegraphics[width=0.32\linewidth]{Figures/SGL-modcloth-N20.png}} \hfill
\subfloat[\textbf{Amazon Baby - SGL}]
{\includegraphics[width=0.32\linewidth]{Figures/SGL-douban-N20.png}} \hfill
\subfloat[\textbf{Amazon Books - SGL}]
{\includegraphics[width=0.32\linewidth]{Figures/SGL-ml-100k-N20.png}}
\vspace{5pt}
\caption{Scale the embedding dimension exponentially by a factor of 2 across different collaborative filtering models and datasets. Each row corresponds to a model (BPR, LightGCN, SGL, NeuMF), and each column represents a dataset (Amazon Beauty, Amazon Baby, Amazon Books).}
\label{scale2}
\end{figure*}

\begin{figure*}[t!]
\centering
\subfloat[\textbf{Yelp - BPR}]
{\includegraphics[width=0.32\linewidth]{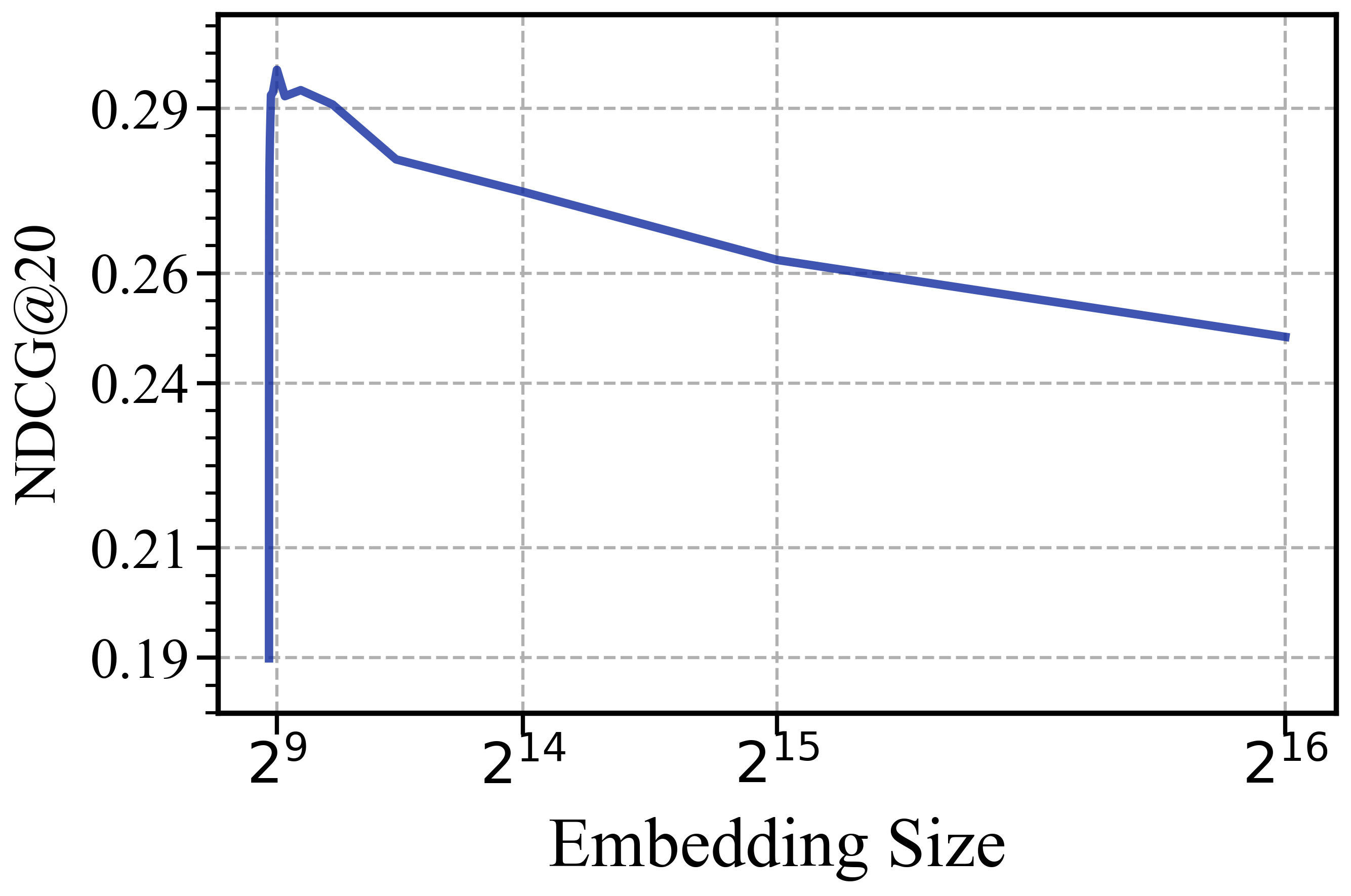}} \hfill
\subfloat[\textbf{Gowalla - BPR}]
{\includegraphics[width=0.32\linewidth]{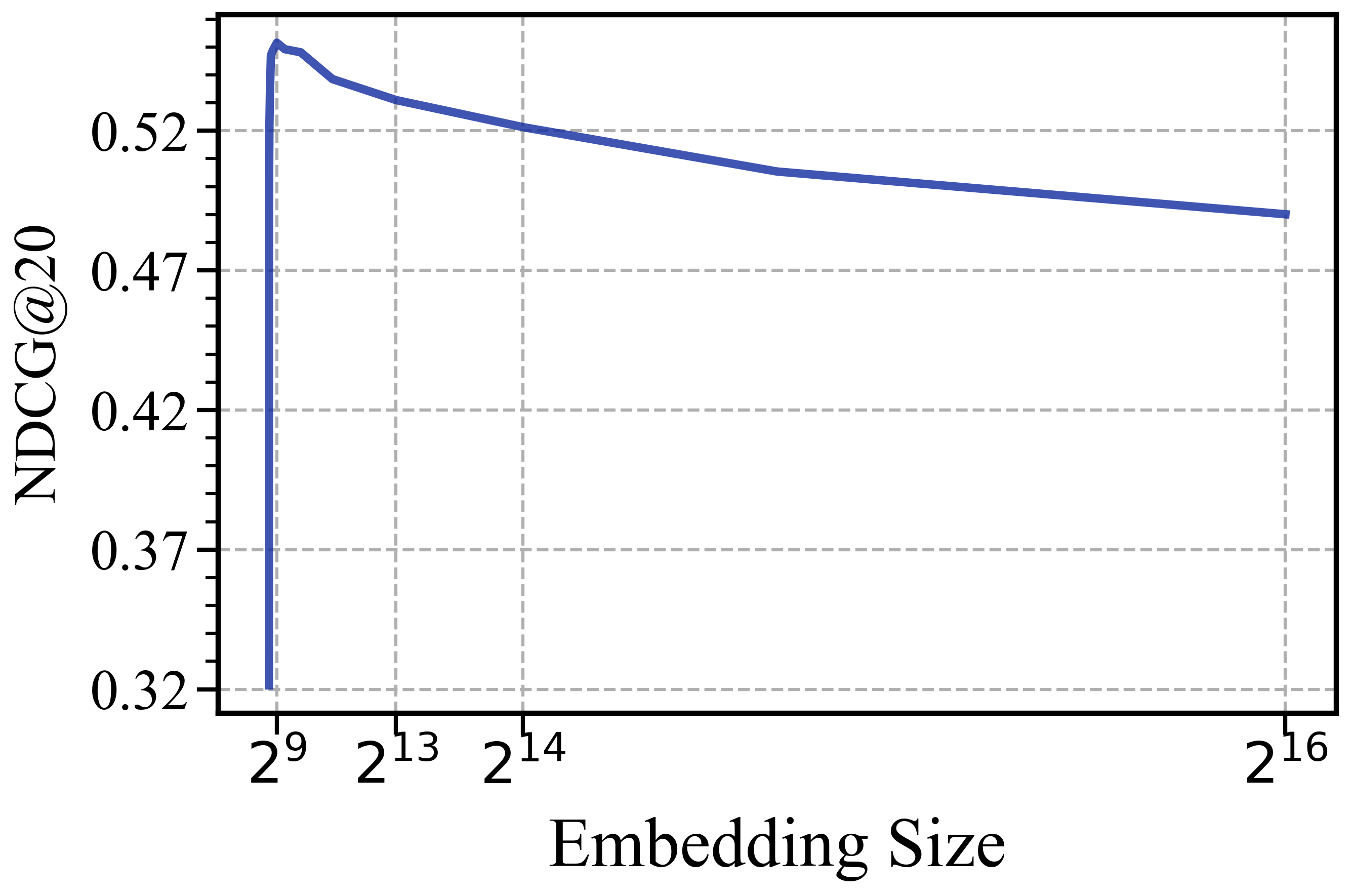}} \hfill
\subfloat[\textbf{Pinterest - BPR}]
{\includegraphics[width=0.32\linewidth]{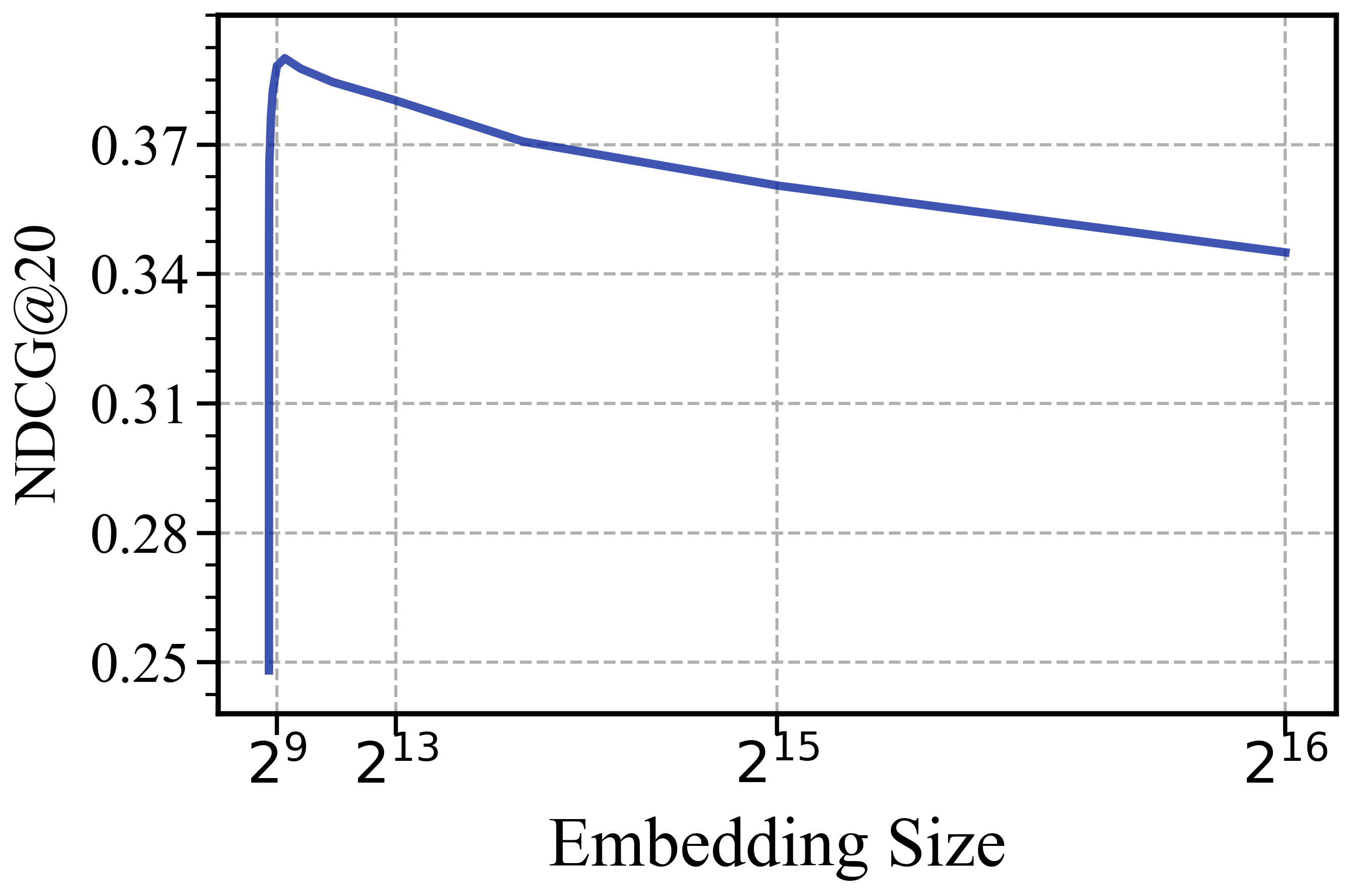}} \
\vspace{-10pt}
\subfloat[\textbf{Yelp - NeuMF}]
{\includegraphics[width=0.32\linewidth]{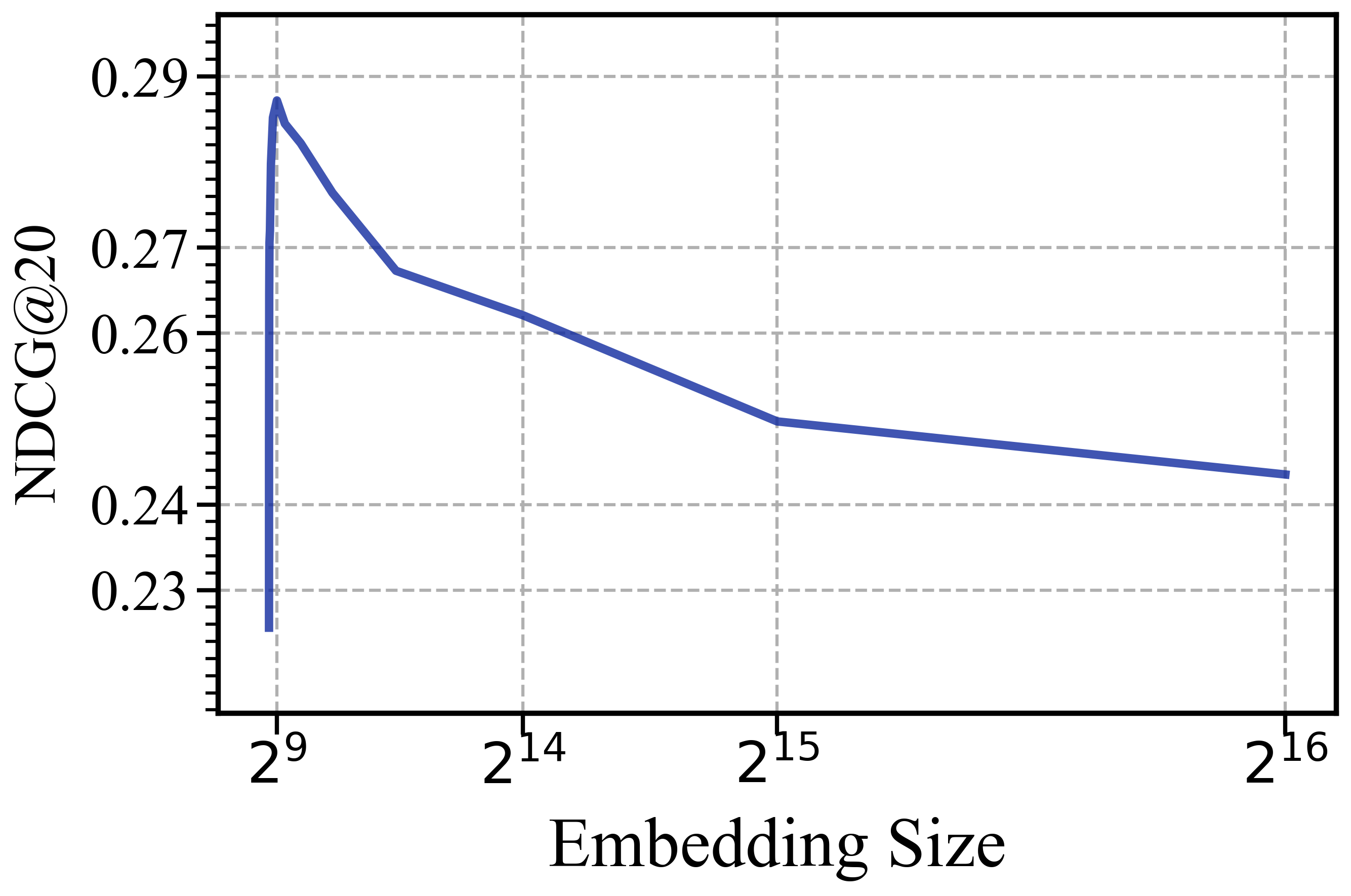}} \hfill
\subfloat[\textbf{Gowalla - NeuMF}]
{\includegraphics[width=0.32\linewidth]{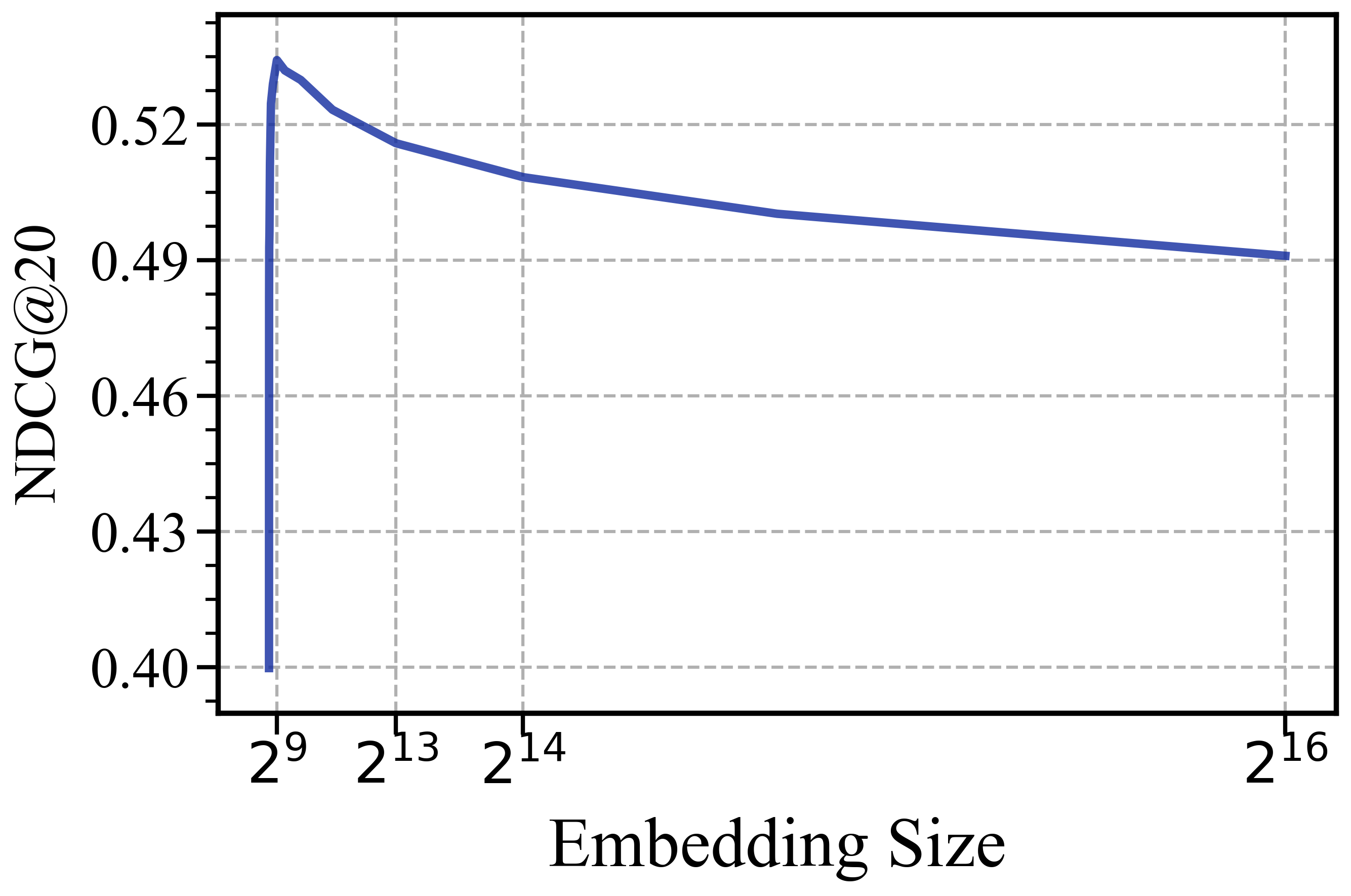}} \hfill
\subfloat[\textbf{Pinterest - NeuMF}]
{\includegraphics[width=0.32\linewidth]{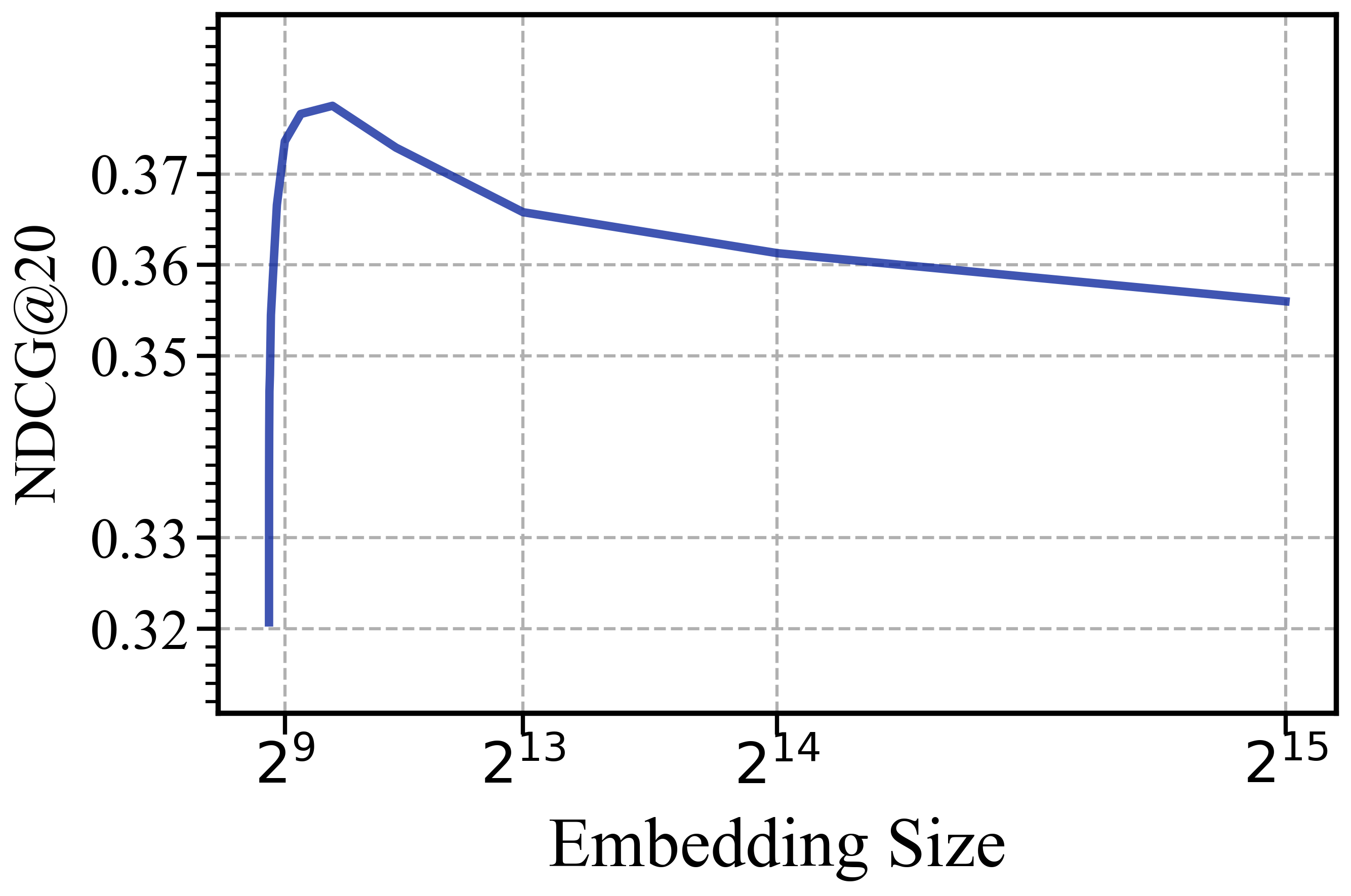}} \
\vspace{-10pt}
\subfloat[\textbf{Yelp - LightGCN}]
{\includegraphics[width=0.32\linewidth]{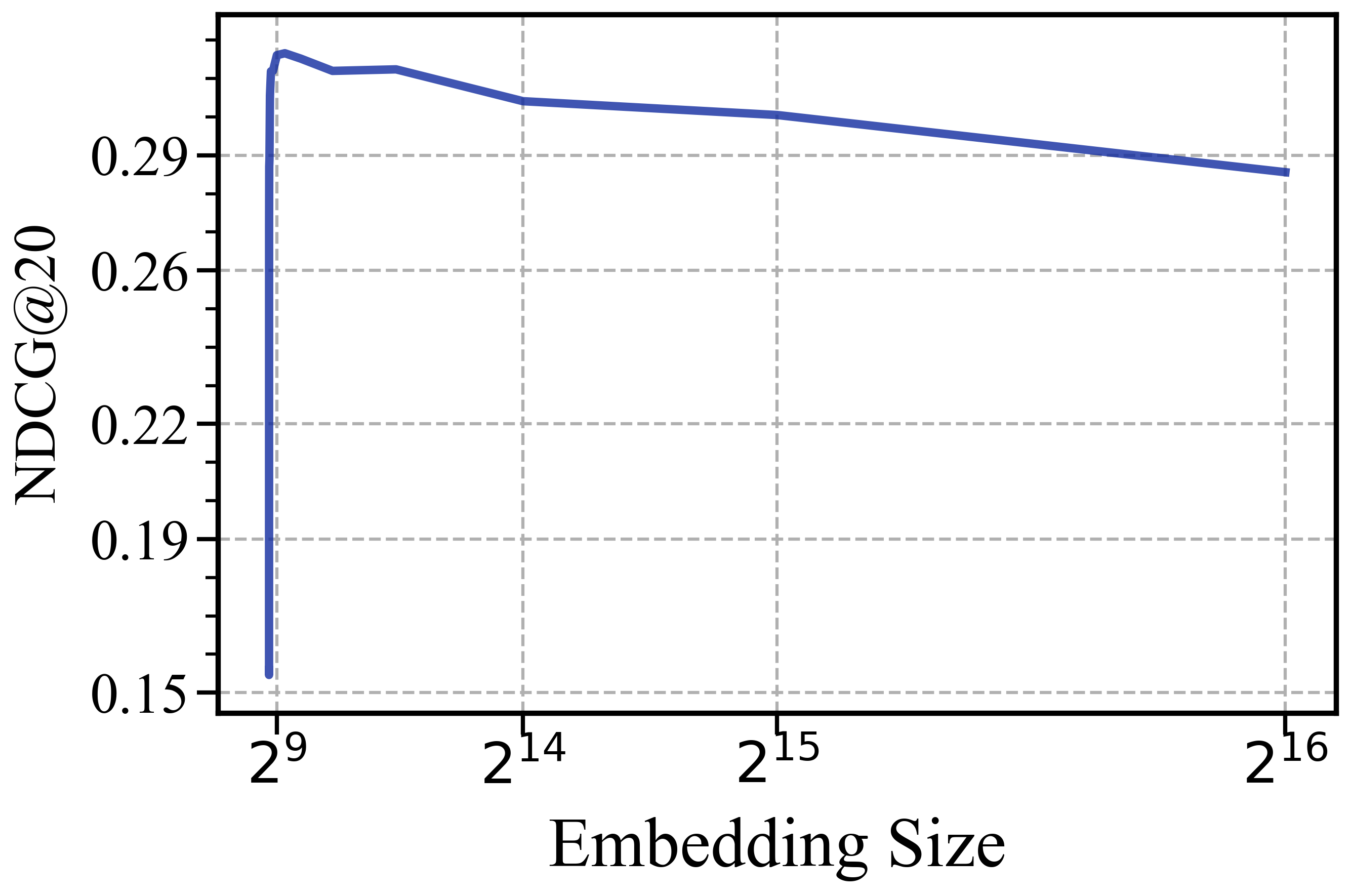}} \hfill
\subfloat[\textbf{Gowalla - LightGCN}]
{\includegraphics[width=0.32\linewidth]{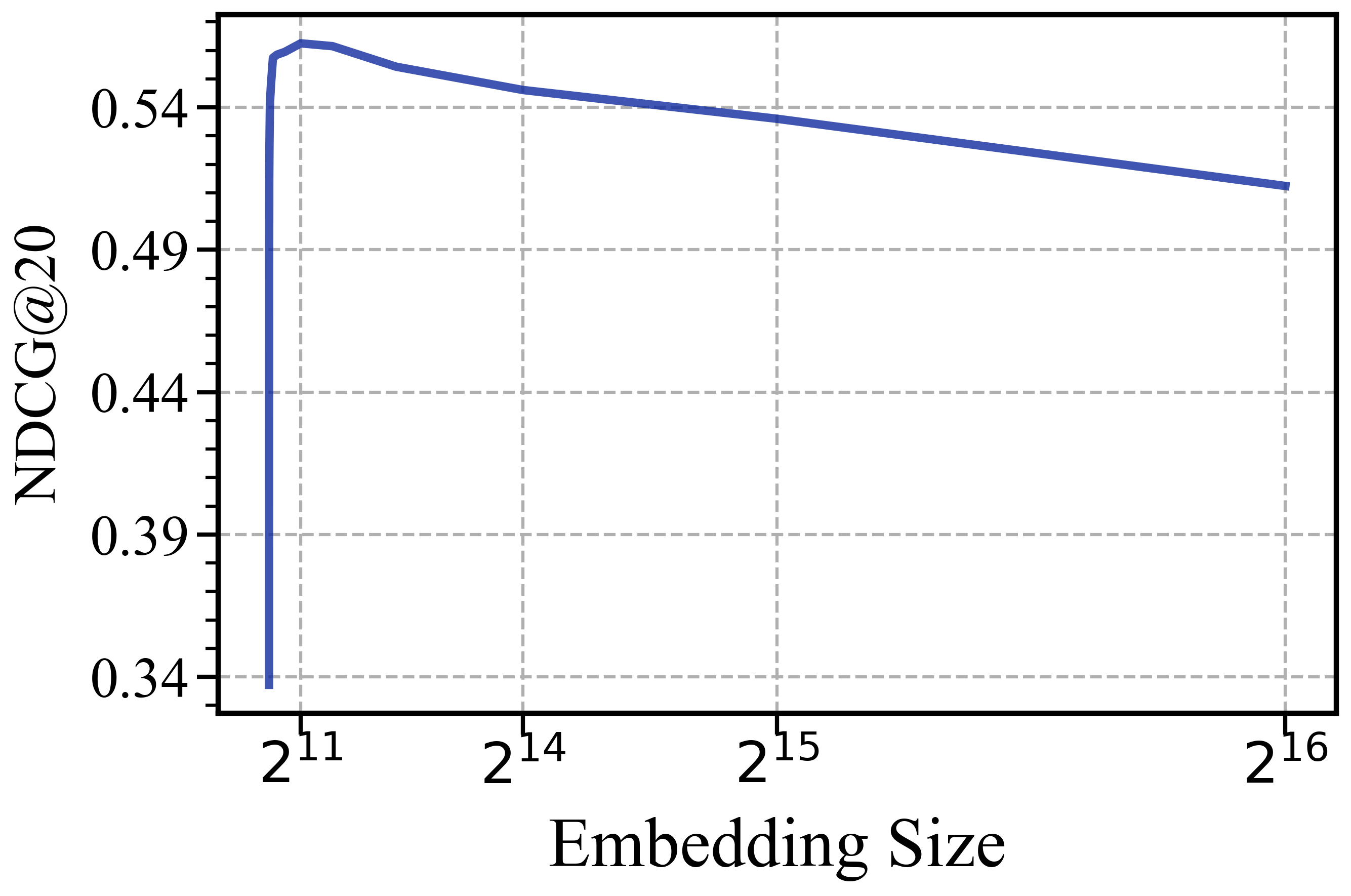}} \hfill
\subfloat[\textbf{Pinterest - LightGCN}]
{\includegraphics[width=0.32\linewidth]{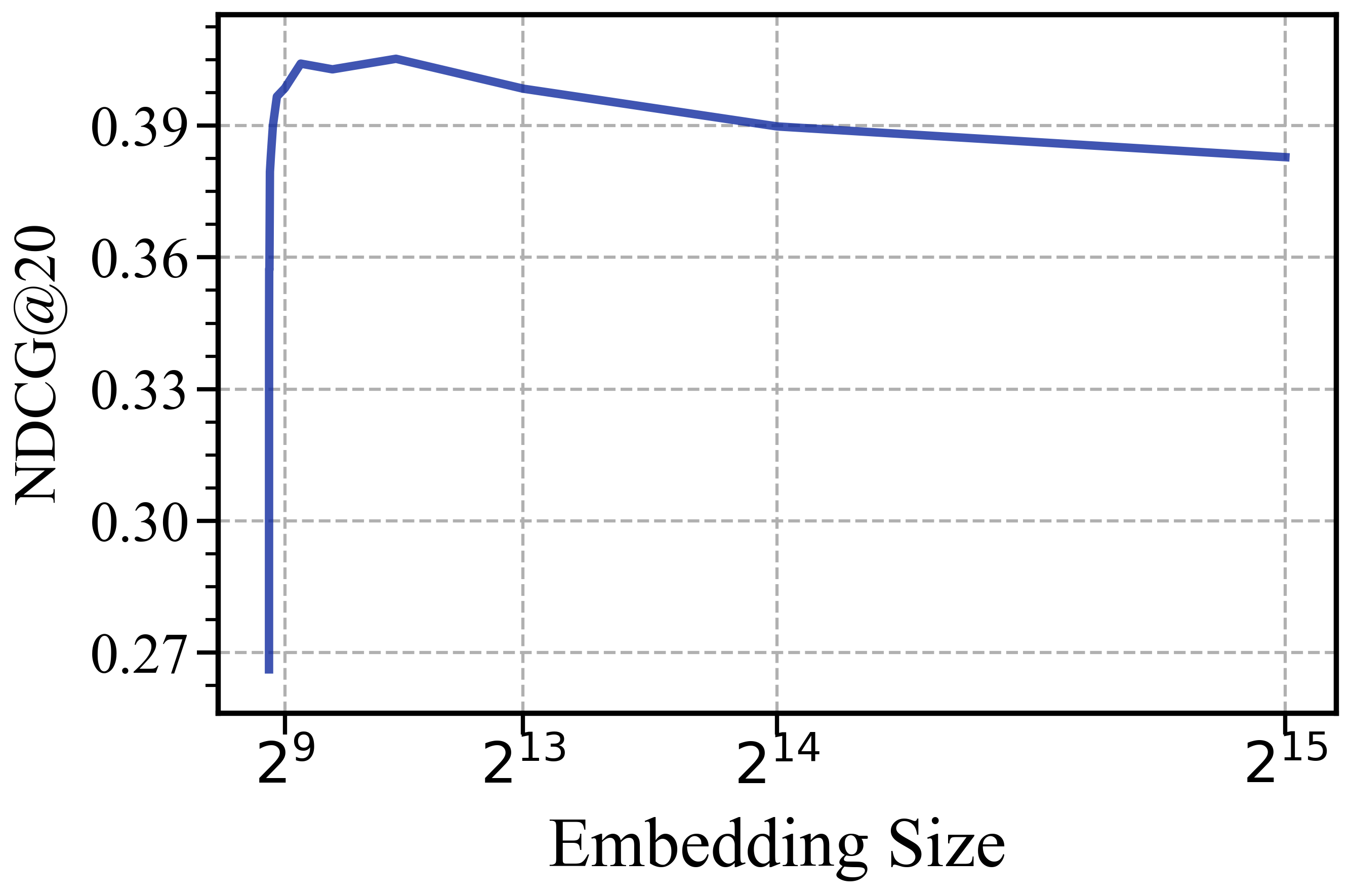}} \
\vspace{-10pt}
\subfloat[\textbf{Yelp - SGL}]
{\includegraphics[width=0.32\linewidth]{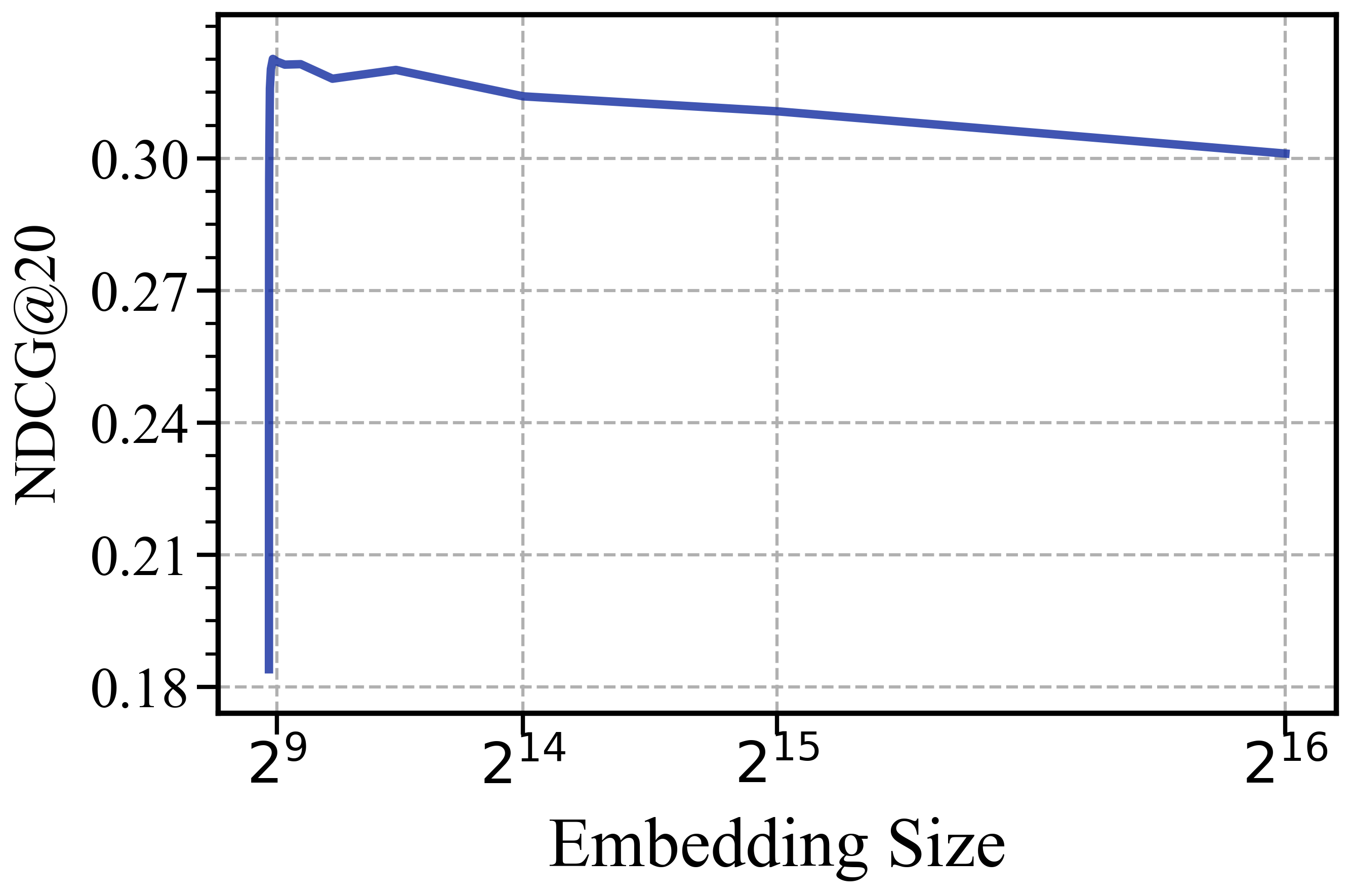}} \hfill
\subfloat[\textbf{Gowalla - SGL}]
{\includegraphics[width=0.32\linewidth]{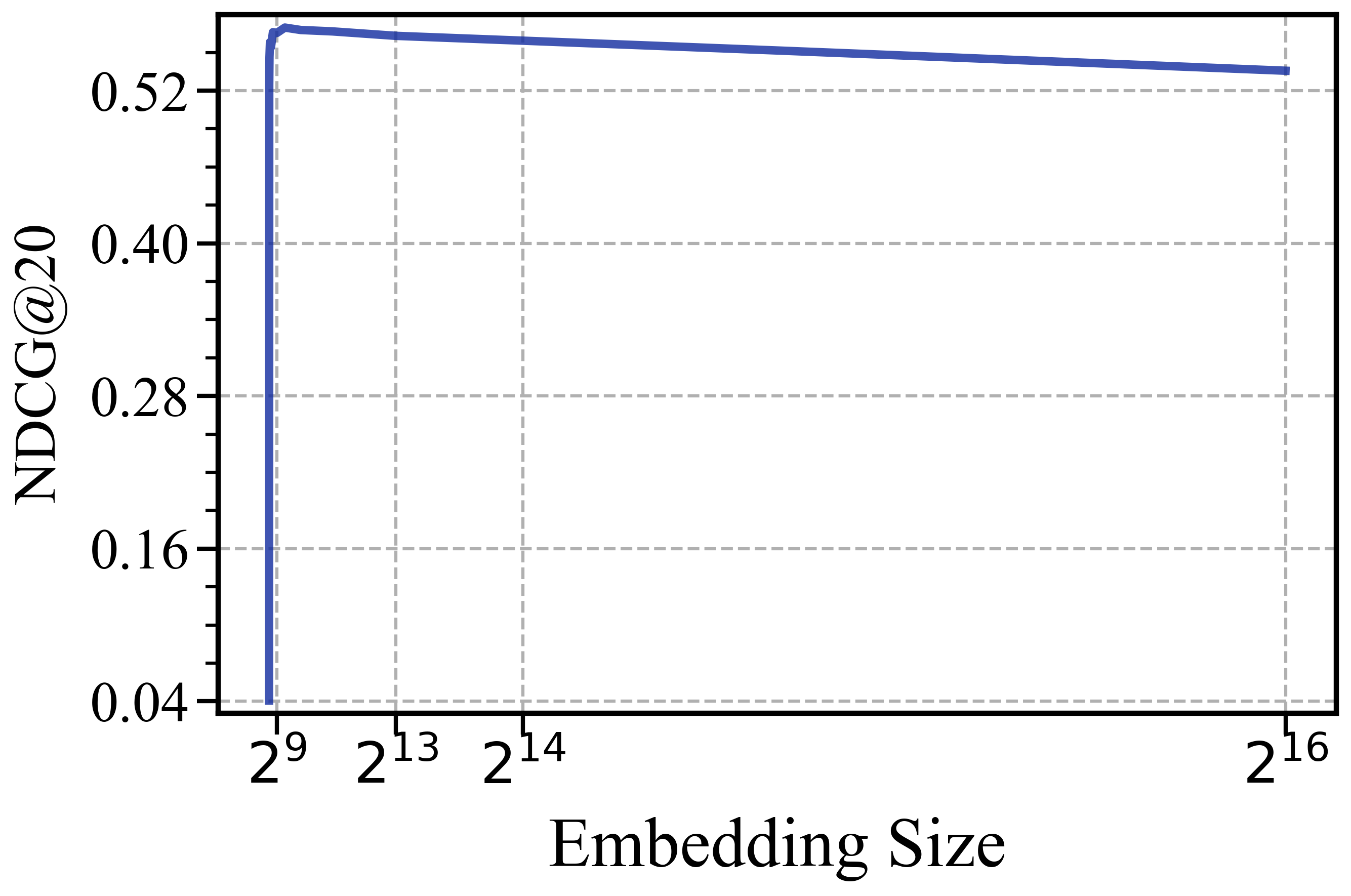}} \hfill
\subfloat[\textbf{Pinterest - SGL}]
{\includegraphics[width=0.32\linewidth]{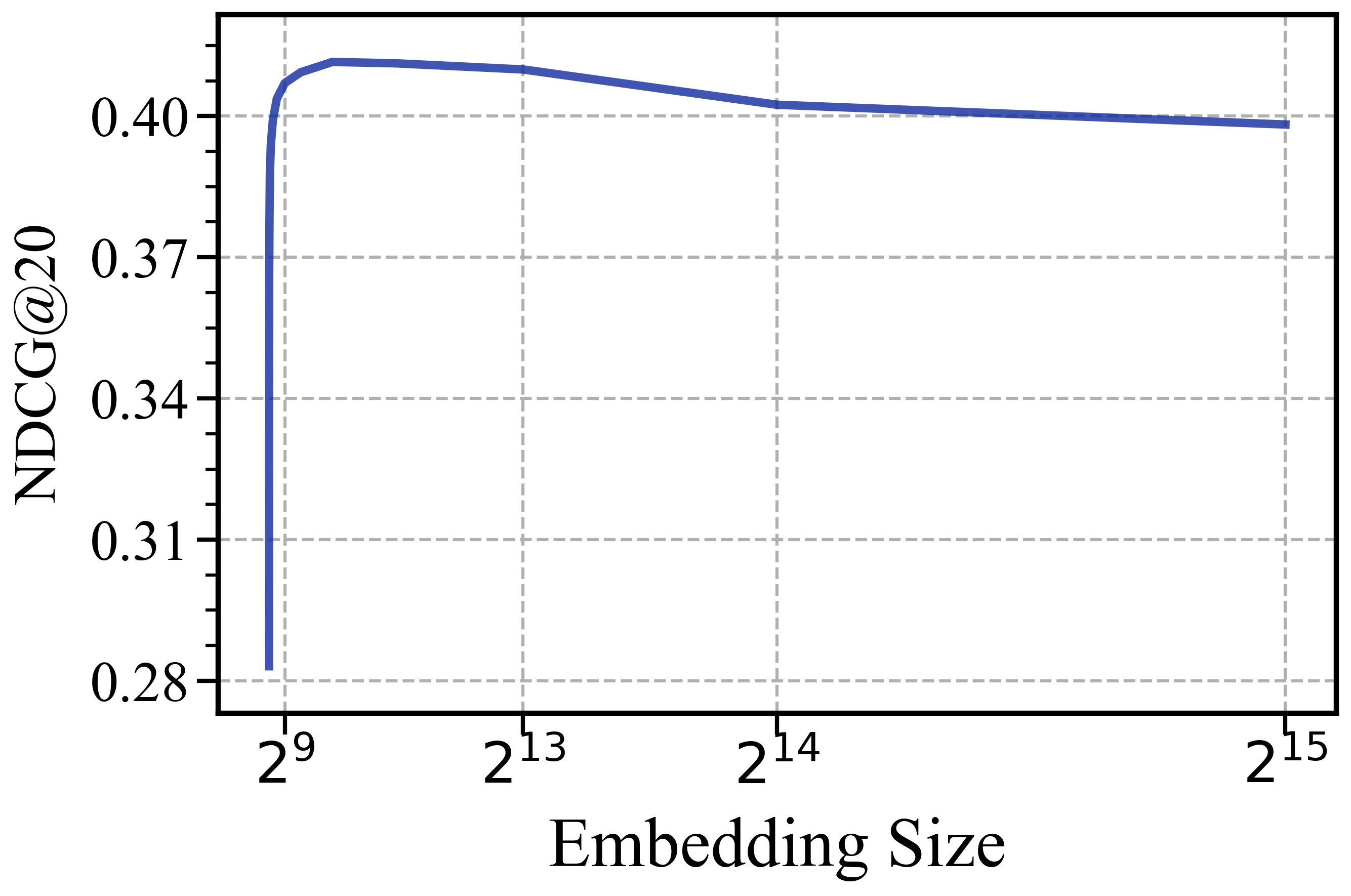}}
\vspace{5pt}
\caption{Scale the embedding dimension exponentially by a factor of 2 across different collaborative filtering models and datasets. Each row corresponds to a model (BPR, LightGCN, SGL, NeuMF), and each column represents a dataset (Yelp, Gowalla, Pinterest).}
\label{scale3}
\end{figure*}

\begin{figure*}[t!]
\centering
\subfloat[\textbf{ML-1M - BPR}]
{\includegraphics[width=0.32\linewidth]{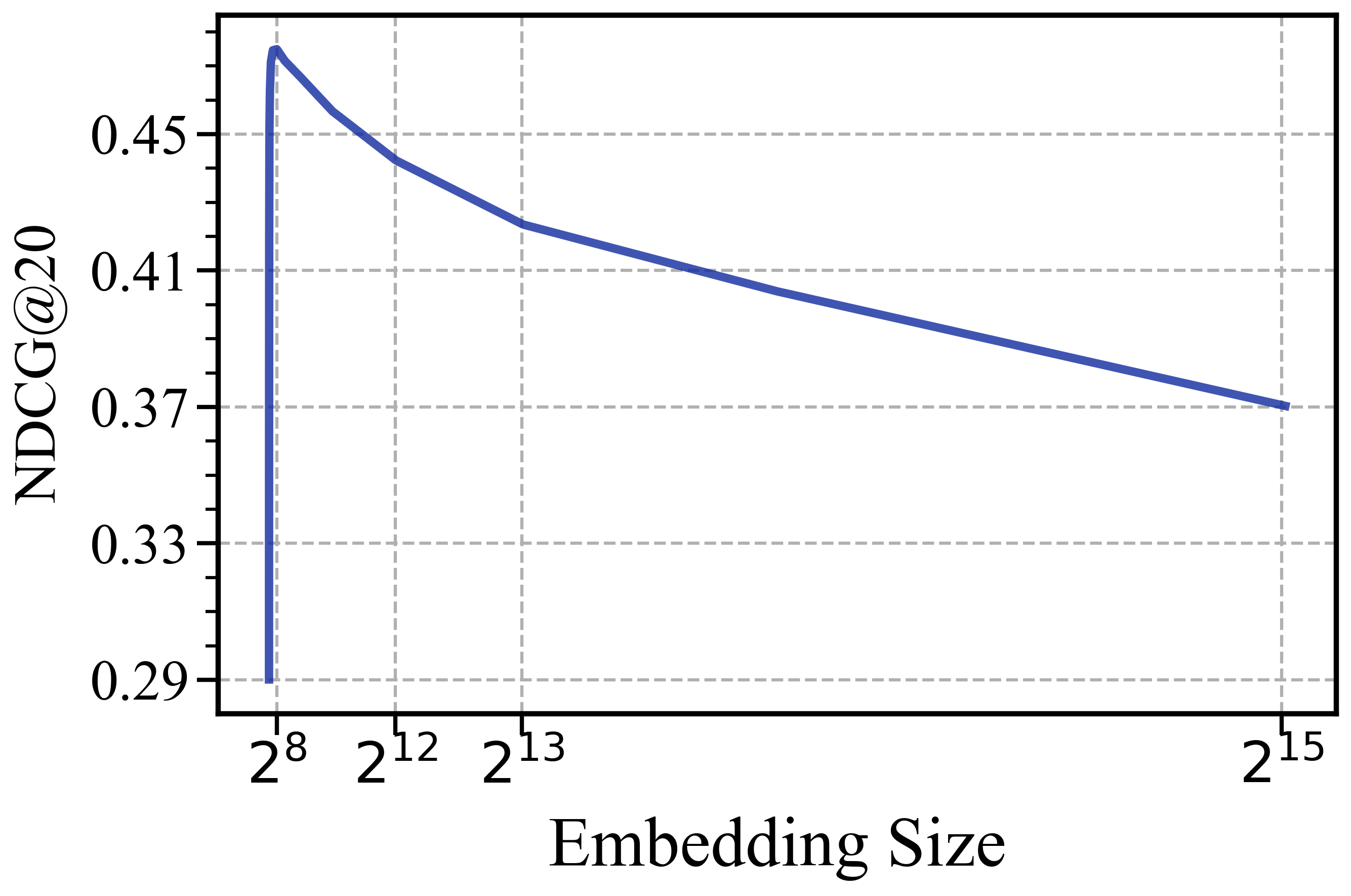}} \hfill
\subfloat[\textbf{ML-1M - LightGCN}]
{\includegraphics[width=0.32\linewidth]{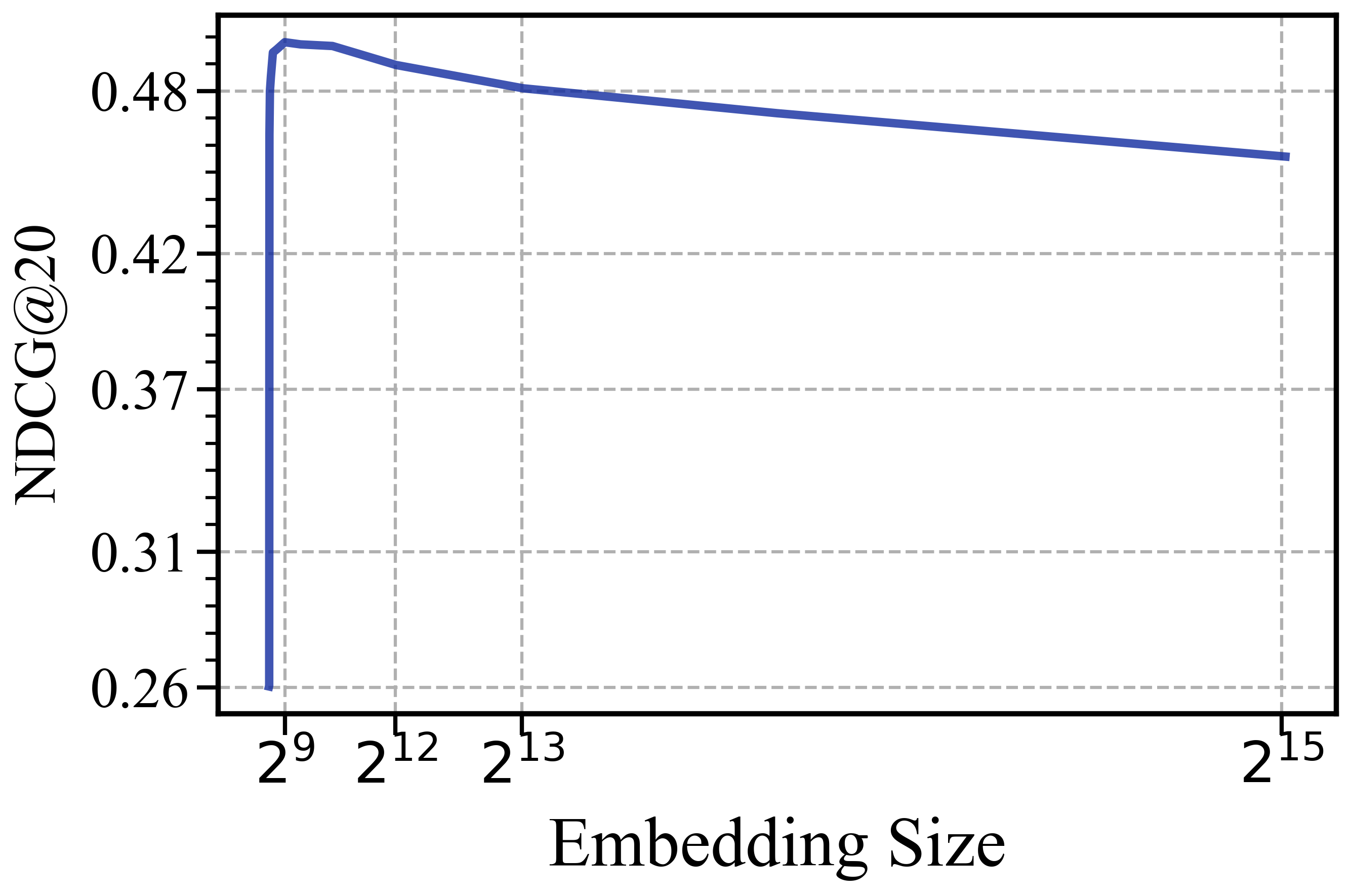}} \hfill
\subfloat[\textbf{ML-1M - SGL}]
{\includegraphics[width=0.32\linewidth]{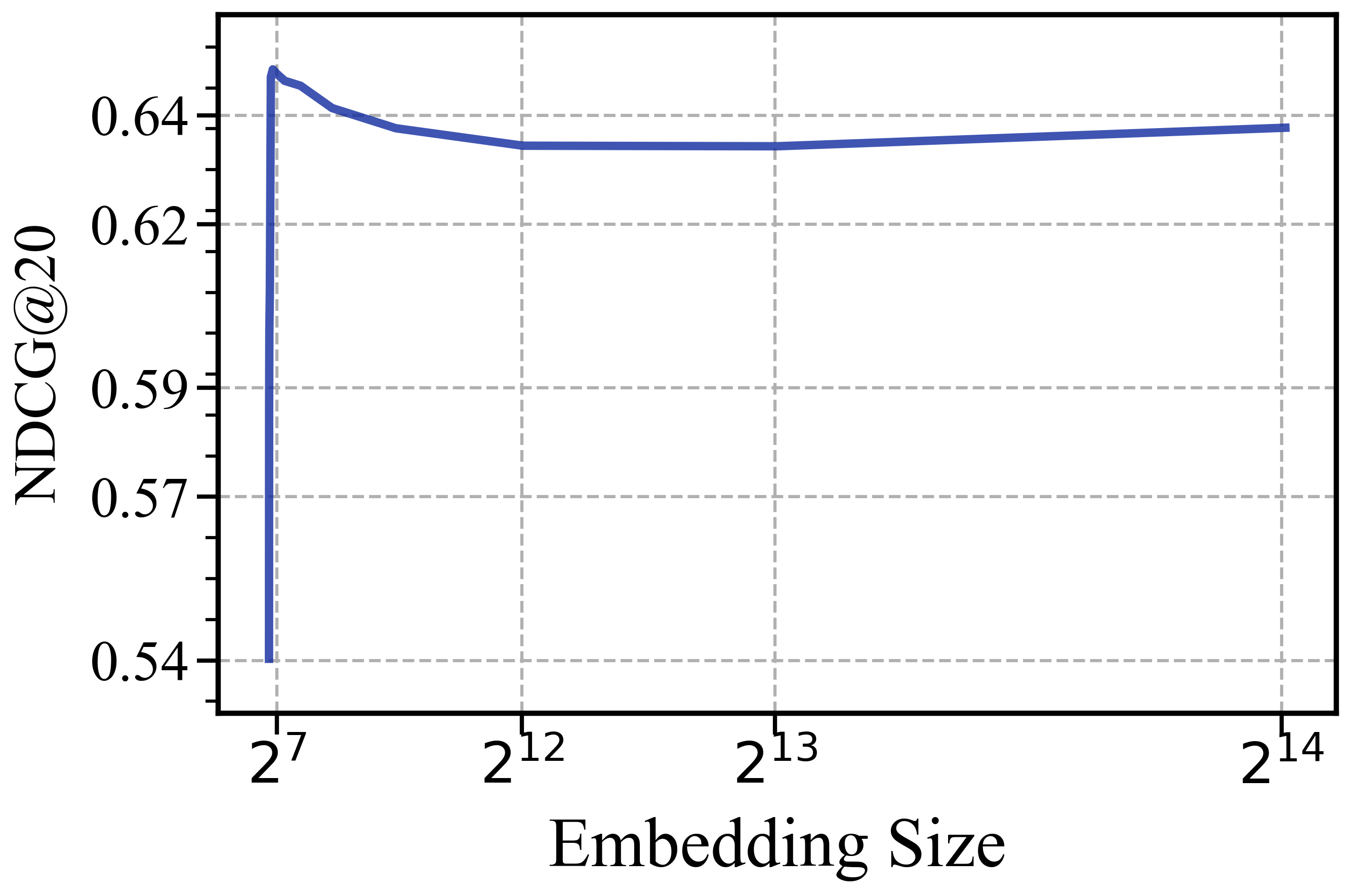}} \
\vspace{5pt}
\caption{Scale the embedding dimension exponentially by a factor of 2 across different collaborative filtering models and datasets.}
\label{scale4}
\end{figure*}

we present additional experimental results on various datasets. As shown in Figures~\ref{scale2},~\ref{scale3}, and~\ref{scale4}, SGL demonstrates a slower performance degradation trend compared to LightGCN, which in turn outperforms NeuMF, while BPR performs similarly to SGL. These results are consistent with our theoretical analysis, confirming the validity of our investigation into the noise robustness of different models. Additionally, we observed a double-peak phenomenon in other datasets as well (e.g., Figure~\ref{scale2} (b, e, h)), along with an logarithm phenomenon (e.g., Figure~\ref{scale2} (j, k, l)). Moreover, the single-peak phenomenon widely discussed in prior work is also clearly present in our experiments (e.g., Figure~\ref{scale3} (a, b, c, etc.)).
We argue that SGL’s ability to maintain or even improve performance as the embedding dimension increases is largely due to the regularization and robustness enhancement effects introduced by contrastive learning. In contrast, LightGCN lacks auxiliary training signals and regularization methods beyond supervised learning, making it more prone to overfitting or performance degradation in high-dimensional settings.
Overall, these experimental results further validate the generality of the newly observed phenomena and confirm that the observed performance trends are well aligned with our theoretical analysis.

\subsection{Effect of Varies Dropping Ratio}
\begin{figure*}[th!]
\centering
\subfloat[\textbf{ML-100K}]
{\includegraphics[width=0.45\linewidth]{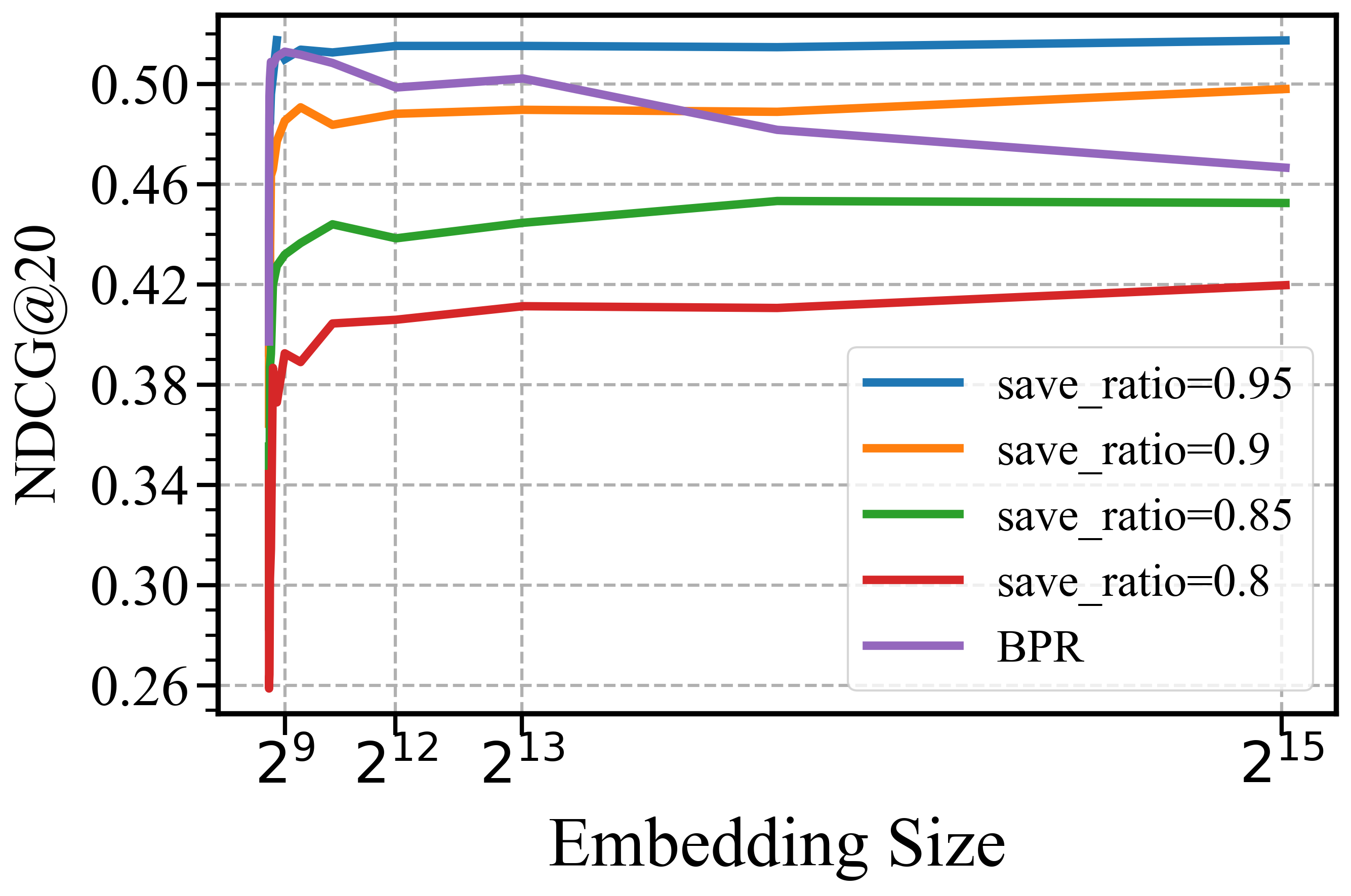}} \hspace{8pt}
\subfloat[\textbf{Douban}]{\includegraphics[width=0.45\linewidth]{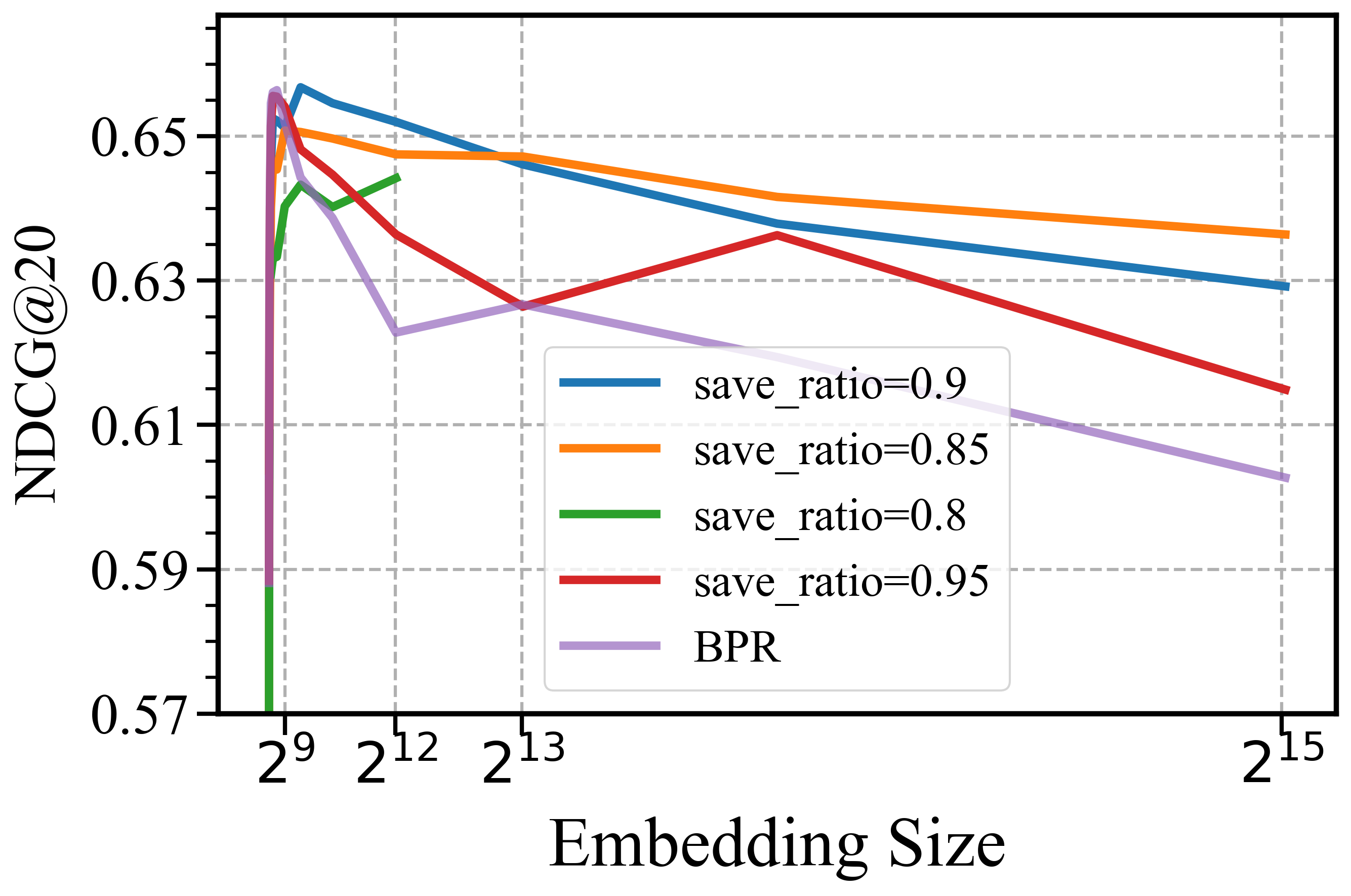}}
\caption{Comparing the standard BPR with the denoising strategy-based BPR\_Drop in ML-100K and Douban, we can clearly observe that the performance degradation has significantly improved.}
\label{scale5}
\end{figure*}

The Figures~\ref{scale5} show the NDCG@20 of BPR\_Drop with different save-ratios and standard BPR on ML - 100K and Douban datasets as embedding size varies.
In Figure (a) for ML-100K, BPR\_Drop with save - ratios 0.95, 0.9, 0.85, 0.8 and BPR are compared as embedding size ranges from $2^9$ to $2^{15}$. BPR\_Drop exhibits distinct trends, indicating sensitivity to scaling.
In Figure (b) for Douban, similar comparisons are made. BPR\_Drop's performance curves show unique patterns with embedding size changes, suggesting better scalability than standard BPR. These plots offer insights into BPR\_Drop's scaling behavior for optimization in different datasets.




%% file: main.bbl
\begin{thebibliography}{34}
\providecommand{\natexlab}[1]{#1}
\providecommand{\url}[1]{\texttt{#1}}
\expandafter\ifx\csname urlstyle\endcsname\relax
  \providecommand{\doi}[1]{doi: #1}\else
  \providecommand{\doi}{doi: \begingroup \urlstyle{rm}\Url}\fi

\bibitem[Vaswani(2017)]{transformer}
A~Vaswani.
\newblock Attention is all you need.
\newblock \emph{Advances in Neural Information Processing Systems}, 2017.

\bibitem[Guo et~al.(2024)Guo, Pan, Wang, Chen, Jiang, and Long]{lms}
Xingzhuo Guo, Junwei Pan, Ximei Wang, Baixu Chen, Jie Jiang, and Mingsheng Long.
\newblock On the embedding collapse when scaling up recommendation models.
\newblock In \emph{Forty-first International Conference on Machine Learning}, 2024.

\bibitem[Ardalani et~al.(2022)Ardalani, Wu, Chen, Bhushanam, and Aziz]{scalingrec1}
Newsha Ardalani, Carole-Jean Wu, Zeliang Chen, Bhargav Bhushanam, and Adnan Aziz.
\newblock Understanding scaling laws for recommendation models.
\newblock \emph{arXiv preprint arXiv:2208.08489}, 2022.

\bibitem[Zhang et~al.(2024)Zhang, Luo, Chen, Nie, Liu, Li, Zhao, Hao, Yao, Wen, et~al.]{wukong}
Buyun Zhang, Liang Luo, Yuxin Chen, Jade Nie, Xi~Liu, Shen Li, Yanli Zhao, Yuchen Hao, Yantao Yao, Ellie~Dingqiao Wen, et~al.
\newblock Wukong: Towards a scaling law for large-scale recommendation.
\newblock In \emph{Forty-first International Conference on Machine Learning}, 2024.

\bibitem[Ye et~al.(2025)Ye, Guo, Chin, Wang, Zhu, Lin, Ye, Liu, Tang, Lian, et~al.]{251}
Yufei Ye, Wei Guo, Jin~Yao Chin, Hao Wang, Hong Zhu, Xi~Lin, Yuyang Ye, Yong Liu, Ruiming Tang, Defu Lian, et~al.
\newblock Fuxi-$\alpha$: Scaling recommendation model with feature interaction enhanced transformer.
\newblock \emph{arXiv preprint arXiv:2502.03036}, 2025.

\bibitem[Xu et~al.(2025)Xu, Lu, Li, Hu, Wen, Chen, Zhang, and Zhang]{252}
Yi~Xu, Zhiyuan Lu, Xiaochen Li, Jinxin Hu, Hong Wen, Zulong Chen, Yu~Zhang, and Jing Zhang.
\newblock Addressing information loss and interaction collapse: A dual enhanced attention framework for feature interaction.
\newblock \emph{arXiv preprint arXiv:2503.11233}, 2025.

\bibitem[Peng et~al.(2025)Peng, Sugiyama, Liu, and Mine]{253}
Shaowen Peng, Kazunari Sugiyama, Xin Liu, and Tsunenori Mine.
\newblock Balancing embedding spectrum for recommendation.
\newblock \emph{ACM Transactions on Recommender Systems}, 3\penalty0 (4):\penalty0 1--25, 2025.

\bibitem[Liu et~al.(2020)Liu, Zhao, Wang, Liu, and Tang]{nas1}
Haochen Liu, Xiangyu Zhao, Chong Wang, Xiaobing Liu, and Jiliang Tang.
\newblock Automated embedding size search in deep recommender systems.
\newblock In \emph{Proceedings of the 43rd International ACM SIGIR Conference on Research and Development in Information Retrieval}, pages 2307--2316, 2020.

\bibitem[Qu et~al.(2023)Qu, Chen, Zhao, Cui, Zheng, and Yin]{nas2}
Yunke Qu, Tong Chen, Xiangyu Zhao, Lizhen Cui, Kai Zheng, and Hongzhi Yin.
\newblock Continuous input embedding size search for recommender systems.
\newblock In \emph{Proceedings of the 46th International ACM SIGIR Conference on Research and Development in Information Retrieval}, pages 708--717, 2023.

\bibitem[Wang et~al.(2024)Wang, Sui, Wu, Zheng, and Xiong]{dim1}
Shuyao Wang, Yongduo Sui, Jiancan Wu, Zhi Zheng, and Hui Xiong.
\newblock Dynamic sparse learning: A novel paradigm for efficient recommendation.
\newblock In \emph{Proceedings of the 17th ACM International Conference on Web Search and Data Mining}, pages 740--749, 2024.

\bibitem[Luo et~al.(2024)Luo, Wang, Zhang, Lai, Mao, Wei, Song, Tsai, Yang, Hu, et~al.]{dim2}
Qinyi Luo, Penghan Wang, Wei Zhang, Fan Lai, Jiachen Mao, Xiaohan Wei, Jun Song, Wei-Yu Tsai, Shuai Yang, Yuxi Hu, et~al.
\newblock Fine-grained embedding dimension optimization during training for recommender systems.
\newblock \emph{arXiv preprint arXiv:2401.04408}, 2024.

\bibitem[Liu et~al.(2021)Liu, Gao, Chen, Jin, and Li]{dim3}
Siyi Liu, Chen Gao, Yihong Chen, Depeng Jin, and Yong Li.
\newblock Learnable embedding sizes for recommender systems.
\newblock In \emph{International Conference on Learning Representations}, 2021.

\bibitem[Rendle et~al.(2009)Rendle, Freudenthaler, Gantner, and Schmidt-Thieme]{bpr}
Steffen Rendle, Christoph Freudenthaler, Zeno Gantner, and Lars Schmidt-Thieme.
\newblock Bpr: Bayesian personalized ranking from implicit feedback.
\newblock UAI '09, 2009.

\bibitem[Wu et~al.(2021)Wu, Wang, Feng, He, Chen, Lian, and Xie]{sgl}
Jiancan Wu, Xiang Wang, Fuli Feng, Xiangnan He, Liang Chen, Jianxun Lian, and Xing Xie.
\newblock Self-supervised graph learning for recommendation.
\newblock In \emph{Proceedings of the 44th international ACM SIGIR conference on research and development in information retrieval}, pages 726--735, 2021.

\bibitem[He et~al.(2022)He, Xie, Zhu, and Qin]{sparse_dscent}
Zheng He, Zeke Xie, Quanzhi Zhu, and Zengchang Qin.
\newblock Sparse double descent: Where network pruning aggravates overfitting.
\newblock In \emph{International Conference on Machine Learning}, pages 8635--8659. PMLR, 2022.

\bibitem[Xu et~al.(2024)Xu, Yang, Xu, Li, Liu, Shankar, Zhang, Liu, Li, Hu, et~al.]{meta}
Rengan Xu, Junjie Yang, Yifan Xu, Hong Li, Xing Liu, Devashish Shankar, Haoci Zhang, Meng Liu, Boyang Li, Yuxi Hu, et~al.
\newblock Enhancing performance and scalability of large-scale recommendation systems with jagged flash attention.
\newblock \emph{arXiv preprint arXiv:2409.15373}, 2024.

\bibitem[He et~al.(2017)He, Liao, Zhang, Nie, Hu, and Chua]{neumf}
Xiangnan He, Lizi Liao, Hanwang Zhang, Liqiang Nie, Xia Hu, and Tat-Seng Chua.
\newblock Neural collaborative filtering.
\newblock In \emph{Proceedings of the 26th international conference on world wide web}, pages 173--182, 2017.

\bibitem[He et~al.(2020)He, Deng, Wang, Li, Zhang, and Wang]{he2020lightgcn}
Xiangnan He, Kuan Deng, Xiang Wang, Yan Li, Yongdong Zhang, and Meng Wang.
\newblock Lightgcn: Simplifying and powering graph convolution network for recommendation.
\newblock In \emph{Proceedings of the 43rd International ACM SIGIR conference on research and development in Information Retrieval}, pages 639--648, 2020.

\bibitem[Xu et~al.(2023)Xu, Tian, Zhang, Zhang, Wang, Zheng, Li, Tang, Zhang, Hou, Pan, Zhao, Chen, and Wen]{recbole}
Lanling Xu, Zhen Tian, Gaowei Zhang, Junjie Zhang, Lei Wang, Bowen Zheng, Yifan Li, Jiakai Tang, Zeyu Zhang, Yupeng Hou, Xingyu Pan, Wayne~Xin Zhao, Xu~Chen, and Ji{-}Rong Wen.
\newblock Towards a more user-friendly and easy-to-use benchmark library for recommender systems.
\newblock In \emph{{SIGIR}}, pages 2837--2847. {ACM}, 2023.

\bibitem[Chen et~al.(2021)Chen, Dong, Qiu, He, Xin, Chen, Lin, and Yang]{bias2}
Jiawei Chen, Hande Dong, Yang Qiu, Xiangnan He, Xin Xin, Liang Chen, Guli Lin, and Keping Yang.
\newblock Autodebias: Learning to debias for recommendation.
\newblock In \emph{Proceedings of the 44th International ACM SIGIR Conference on Research and Development in Information Retrieval}, pages 21--30, 2021.

\bibitem[Chen et~al.(2023)Chen, Dong, Wang, Feng, Wang, and He]{bias1}
Jiawei Chen, Hande Dong, Xiang Wang, Fuli Feng, Meng Wang, and Xiangnan He.
\newblock Bias and debias in recommender system: A survey and future directions.
\newblock \emph{ACM Transactions on Information Systems}, 41\penalty0 (3):\penalty0 1--39, 2023.

\bibitem[Wang et~al.(2021)Wang, Feng, He, Nie, and Chua]{tce}
Wenjie Wang, Fuli Feng, Xiangnan He, Liqiang Nie, and Tat-Seng Chua.
\newblock Denoising implicit feedback for recommendation.
\newblock In \emph{Proceedings of the 14th ACM international conference on web search and data mining}, pages 373--381, 2021.

\bibitem[He et~al.(2024)He, Wang, Yang, Sun, Wu, Bai, Gong, Hong, and Zhang]{dcf}
Zhuangzhuang He, Yifan Wang, Yonghui Yang, Peijie Sun, Le~Wu, Haoyue Bai, Jinqi Gong, Richang Hong, and Min Zhang.
\newblock Double correction framework for denoising recommendation.
\newblock In \emph{Proceedings of the 30th ACM SIGKDD Conference on Knowledge Discovery and Data Mining}, pages 1062--1072, 2024.

\bibitem[Zhang et~al.(2017)Zhang, Bengio, Hardt, Recht, and Vinyals]{noise_first}
Chiyuan Zhang, Samy Bengio, Moritz Hardt, Benjamin Recht, and Oriol Vinyals.
\newblock Understanding deep learning requires rethinking generalization.
\newblock In \emph{International Conference on Learning Representations}, 2017.

\bibitem[Zhang et~al.(2018)Zhang, Ciss{\'{e}}, Dauphin, and Lopez{-}Paz]{mixup}
Hongyi Zhang, Moustapha Ciss{\'{e}}, Yann~N. Dauphin, and David Lopez{-}Paz.
\newblock mixup: Beyond empirical risk minimization.
\newblock \emph{International Conference on Learning Representations}, 2018.

\bibitem[Han et~al.(2024)Han, Zeng, Chen, Nie, Liu, Narang, Shakeri, Sankararaman, Jiang, Khabsa, Wang, and Hu]{gcnemixup}
Xiaotian Han, Hanqing Zeng, Yu~Chen, Shaoliang Nie, Jingzhou Liu, Kanika Narang, Zahra Shakeri, Karthik~Abinav Sankararaman, Song Jiang, Madian Khabsa, Qifan Wang, and Xia Hu.
\newblock On the equivalence of graph convolution and mixup.
\newblock \emph{Transactions on Machine Learning Research}, 2024.

\bibitem[Zhang et~al.(2023)Zhang, Hou, Lu, Chen, Zhao, and Wen]{scaling4seqrec}
Gaowei Zhang, Yupeng Hou, Hongyu Lu, Yu~Chen, Wayne~Xin Zhao, and Ji-Rong Wen.
\newblock Scaling law of large sequential recommendation models.
\newblock \emph{arXiv preprint arXiv:2311.11351}, 2023.

\bibitem[Loveland et~al.(2025)Loveland, Wu, Zhao, Koutra, Shah, and Ju]{jumingxuan}
Donald Loveland, Xinyi Wu, Tong Zhao, Danai Koutra, Neil Shah, and Mingxuan Ju.
\newblock Understanding and scaling collaborative filtering optimization from the perspective of matrix rank.
\newblock In \emph{Proceedings of the ACM on Web Conference 2025}, pages 436--449, 2025.

\bibitem[Tran et~al.(2025)Tran, Chen, Quoc Viet~Hung, Huang, Cui, and Yin]{ef25}
Hung~Vinh Tran, Tong Chen, Nguyen Quoc Viet~Hung, Zi~Huang, Lizhen Cui, and Hongzhi Yin.
\newblock A thorough performance benchmarking on lightweight embedding-based recommender systems.
\newblock \emph{ACM Transactions on Information Systems}, 2025.

\bibitem[Ginart et~al.(2021)Ginart, Naumov, Mudigere, Yang, and Zou]{mixed}
Antonio~A Ginart, Maxim Naumov, Dheevatsa Mudigere, Jiyan Yang, and James Zou.
\newblock Mixed dimension embeddings with application to memory-efficient recommendation systems.
\newblock In \emph{2021 IEEE International Symposium on Information Theory (ISIT)}, pages 2786--2791. IEEE, 2021.

\bibitem[Chen et~al.(2025)Chen, He, and Liu]{chen2025sparsemoe}
Weipu Chen, Zhuangzhuang He, and Fei Liu.
\newblock When sparsemoe meets noisy interactions: An ensemble view on denoising recommendation.
\newblock In \emph{ICASSP 2025-2025 IEEE International Conference on Acoustics, Speech and Signal Processing (ICASSP)}, pages 1--5. IEEE, 2025.

\bibitem[Yang et~al.(2024)Yang, Wu, Wang, He, Hong, and Wang]{yang2024graph}
Yonghui Yang, Le~Wu, Zihan Wang, Zhuangzhuang He, Richang Hong, and Meng Wang.
\newblock Graph bottlenecked social recommendation.
\newblock In \emph{Proceedings of the 30th ACM SIGKDD Conference on Knowledge Discovery and Data Mining}, pages 3853--3862, 2024.

\bibitem[Yang et~al.(2025)Yang, Wu, Liao, He, Shao, Hong, and Wang]{yang2025invariance}
Yonghui Yang, Le~Wu, Yuxin Liao, Zhuangzhuang He, Pengyang Shao, Richang Hong, and Meng Wang.
\newblock Invariance matters: Empowering social recommendation via graph invariant learning.
\newblock In \emph{Proceedings of the 48th International ACM SIGIR Conference on Research and Development in Information Retrieval}, pages 2038--2047, 2025.

\bibitem[Lin et~al.(2023)Lin, Zhao, Wang, Zhu, and Wang]{AutoDenoise-reinforce}
Weilin Lin, Xiangyu Zhao, Yejing Wang, Yuanshao Zhu, and Wanyu Wang.
\newblock Autodenoise: Automatic data instance denoising for recommendations.
\newblock In \emph{Proceedings of the ACM Web Conference 2023}, pages 1003--1011, 2023.

\end{thebibliography}
